\documentclass[longauth]{aa}  

\usepackage{graphicx}
\usepackage{txfonts}
\usepackage{hyperref}
\makeatletter
\renewcommand*\aa@pageof{, page \thepage{} of \pageref*{LastPage}}
\makeatother
\usepackage{booktabs}
\usepackage{amsmath}
\usepackage{adjustbox,lipsum}
\usepackage{natbib}

\usepackage{xcolor}
\usepackage{array}
\newcolumntype{P}[1]{>{\centering\arraybackslash}p{#1}}
\usepackage[caption=false]{subfig} 
\usepackage{graphicx}
\usepackage{txfonts}

\usepackage{upgreek}
\usepackage{geometry}
\usepackage{multirow}
\usepackage{amsfonts}
\usepackage{amsmath}
\usepackage{amssymb}
\usepackage{booktabs}
\usepackage{siunitx}
\usepackage{color, colortbl}
\usepackage[flushleft]{threeparttable}
\usepackage{natbib}

\usepackage{threeparttable}

\graphicspath{{./}{figures/}}

\newcommand{\pymusepipe}{\textsc{pymusepipe}}
\newcommand{\pyneb}{\textsc{pyneb}}
\newcommand{\ppxf}{\textsc{pPXF}}
\newcommand{\powerlaw}{\textsc{powerlaw}}

\newcommand{\DrSFMS}{$\Delta$\,rSFMS}
\newcommand{\DrKS}{$\Delta$\,rKS}
\newcommand{\DrMGMS}{$\Delta$\,rMGMS}

\newcommand{\rSFMS}{rSFMS}
\newcommand{\rKS}{rKS}
\newcommand{\rMGMS}{rMGMS}

\newcommand{\OIII}{\textup{[O\,\textsc{iii}]}}

\newcommand{\OI}{\textup{[O\,\textsc{i}]}}
\newcommand{\NII}{\textup{[N\,\textsc{ii}]}}
\newcommand{\SII}{\textup{[S\,\textsc{ii}]}}
\newcommand{\SIII}{\textup{[S\,\textsc{iii}]}}
\newcommand{\HII}{\textup{H\,\textsc{ii}}}
\newcommand{\Ha}{\textup{H}\ensuremath{\alpha}}
\newcommand{\Hb}{\textup{H}\ensuremath{\beta}}

\newcommand{\Rtwentyfive}{$R_{25}$}
\newcommand{\SHa}{$\Sigma_{\mathrm{H\alpha}}$}
\newcommand{\SHtwo}{$\Sigma_{\mathrm{H_{2}}}$}
\newcommand{\SMstar}{$\Sigma_{\mathrm{M_{*}}}$}
\newcommand{\Ssfr}{\ensuremath{\Sigma_{\mathrm{SFR}}}}
\newcommand{\HIIphot}{\textsc{HIIphot}}
\newcommand{\Lmin}{$L_\mathrm{min}$}

\def\kms{km~s$^{-1}$}
\def\ergs{erg~s$^{-1}$}

\begin{document}

   \title{PHANGS--MUSE: the \HII\ region luminosity function of local star-forming galaxies }
      \author{
   Francesco~Santoro\inst{\ref{mpia}}\fnmsep\thanks{santoro@mpia.de} \and 
    Kathryn~Kreckel\inst{\ref{rechen}} \and
    Francesco~Belfiore\inst{\ref{inaf}} \and
    Brent~Groves\inst{\ref{ANU}} \and 
    Enrico~Congiu\inst{\ref{uch}} \and
    David~A.~Thilker\inst{\ref{jhu}} \and
    Guillermo~A.~Blanc\inst{\ref{uch},\ref{carn}} \and
    Eva~Schinnerer\inst{\ref{mpia}} \and    
    I-Ting~Ho\inst{\ref{mpia}} \and 
%useful comments and/or data products    
    J.~M.~Diederik~Kruijssen\inst{\ref{rechen}} \and    
    Sharon~Meidt\inst{\ref{Gent}} \and
    Ralf~S.~Klessen\inst{\ref{zah},\ref{zw}} \and
    Andreas~Schruba \inst{\ref{mpe}} \and
    Miguel~Querejeta\inst{\ref{oan}} \and
    Ismael~Pessa\inst{\ref{mpia}} \and 
    M\'elanie~Chevance\inst{\ref{rechen}} \and
    Jaeyeon~Kim\inst{\ref{rechen}} \and
    Eric~Emsellem\inst{\ref{eso},\ref{lyon}} \and
    Rebecca~McElroy \inst{\ref{usyd}} \and 
%comments
    Ashley~T.~Barnes\inst{\ref{UBonn}} \and
    Frank~Bigiel\inst{\ref{UBonn}} \and
    Médéric~Boquien\inst{\ref{UA}} \and
    Daniel~A.~Dale \inst{\ref{wyo}} \and
    Simon~C.~O.~Glover\inst{\ref{zah}} \and 
    Kathryn~Grasha \inst{\ref{ANU}} \and
    Janice~Lee\inst{\ref{gem}} \and   
    Adam~K.~Leroy\inst{\ref{ohio}} \and 
    Hsi-An~Pan\inst{\ref{mpia},\ref{Tai}} \and
    Erik~Rosolowsky\inst{\ref{alb}} \and
    Toshiki~Saito\inst{\ref{cenu},\ref{NAOJ}} \and
    Patricia~Sanchez-Blazquez\inst{\ref{ucm}} \and 
    Elizabeth~J.~Watkins\inst{\ref{rechen}} \and
    Thomas~G.~Williams\inst{\ref{mpia}} 
    }
   
   \institute{
   Max-Planck-Institute for Astronomy, K\"onigstuhl 17, D-69117 Heidelberg, Germany\label{mpia} 
    \and Astronomisches Rechen-Institut, Zentrum f\"ur Astronomie der Universit\"at Heidelberg, M\"onchhofstra{\ss}e 12-14, D-69120 Heidelberg, Germany\label{rechen}
    \and INAF — Osservatorio Astrofisico di Arcetri, Largo E. Fermi 5, I-50125, Florence, Italy\label{inaf}
    \and Research School of Astronomy and Astrophysics, Australian National University, Canberra, ACT 2611, Australia\label{ANU}
    \and Departamento de Astronom\'ia, Universidad de Chile, Santiago,Chile\label{uch}
    \and Department of Physics and Astronomy, Johns Hopkins University, Baltimore, MD 21218, USA\label{jhu}
    \and Observatories of the Carnegie Institution for Science, Pasadena, CA, USA\label{carn} 
    \and Sterrenkundig Observatorium, Universiteit Gent, Krijgslaan 281 S9, B-9000 Gent, Belgium\label{Gent}
    \and Universit\"at Heidelberg, Zentrum f\"ur Astronomie, Institut f\"ur theoretische Astrophysik, Albert-Ueberle-Stra{\ss}e 2, D-69120, Heidelberg, Germany\label{zah}
    \and Universit\"at Heidelberg, Interdisziplin\"ares Zentrum f\"ur Wissenschaftliches Rechnen, Im Neuenheimer Feld 205, D-69120 Heidelberg, Germany\label{zw}
    \and Max-Planck-Institute for extraterrestrial Physics, Giessenbachstra{\ss}e 1, D-85748 Garching, Germany\label{mpe}
    \and Observatorio Astronómico Nacional (IGN), C/Alfonso XII, 3, E-28014 Madrid, Spain\label{oan} 
   \and European Southern Observatory, Karl-Schwarzschild-Stra{\ss}e 2, 85748 Garching, Germany\label{eso}
    \and Univ Lyon, Univ Lyon1, ENS de Lyon, CNRS, Centre de Recherche Astrophysique de Lyon UMR5574, F-69230 Saint-Genis-Laval France\label{lyon}
   \and Sydney Institute for Astronomy, School of Physics, Physics Road, The University of Sydney, Darlington 2006, NSW, Australia \label{usyd}
    \and Argelander-Institut f\"ur Astronomie, Universit\"at Bonn, Auf dem H\"ugel 71, D-53121 Bonn, Germany\label{UBonn}
    \and Centro de Astronomía (CITEVA), Universidad de Antofagasta, Avenida Angamos 601, Antofagasta, Chile\label{UA}
    \and Department of Physics and Astronomy, University of Wyoming, Laramie, WY 82071, USA\label{wyo}
   \and Gemini Observatory/NSF’s NOIRLab, 950 N. Cherry Avenue, Tucson, AZ, 85719, USA\label{gem}
    \and Department of Astronomy, The Ohio State University, 140 West 18th Avenue, Columbus, OH 43210, USA\label{ohio}
    \and Department of Physics, Tamkang University, No.151, Yingzhuan Rd., Tamsui Dist., New Taipei City 251301, Taiwan\label{Tai}
    \and Department of Physics, University of Alberta, Edmonton, AB T6G 2E1, Canada\label{alb}
    \and Department of Physics, General Studies, College of Engineering, Nihon University, 1 Nakagawara, Tokusada, Tamuramachi, Koriyama, Fukushima, 963-8642, Japan \label{cenu}         
    \and National Astronomical Observatory of Japan, 2-21-1 Osawa, Mitaka, Tokyo, 181-8588, Japan \label{NAOJ}
    \and Departamento de F\'isica de la Tierra y Astrof\'isica, Universidad Complutense de Madrid, E-28040 Madrid, Spain \label{ucm}
    }

    \titlerunning{PHANGS--MUSE: \HII\ region LF}
   \authorrunning{F.~Santoro et al.}
 
  \date{Received August 30, 2021; accepted November 11, 2021}

% \abstract{}{}{}{}{} 
% 5 {} token are mandatory
 
  \abstract{
  We use an unprecedented sample of about $23\,000$ \HII\ regions detected at an average physical resolution of $67$~pc in the PHANGS--MUSE sample to study the extragalactic \HII\ region \Ha\ luminosity function (LF). Our observations probe the star-forming disk of 19 nearby spiral galaxies with low inclination and located close to the star formation main sequence at $z=0$. The mean LF slope $\alpha$ in our sample is $=1.73$ with a $\sigma$ of $0.15$. We find that $\alpha$ decreases with the galaxy's star formation rate surface density \Ssfr and argue that this is driven by an enhanced clustering of young stars at high gas surface densities.  
  Looking at the \HII\ regions within single galaxies we find that no significant variations occur between the LF of the inner and outer part of the star-forming disk, whereas the LF in the spiral arm areas is shallower than in the inter-arm areas for six out of the 13 galaxies with clearly visible spiral arms. We attribute these variations to the spiral arms increasing the molecular clouds arm--inter-arm mass contrast and find suggestive evidence that they are more evident for galaxies with stronger spiral arms. 
  Furthermore, we find systematic variations in $\alpha$ between samples of \HII\ regions with high and low ionization parameter~$q$ and argue that they are driven by the aging of \HII\ regions.}
  %
%   % context heading (optional)
%   % {} leave it empty if necessary  
%   {}
%   % aims heading (mandatory)
%   {}
%   % methods heading (mandatory)
%   {}
%   % results heading (mandatory)
%   {}
%   % conclusions heading (optional), leave it empty if necessary 
%   {}


   \maketitle
%
%-------------------------------------------------------------------
\section{Introduction}\label{sec:Introduction}

\Ha\ emission is one of the most effective tracers of young stars that formed within the last $10$~Myr \citep[e.g.][]{Kennicutt2012R,Haydon2020}. In star-forming regions, it originates from massive OB-type stars which energetic (hard~UV) radiation is able to heat and ionize the surrounding gas to form an \HII\ region. 
This is why \HII\ regions have long been considered as the optimal probes of massive star formation in galaxies \citep[e.g.][]{Kennicutt1989,Thilker2000,Lawton2010}, even though we should not forget that star formation currently ongoing within dust enshrouded dense cores may remain hidden for about $3$~Myr \citep{Kim2021}.
Since the \Ha\ luminosity of an \HII\ region is directly related to the amount of ionizing radiation emitted by the OB~stars at its center, the \HII\ region luminosity function~(LF) allows us to constrain the mass function (MF) of young stellar regions. 

Similarities between the spatial distribution of stars, stellar associations and clusters on different physical scales indicate that star formation is a scale-free process \citep[e.g.][and references therein]{Efremov1998,Bastian2009,Kruijssen2012,Hopkins2013,Krumholz2014}. For this reason, the mass/\linebreak[0]{}luminosity function of star-forming regions is expected to follow a power-law with slope of about~$-2$ \citep{Elmegreen1996b}. 
Studies probing the \HII\ region LFs using different tracers largely agree with this expectation, measuring LF slopes close to $-2$ with minor variations of $\pm0.2{-}0.5$ among galaxies (e.g.\ UV: \citealp{Cook2016}, \Ha: \citealp{Kennicutt1989}, Pa$\alpha$: \citealp{Liu2013}, infrared and radio: \citealp{Mascoop2021}).
It is worth noticing that the main physical driver behind this is hierarchical growth under the influence of gravity. 

The accessibility of the \Ha\ emission line to ground-based observations favored the study of the so-called nebular LF (i.e.\ built using the \Ha\ luminosity of \HII\ regions) in numerous extra-galactic \Ha\ surveys \citep[e.g.][]{Knapen1998,Thilker2002,Kennicutt2008}.
Thanks to studies in the Milky Way~(MW) and Local Group, we know that \HII\ regions can be extremely diverse in their nature, ranging from the least bright and sub-parsec-sized regions (e.g.\ ultra-compact \HII\ regions observed within the MW; see e.g.\ \citealp{Hoare2005}) to the most bright and ${\sim}400$~pc sized superbubbles (e.g.\ resembling 30~Doradus in the Large Magellanic Cloud; see \citealp{Oey1996,Pellegrini2010}).
Studies of nebular LFs in nearby external galaxies commonly probe \HII\ regions ionized by star clusters and associations while only in a few cases does the sensitivity of the observations allow probing \HII\ regions ionized by single OB~stars \citep[e.g.][]{Azimlu2011}. 

Variations in the slope of the LF can unveil whether global properties of galaxies, as well as local parameters such as chemical abundance, dust content, gas dynamics (e.g.\ spiral arm perturbations) influence the star formation process.
Only a handful of studies have attempted to examine the relation between the LF slope and global galaxy parameters mostly finding weak or statistically insignificant correlations \citep{Kennicutt1989,Elmegreen1999,Youngblood1999,vanZee2000,Thilker2002,Liu2013,Cook2016}.
On the other hand, studies of individual star-forming galaxies with larger \HII\ region samples mainly focused on variations of the LF between the inner and outer disk and between the spiral arm and inter-arm areas. These studies showed that in the inter-arm areas and in the outer disk the \HII\ region LF becomes steeper, however such variations have not been found to be universal \citep[see, e.g.\ ][]{Cepa1990,Rand1992,Banfi1993,Knapen1993,Rozas1996,Knapen1998,Lelievre2000,Thilker2000,Helmboldt2005,Gutierrez2011,Scoville2001,Azimlu2011}. Quantifying variations of the LF slope with the galaxy's SFR, morphology, or local environmental properties can elucidate whether e.g.\ environment affects the demographics of young stellar populations and may provide new clues for our understanding of the star formation process.

New observations with integral field units (IFUs) are now allowing us to characterize the physical properties of \HII\ regions in external galaxies in greater detail, with higher sensitivity and spatial resolution, and unprecedented statistics \citep[see e.g.][]{Rousseau2018,Kreckel2019,Espinosa2020}.
The PHANGS--MUSE\footnote{\url{http://www.phangs.org}} data set \citep{Emsellem2021} is now starting to unveil its full potential in resolving and studying \HII\ region properties and connecting them to galactic structure and galaxy evolution \citep[see e.g.][]{Ho2019,Kreckel2019,Kreckel2020}. With its 19~nearby star-forming galaxies and the detection of tens of thousands of \HII\ regions at a spatial resolution of about $70$~pc, the PHANGS--MUSE data are ideal to study the \HII\ region LF, marking a turning point in terms of statistics and ability to deblend/\linebreak[0]{}resolve \HII\ regions. Furthermore, the availability of PHANGS--ALMA \citep{Leroy2021} and PHANGS--HST \citep{Lee2021} observations opens up the possibility to compare the \HII\ region LFs to the MFs of giant molecular clouds (GMCs) and young stellar regions \citep{Wei2020,Rosolowsky2021,Thilker2021,Whitmore2021} and to investigate how the demographics of substructure changes over the course of the star formation and feedback process. 

This work is focused on the study of nebular LFs and their variations as a function of galaxy global properties (e.g.\ galaxy morphology, star formation rate, stellar mass, and gas-phase metallicity) and galactic environment (e.g.\ spiral arm vs.\ inter-arm) across the entire PHANGS--MUSE sample. 
The paper is structured as follows: Section~\ref{sec:Observations and data} briefly describes the PHANGS--MUSE data set, the data reduction strategy, and the ancillary data that have been used. In Section~\ref{sec:Data analysis}, we describe the source identification technique and present our \HII\ region catalogs. In Section~\ref{sec:Results}, we present the nebular LF of our galaxies and the fitting technique (Section~\ref{sec:HII regions luminosity function}), and we investigate variations of the LF slope as a function of global properties across the sample (Section~\ref{sec:LF variations between galaxies}) or for different populations of \HII\ regions within individual galaxies (Section~\ref{sec:LF variations within galaxies}). In Section~\ref{subsec:complcrow}, we assess the effects of completeness and crowding on our results. In Section~\ref{sec:Discussion}, we discuss our results and what drives variations in the LF slope. We finally summarize our main findings in Section~\ref{sec:conclusions} and highlight future prospects in Section~\ref{sec:future prospects}.

\section{Observations and data}\label{sec:Observations and data}

For a complete and detailed description of the PHANGS--MUSE sample and data we redirect the reader to \cite{Emsellem2021}, while in this section we summarize the overall properties of the sample and provide essential details of the data reduction and analysis pipelines. In the last part of this section, we also describe the ancillary data products that we use in this work.

\subsection{PHANGS--MUSE sample}\label{sec:The PHANGS--MUSE sample}

The PHANGS--MUSE sample includes 19~nearby ($D < 20$~Mpc) star-forming galaxies (Sa--Sc Hubble morphological type) with relatively low inclination ($i \lesssim 55$~degrees). We summarize their main properties in Table~\ref{tab:Table1}. The galaxies are located close to the so-called star formation main sequence and span a stellar mass range of $\log(M_\star/M_{\odot}) = 9.4 {-} 11$. 
The galactocentric radii used in this paper have been deprojected according to the inclinations and position angles reported in Table~\ref{tab:Table1}, which are taken from \cite{Lang2020} who performed modeling of the \mbox{CO(2--1)} kinematics using the PHANGS--ALMA data.
The area covered by the MUSE observations mainly samples the star-forming disk of the galaxies. The mean maximum galactocentric radius covered across the sample is $0.86 R_{25}$, the exact coverage for each galaxy is reported in Table~\ref{tab:Table1}.

\subsection{PHANGS--MUSE data reduction and analysis}\label{sec:PHANGS--MUSE data reduction and analysis}

The PHANGS--MUSE data set \citep{Emsellem2021} is the result of an extensive (${\sim}170$~h) observational campaign (PI E.\,Schinnerer) with the MUSE IFU \citep{Bacon2010} mounted on the Unit 4 telescope (UT4) at ESO's VLT, with the addition of archival observations for \object{NGC0628} and the centers of 5~galaxies in the sample. Each galaxy has been covered by several pointings (from~3 up to~15) to obtain a contiguous mosaic of the star-forming disk. All the observations have been performed in wide-field mode (WFM with $1\arcmin \times 1\arcmin$ field of view), either in natural seeing (WFM-noAO, for 11~galaxies) or taking advantage of the GALACSI adaptive optics system (WFM-AO, for eight galaxies). 

The data have been reduced using a version of the \pymusepipe\ python package,\footnote{Available from https://github.com/emsellem/pymusepipe} a wrapper of the \textsc{esorex} MUSE reduction recipes, tailored to the PHANGS--MUSE program \citep{Emsellem2021}.
Broad-band images taken with the ESO/MPG 2.2m Wide Field Imager and with the Du~Pont DirectCCD camera (Razza et al. in preparation) were used as a reference to astrometrically align and flux calibrate the MUSE data.
The complete data reduction provides a sky-subtracted, flux-calibrated, mosaicked data cube for each galaxy with a spaxel size of $0.2\arcsec$.  
The point spread function (PSF) of the observations was determined for each pointing, and for each galaxy the largest value (reported in Table~\ref{tab:Table1}) was used to create PSF-homogenized data cubes. We refer to these as the \textit{copt} data cubes to distinguish them from the \textit{native} data cubes maintaining the native resolution of the observations.   

The data analysis of the mosaicked data cubes has been performed with a data analysis pipeline (DAP) based on the \textsc{gist} (Galaxy IFU Spectroscopy Tool; \citealt{Bittner2019}) software package and tailored to the PHANGS--MUSE program which provides maps of stellar kinematics, stellar population properties (including stellar ages and mass surface density), and emission line fluxes and kinematics \citep{Emsellem2021}.
The DAP employs the penalized pixel fitting method via the \ppxf\ package \citep{Cappellari2017} to extract information on the stellar population and the ionized gas from the MUSE spectra over the wavelength range $4850{-}7000$~\r{A}\ taking into account the MUSE spectral resolution as parametrized in \cite{Bacon2017}.
First, the stellar kinematics and the stellar population information are extracted in two subsequent steps using binned spectra with a continuum S/N target of~35, and \mbox{E-MILES} simple stellar population models \citep{Vazdekis2016} generated with a \cite{Chabrier2003} initial mass function, BaSTI isochrones \citep{Pietrinferni2004}, eight ages ($0.15{-}14$~Gyr), and four metallicities ($\textrm{[Z/H]} = [-1.5, -0.35, 0.06, 0.4]$). 
Finally, the ionized gas analysis is performed by leveraging the previous steps and simultaneously fitting the stellar continuum and the emission lines using single spaxel spectra. Emission lines are modeled with a single Gaussian function and emission line flux maps are corrected for foreground Galactic extinction, using the \cite{ODonnell1994} extinction law and the ${E(B-V)}$ values from \cite{Schlafly2011}. We refer the reader to the PHANGS-MUSE survey paper \citep{Emsellem2021} for further details on the choice of models and the spectral fitting procedure.
In this paper, we will make use of the \textit{copt} \Ha\ maps to localize \HII\ regions and the \textit{native} MUSE mosaicked data cubes to extract their integrated spectra (see Section~\ref{sec:Data analysis}).

\subsection{Environmental and foreground stars masks}\label{sec:Environmental and foreground stars masks}

Morphological masks delimiting stellar structures by using \textit{Spitzer} NIR $3.6~\mu$m imaging are available for the PHANGS--MUSE galaxies from \cite{Querejeta2021}. We use these masks to define galaxy centers, bars, spiral arms, and inter-arms regions.
In the work of \cite{Querejeta2021}, spiral arms have been defined only when they are dominant features across the galaxy disk (omitting flocculent galaxies) by fitting a log-spiral function to bright regions in the NIR images along each spiral arm. These modeled log-spiral curves are then assigned an empirically determined width based on the overlap with \mbox{CO(2--1)} emission from PHANGS--ALMA data and as a last step the starting and ending azimuth of each spiral segment is adjusted to define the final spiral arm footprint. The outer edges of bars have been defined as ellipses where the bar length, ellipticity, and position angle come from a compilation of measurements from the literature based on NIR imaging \citep[mostly from][]{Herrera-Endoqui2015}.

We make use of the \cite{Querejeta2021} environmental masks to define five environments -- namely galaxy center, bar, spiral arm, inter-arm, and disk -- according to the criteria described below. 
Our galaxy ``centers'' include small bulges, inner star-forming rings or nuclei. We define ``bars'' in the same way they are defined in the \cite{Querejeta2021} masks. For the galaxies with spiral arms, we define as ``spiral arms'' the regions flagged as spiral arms with the addition of the bar ends regions (when spiral arms overlap with bar ends). All remaining regions are defined as ``inter-arm'' for the galaxies with spiral arms and ``disc'' for the galaxies without spiral arms.
These five environments can be visualized for \object{NGC4321} in Fig.~\ref{Fig:Map} and for the remaining galaxies in Appendix~\ref{appendix1}, in Fig.~\ref{Fig:Map} we also show a sketch of the different environments to guide the reader.

There are a number of foreground stars across the field of view (FoV) of our observations which need to be excluded from our analysis. Masks of foreground stars for the PHANGS--MUSE galaxies are described in \cite{Emsellem2021}. They have been generated by matching the positions of \textit{Gaia} point sources. To avoid masking compact \HII\ regions and galactic nuclei, which may be included as point sources in the \textit{Gaia} catalog, a further check was performed to identify the rest-frame Ca\,\textsc{ii} triplet absorption features at $8498$, $8542$, and $8662$~\r{A}\ in the MUSE spectra. 

\begin{table*}
    \caption{Main properties of the galaxies in the PHANGS--MUSE sample.}
    \centering
    \adjustbox{max width=\textwidth}{%
        \begin{tabular}{cccccccccccccc}
\toprule
Galaxy & RA & Dec & Morph$^{1}$ & Distance$^{2}$ & $i^{3}$ & PA$^{3}$ & Log\,$M_{*}^{4}$ & SFR$^{4}$ & $\Delta_\mathrm{SFMS}^{4}$ & $R_{25}^{1}$ & $\mathrm{FWHM}_\mathrm{PSF}^{5}$ & pc/$''$ & r$_\mathrm{max}$ \\
 & ${}^{\circ}$ & ${}^{\circ}$ &  & Mpc & ${}^{\circ}$ & $\mathrm{{}^{\circ}}$ & $M_{\odot}$ & $M_{\odot}~\mathrm{yr}^{-1}$ &  & $''$ & $''$ &  &  \\
\toprule
IC5332 & 353.61453 & -36.10108 & SABc & 9.01 & 26.9 & 74.4 & 9.67 & 0.41 & 0.01 & 182 & 0.87 & 43.6 & 0.68 \\
NGC0628 & 24.173855 & 15.783643 & Sc & 9.84 & 8.9 & 20.7 & 10.34 & 1.75 & 0.18 & 297 & 0.92 & 47.6 & 0.51 \\
NGC1087 & 41.60492 & -0.498717 & Sc & 15.85 & 42.9 & 359.1 & 9.93 & 1.31 & 0.33 & 89 & 0.92 & 76.5 & 1.37 \\
NGC1300 & 49.920815 & -19.411114 & Sbc & 18.99 & 31.8 & 278.0 & 10.62 & 1.17 & -0.18 & 178 & 0.89 & 91.6 & 0.96 \\
NGC1365 & 53.40152 & -36.140404 & Sb & 19.57 & 55.4 & 201.1 & 10.99 & 16.90 & 0.72 & 361 & 1.15 & 94.4 & 0.73 \\
NGC1385 & 54.369015 & -24.501162 & Sc & 17.22 & 44.0 & 181.3 & 9.98 & 2.09 & 0.50 & 102 & 0.67 & 83.1 & 1.21 \\
NGC1433 & 55.506195 & -47.221943 & SBa & 18.63 & 28.6 & 199.7 & 10.87 & 1.13 & -0.36 & 186 & 0.91 & 89.9 & 1.05 \\
NGC1512 & 60.975574 & -43.348724 & Sa & 18.83 & 42.5 & 261.9 & 10.71 & 1.28 & -0.21 & 253 & 1.25 & 90.8 & 0.62 \\
NGC1566 & 65.00159 & -54.93801 & SABb & 17.69 & 29.5 & 214.7 & 10.78 & 4.54 & 0.29 & 217 & 0.8 & 85.3 & 0.73 \\
NGC1672 & 71.42704 & -59.247257 & Sb & 19.4 & 42.6 & 134.3 & 10.73 & 7.60 & 0.56 & 185 & 0.96 & 93.5 & 1.00 \\
NGC2835 & 139.47044 & -22.35468 & Sc & 12.22 & 41.3 & 1.0 & 10.00 & 1.24 & 0.26 & 192 & 1.15 & 59.0 & 0.69 \\
NGC3351 & 160.99065 & 11.70367 & Sb & 9.96 & 45.1 & 193.2 & 10.36 & 1.32 & 0.05 & 217 & 1.05 & 48.2 & 0.76 \\
NGC3627 & 170.06252 & 12.9915 & Sb & 11.32 & 57.3 & 173.1 & 10.83 & 3.84 & 0.19 & 308 & 1.05 & 54.7 & 0.66 \\
NGC4254 & 184.7068 & 14.416412 & Sc & 13.1 & 34.4 & 68.1 & 10.42 & 3.07 & 0.37 & 151 & 0.89 & 63.3 & 1.28 \\
NGC4303 & 185.47888 & 4.473744 & Sbc & 16.99 & 23.5 & 312.4 & 10.52 & 5.33 & 0.54 & 207 & 0.78 & 82.0 & 0.66 \\
NGC4321 & 185.72887 & 15.822304 & SABb & 15.21 & 38.5 & 156.2 & 10.75 & 3.56 & 0.21 & 183 & 1.16 & 73.4 & 0.97 \\
NGC4535 & 188.5846 & 8.197973 & Sc & 15.77 & 44.7 & 179.7 & 10.53 & 2.16 & 0.14 & 244 & 0.56 & 76.1 & 0.56 \\
NGC5068 & 199.72807 & -21.038744 & Sc & 5.2 & 35.7 & 342.4 & 9.40 & 0.28 & 0.02 & 224 & 1.04 & 25.2 & 0.89 \\
NGC7496 & 347.44702 & -43.42785 & Sb & 18.72 & 35.9 & 193.7 & 10.00 & 2.26 & 0.53 & 100 & 0.89 & 90.3 & 1.06 \\
\bottomrule
\end{tabular}

    }
    \tablebib{
            $^{1}$ HyperLEDA, \citet{Makarov2014},
            $^{2}$ \citet{Anand2021},
            $^{3}$ \citet{Lang2020},
            $^{4}$ \citet{Leroy2021},
            $^{5}$ \citet{Emsellem2021}
        }

    \tablefoot{%
         The table reports the galaxy name (col\,1), galaxy center sky coordinates (col\,2--3), Hubble morphological type (col\,4), distance (col\,5), inclination (col\,6), position angle (col\,7), total stellar mass (col\,8), total star formation rate (col\,9), offset from the star formation main sequence (col\,10), \Rtwentyfive\ $B$-band isophotal radius  (col\,11), angular resolution of the \textit{copt} data (col\,12), the parsec per arcsec scaling factor (col\,13), and the maximum deprojected galactocentric radius in units of \Rtwentyfive\ (col\,14).
        \label{tab:Table1}
    }
\end{table*}

\section{Data analysis}\label{sec:Data analysis}

\subsection{Ionized nebulae identification}\label{sec:Ionized nebulae identification}

To isolate \HII\ regions across our sample, we make use of the PHANGS--MUSE \textit{copt} \Ha\ emission line maps provided by the DAP procedure summarized in Sec.~\ref{sec:PHANGS--MUSE data reduction and analysis}. We chose the \textit{copt} rather than the \textit{native} resolution maps to mitigate the effect of PSF spatial variations across the FoV of single targets.
Ionized nebulae are identified using an adaptation of the \HIIphot\ software \citep{Thilker2000} originally implemented to work with \Ha\ narrow-band images. The software first detects what are called ``seed regions'' and then grows them up to a given contrast (i.e.\ the termination gradient) that is selected by the user.
We set the background level of the \Ha\ maps to the mean \Ha\ flux of the pixels with surface brightness below $\Sigma_\mathrm{H\alpha} < 1\!\times\!10^{-17}$ erg~s$^{-1}$ arcsec$^{-2}$ and the detection threshold to $3\sigma$ above the background, with $\sigma$ being the standard deviation of the background pixels.  
Once the seed regions have been identified, we perform further cleaning based on visual inspection to avoid artifacts due to noise in the \Ha\ map. We consider the total integrated flux of the seed region and impose a S/N cut of~$50$ (using the \Ha\ emission line error maps to assess the noise) and a $\Sigma_\mathrm{H\alpha}$ cut above $3\sigma$ of the background.
To avoid the detection of regions with unphysical sizes, we find that the best solution is to limit the spatial smoothing operated by \HIIphot\ on the input maps for the iterative search of seed regions to three iterations, each time increasing the starting PSF by $10$\%.  

Once the seed regions have been identified, \HIIphot\ starts to grow them.
Different methods have been used in the literature to define \HII\ region boundaries: emission line ratios, \Ha\ equivalent width, \Ha\ spatial gradients, or a combination thereof \citep[see e.g.][]{Blanc2009, Thilker2000, Sanchez2012, Zhang2017}.  
In this work, we rely on \HIIphot, which uses a user-selected termination gradient of the \Ha\ surface brightness in units of emission measure~(EM), to stop the growth of seed regions -- with a lower termination gradient the seed regions are allowed to grow more and vice versa \citep[see][for an example]{Thilker2000}.
In principle, in the classical scenario of an \HII\ region represented as a Str{\"o}mgren sphere, closer galaxies and/or better seeing conditions lead to the detection of steeper gradients and vice versa \citep[see e.g.][]{Oey2007}. Considering the distances of the galaxies in our sample ($D < 20$~Mpc) and the seeing of our observations ($\mathrm{FWHM}_\mathrm{PSF} \leq 1.2\arcsec$), after visual inspection we select a termination gradient of $2.43\!\times\!10^{-16}$ erg~s$^{-1}$ arcsec$^{-2}$~pc$^{-1}$ (i.e.\ corresponding to $5$~EM~pc$^{-1}$) to also ensure coherence with results presented in the literature \citep[see][]{Oey2007}.
In general a higher/\linebreak[0]{}lower termination gradient results in a steeper/\linebreak[0]{}shallower LF. We checked that varying the termination gradient in the interval $1.5{-}10$~EM~pc$^{-1}$ gives LF with slopes that are, within the errors, fully compatible with what we discuss in Sec.~\ref{sec:HII regions luminosity function} of this paper.  
Excluding the regions overlapping with foreground stars (see Sec.~\ref{sec:Environmental and foreground stars masks}), we detect a total number of $31\,399$ ionized nebulae across the 19~galaxies of the PHANGS--MUSE sample. We use the nebulae spatial masks for each galaxy to extract the nebula integrated spectra from the MUSE \textit{native} mosaicked data cubes. These spectra are then fitted using the DAP of the PHANGS--MUSE data as described in Sec.~\ref{sec:PHANGS--MUSE data reduction and analysis}. The only difference at this level is that we fit the MUSE spectra across their entire wavelength range so to include also the modeling of the \SIII$\lambda9069$\r{A} emission line.
It is worth noting that in this work we do not correct the emission line fluxes for the diffuse background (i.e. the emission of the diffuse ionized gas DIG component permeating star-forming disks, see \citealp{Haffner2009}). However, we checked that such a correction starts to be more relevant only for the \Ha\ emission of the fainter \HII\ regions and, therefore, does not affect the LF slopes presented in Sect.~\ref{sec:Results}. Depending on the galaxy, the \Ha\ emission outside the ionized nebulae footprints can represents $20{-}50\%$ of the total \Ha\ emission \citep[see][]{Belfiore2021}.

\subsection{The final \texorpdfstring{\HII}{HII} region catalog}\label{sec:The final HII region catalogs}

\begin{figure*}
\begin{center}
\includegraphics[width=1.0\textwidth]{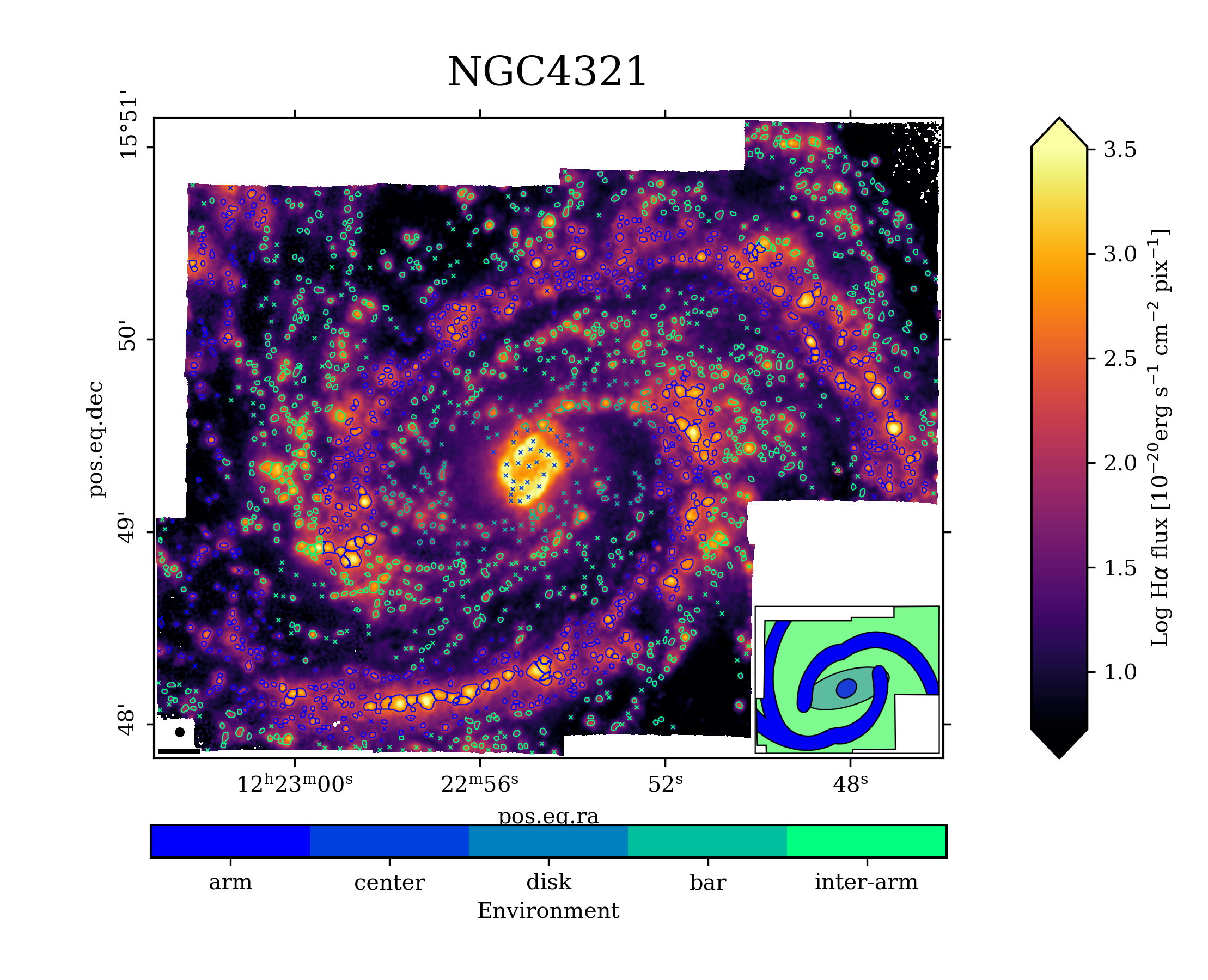}
\caption{\HII\ regions and environments for \object{NGC4321}. The figure shows the H$\alpha$ emission in the background, color coded according to the color scheme on the right, with overlaid the borders of the \HII\ regions in our catalog. The centers of the nebulae which have been discarded by our selection criteria are marked with crosses. In the lower left corner, the black circle indicates the PSF of the MUSE observations while the black line marks a physical scale corresponding to $1$~kpc. Both the \HII\ regions and the discarded nebulae are color coded according to our definition of environments as outlined by the color scheme at the bottom and shown in the bottom-right sketch.
\label{Fig:Map}}
\end{center}
\end{figure*}

In order to obtain our fiducial \HII\ regions catalog we exclude:
\begin{itemize}
    \item Nebulae located in galaxy centers (as defined in Sec.~\ref{sec:Environmental and foreground stars masks}); this is meant to avoid areas with a high SFR surface density (e.g.\ starburst rings in \object{NGC1300}, \object{NGC1512}, \object{NGC1672}, \object{NGC3351}, \object{NGC4303}, and \object{NGC4321}), where the deblending of \HII\ regions is not possible at our resolution \citep[i.e.\ typically few parsec in radius; see e.g.][]{Barnes2020}, and contamination by active galactic nuclei (AGN) (i.e.\ \object{NGC1365}, \object{NGC1512}, \object{NGC1566}, and \object{NGC1672} show evidence of hosting low-luminosity AGN when looking at emission line maps). This cut comprises about $1.5$\% of the nebulae. 
    
    \item Nebulae located outside the ``\HII\ region'' area in at least one of the classical Baldwin--Phillips--Terlevich (BPT) diagrams \citep{Baldwin1981}; more specifically, these are the regions located above the \cite{Kauffmann2003} line in the \OIII /\Hb\ vs.\ \NII /\Ha\ diagnostic, and above the \cite{Kewley2001} line in the \OIII /\Hb\ vs.\ \SII /\Ha, as well as the \OIII /\Hb\ vs.\ \OI /\Ha\ diagnostic. To build the BPT diagrams we require a $\mathrm{S/N} > 3$ for the \Hb, \OIII$\lambda5007$\r{A}, \OI$\lambda6300$\r{A}, \Ha, \NII$\lambda6584$\r{A}, \SII$\lambda6716$\r{A}, and \SII$\lambda6731$\r{A} emission lines; this is meant to ensure a robust estimate of both the dust correction based on the \Ha/\Hb\ Balmer decrement and the quantities relying on line ratios (e.g.\ ionization parameter and gas-phase metallicity) used in this work. This cut comprises about $23$\% of the nebulae.
    \item Nebulae with velocity dispersion $\sigma > 400$~\kms\ for all three emission line groups (i.e.\ hydrogen, high ionization and low ionization lines) fitted separately by the DAP \citep[see][]{Emsellem2021} in order to avoid cases where the emission line fitting performs poorly and to exclude objects that are potentially supernovae remnant (SNR) candidates. This cut comprises about $1$\% of the nebulae.
    \item Regions whose geometric center matches within $0.5\arcsec$ the positions of planetary nebulae (PNe) identified using \OIII\ emission line maps (Scheuermann et al. in preparation). This cut comprises about $0.6$\% of the nebulae.
    \item Regions where the distance of the
    geometric center from the edges of the MUSE mosaic FoV is less than one $\mathrm{FWHM}_\mathrm{PSF}$. This cut comprises about $2$\% of the nebulae.
\end{itemize}

We note that the S/N of the \Ha\ and \Hb\ lines is usually high, more specifically, $99$\% of the region spectra have a $\mathrm{S/N} \gtrsim 32$ for the \Ha\ line and ${\gtrsim} 12$ for the \Hb\ line. All together, the cuts remove about $26$\% of the original nebulae and leave us with a final joined catalog of $23\,301$ \HII\ regions.
A discussion on how selection effects may influence our results is presented in Appendix~\ref{appendix3}. 
 
In Fig.~\ref{Fig:Map}, we show the spatial masks of the \HII\ regions found in \object{NGC4321}, highlighting also the different galactic environments defined in Sec.~\ref{sec:Environmental and foreground stars masks}. Analogous figures for the remaining galaxies in our sample are presented in Appendix~\ref{appendix1}. 

We assume a screen geometry and use \pyneb\footnote{https://pypi.org/project/PyNeb/} \citep{Luridiana2015} to correct line fluxes for dust extinction via the H$\alpha$/H$\beta$ ratio, adopting the \cite{ODonnell1994} reddening law with $R_V = 3.1$ and a theoretical H$\alpha$/H$\beta = 2.86$. The extinction corrected emission line luminosities of the \HII\ regions are then computed using the distances reported in Table~\ref{tab:Table1}.
For every \HII\ region in our catalog, we also estimate the gas-phase metallicity and the gas ionization parameter by using extinction corrected emission line fluxes. The gas-phase metallicity $\mathrm{O/H}$ is calculated using the \cite{Pilyugin2016} S\nobreakdash-\hspace{0pt}calibration (Scal hereafter). This calibration relies on three diagnostic line ratios (i.e.\ \NII/\Hb, \SII/\Hb, and \OIII/\Hb) and provides an empirical calibration against \HII\ regions that have direct constraints on their nebular temperatures and hence their abundances. This calibration is relatively insensitive to changes in gas pressure or ionization parameter and we adopted it as fiducial approach in this paper \citep[see][for a discussion]{Kreckel2019}. In addition, for each galaxy we fit the radial metallicity gradient by using an unweighted least-square linear fitting of the trend between $12+\log(\mathrm{O/H})$ and the deprojected galactocentric radius (see Fig.~\ref{fig:met_radial_gradients} in Appendix~\ref{appendix1}).
The gas ionization parameter represents the ratio between the ionizing photon flux density and the gas hydrogen density. In this paper, we express the ionization parameter as $q = U \times c = Q(H^0)/4\pi R^{2}n$, where $c$ is the speed of light, $U$ is the dimensionless ionization parameter, $Q(H^0)$ is the number of hydrogen ionizing photons ($E > 13.6$~eV) emitted per second, $R$ is the empty (wind-blown) radius of the \HII\ region, and $n$ its hydrogen density. The ionization parameter is ultimately defined by the structure of an \HII\ region (e.g.\ size, gas density, filling factor) and the properties of its ionizing source. Photoionization models show that it can be extracted via different diagnostic lines \citep{Kewley2002,Dors2011}.
In this paper, we use the calibration proposed by \cite{Diaz1991} based on the \SIII(9069+9532)/\SII(6717+6713) line ratio. As the \SIII$\lambda9532$\r{A} line falls outside the wavelength range covered by MUSE, we assume that $\text{\SIII}\lambda9532$\r{A}$ = 2.47 \text{\SIII}\lambda9069$\r{A} according to default atomic data in \pyneb\ \citep{Luridiana2015}. It should be noted that for about $3000$ \HII\ regions we are not able to estimate the ionization parameter due to lack of detection of the \SIII$\lambda9532$\r{A} emission line. 

The \HII\ regions catalog containing the host galaxy name, the sky coordinates of the geometrical center, the deprojected galactocentric distance from the host galaxy center, the \Ha\ and \Hb\ observed fluxes, the \Ha\ extinction corrected flux, the ionization parameter, and an environmental flag for all the \HII\ regions discussed in the current paper is only available in electronic form at the CDS via anonymous ftp to \url{cdsarc.u-strasbg.fr} (130.79.128.5) or via \url{http://cdsweb.u-strasbg.fr/cgi-bin/qcat?J/A+A/}.

\section{Results}\label{sec:Results}

\subsection{The \texorpdfstring{\HII}{HII} region luminosity function}\label{sec:HII regions luminosity function}

\begin{figure*}[ht]
    \centering
    \includegraphics[trim=3cm 3cm 3cm 3cm, clip, width=1\textwidth]{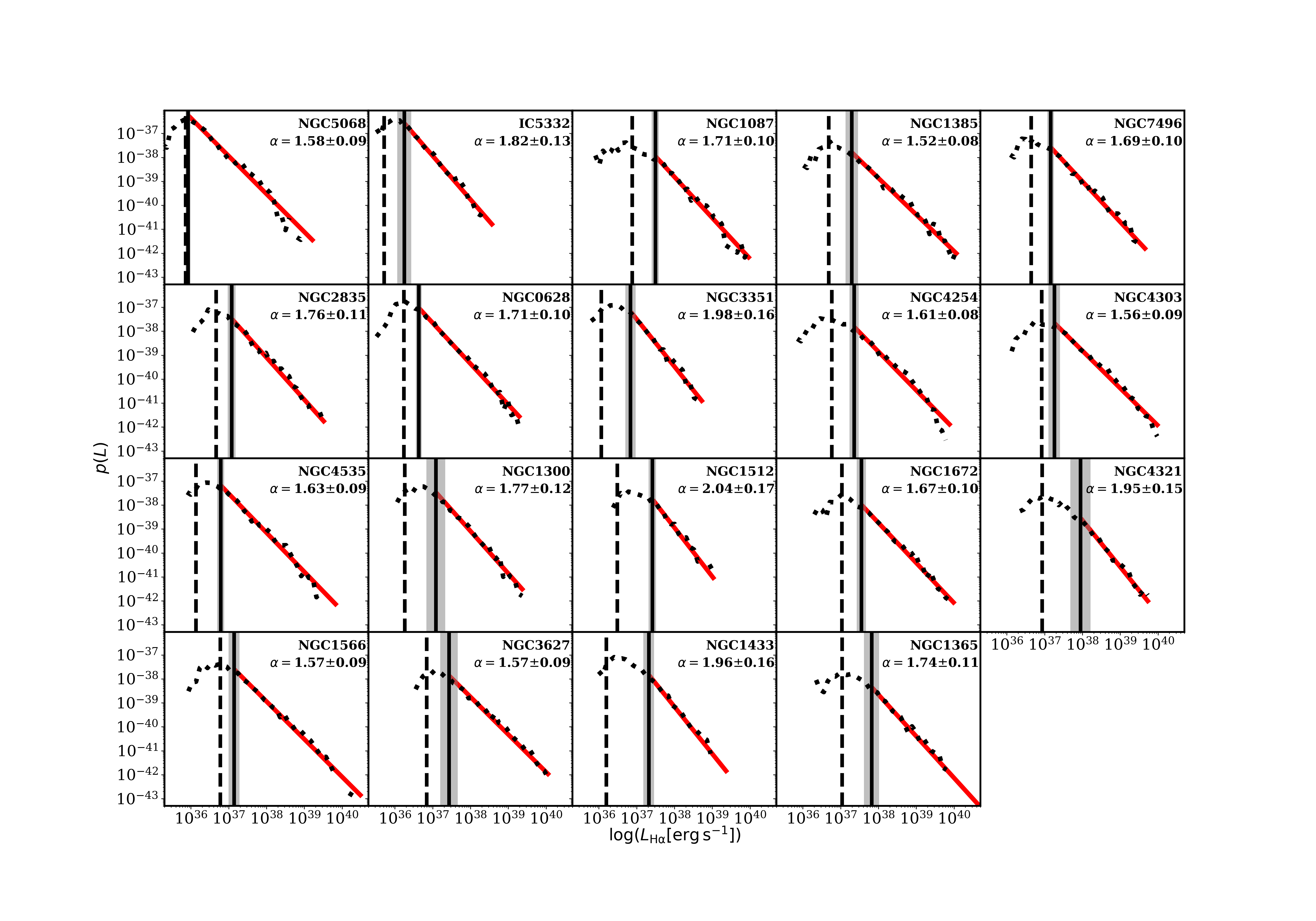}
    \caption{Best-fitting models of the \HII\ region LF for the galaxies of our sample. Galaxies are ordered by increasing stellar mass from top left to bottom right and their names are indicated within each panel. The dotted black line indicates the empirical PDF, plotted as $p(L)$ versus the logarithm of the \HII\ regions \Ha\ luminosity expressed in \ergs. The vertical dashed line indicates the estimated completeness limit. The vertical solid black line and the shaded gray area indicate the best-fitting \Lmin\ and its uncertainty. The LF best-fitting model is marked by the solid red line while the associated slope $\alpha$ and its uncertainty are indicated in the upper right corner of each panel.
    \label{Figure:global_LF}}
\end{figure*}

We make use of the extinction corrected \Ha\ luminosities of our \HII\ regions to obtain the nebular luminosity function~(LF) of each galaxy in our sample.
\HII\ region LFs have long been known to be well-described by a power-law \citep{Kennicutt1989}. 
The key parameter of the LF is its slope~$\alpha$. Several different methods for deriving this from the empirical data have been employed.
Past studies have often used least-squares linear regression of binned data (i.e.\ histograms) in log--log space. This method, despite being widespread, can give incorrect results under relatively common conditions, as discussed in \cite{Clauset2009}. In the context of LFs, this is mainly related to the fact that linear regression assumes Gaussian noise in the dependent variable while the noise of the logarithm of a histogram is not Gaussian. Furthermore, the choice of the binning scheme (e.g.\ equal-luminosity bins versus equal-number bins) and the definition of the bin center can affect the LF modeling and add uncertainties to~$\alpha$ \citep[see e.g.][]{2005ApJ...629..873M,Cook2016}.

\begin{table*}
    \caption{Properties of the \HII\ region LF in the PHANGS--MUSE galaxies.}
    \centering
    \begin{tabular}{ccccccccccc}
\toprule
Galaxy & Slope~$\alpha$ & $\log L_\mathrm{min}$ & $\log L_\mathrm{compl}$ & O7V~eq & $\log L_\mathrm{max}$ & $\log L_\mathrm{tot}$ & $N_\mathrm{tot}$ & $N_\mathrm{LFfit}$ & $N_\mathrm{DB}$ & $\mathrm{FWHM}_\mathrm{seeing}$ \\
 &  & $\mathrm{erg~s^{-1}}$ & $\mathrm{erg~s^{-1}}$ &  & $\mathrm{erg~s^{-1}}$ & $\mathrm{erg~s^{-1}}$ &  &  &  & $\mathrm{pc}$ \\
\toprule
NGC5068 & 1.58$\pm$0.09 & 35.92$\pm$0.06 & 35.86 & 0.06 & 39.26 & 40.38 & 1465 & 1206 & 9 & 26.18 \\
IC5332 & 1.82$\pm$0.13 & 36.25$\pm$0.19 & 35.71 & 0.12 & 38.62 & 39.71 & 612 & 357 & 1 & 37.91 \\
NGC1087 & 1.71$\pm$0.10 & 37.49$\pm$0.10 & 36.88 & 2.04 & 40.01 & 41.24 & 891 & 439 & 101 & 70.38 \\
NGC1385 & 1.52$\pm$0.08 & 37.29$\pm$0.17 & 36.68 & 1.29 & 40.11 & 41.47 & 914 & 556 & 157 & 55.67 \\
NGC7496 & 1.69$\pm$0.10 & 37.16$\pm$0.09 & 36.64 & 0.95 & 39.70 & 40.75 & 547 & 300 & 39 & 80.35 \\
NGC2835 & 1.76$\pm$0.11 & 37.08$\pm$0.10 & 36.67 & 0.78 & 39.56 & 40.72 & 819 & 432 & 35 & 67.90 \\
NGC0628 & 1.71$\pm$0.10 & 36.63$\pm$0.08 & 36.24 & 0.28 & 39.34 & 40.87 & 2230 & 1273 & 42 & 43.77 \\
NGC3351 & 1.98$\pm$0.16 & 36.84$\pm$0.13 & 36.06 & 0.45 & 38.77 & 40.14 & 784 & 369 & 3 & 50.56 \\
NGC4254 & 1.61$\pm$0.08 & 37.36$\pm$0.12 & 36.77 & 1.49 & 39.93 & 41.62 & 2536 & 1430 & 333 & 56.32 \\
NGC4303 & 1.56$\pm$0.09 & 37.26$\pm$0.15 & 36.92 & 1.18 & 40.03 & 41.75 & 2208 & 1567 & 353 & 63.94 \\
NGC4535 & 1.63$\pm$0.09 & 36.79$\pm$0.09 & 36.13 & 0.40 & 39.87 & 41.00 & 1444 & 934 & 65 & 42.63 \\
NGC1300 & 1.77$\pm$0.12 & 37.08$\pm$0.25 & 36.26 & 0.79 & 39.42 & 40.84 & 1169 & 632 & 49 & 81.50 \\
NGC1512 & 2.04$\pm$0.17 & 37.41$\pm$0.10 & 36.48 & 1.71 & 39.07 & 40.43 & 472 & 209 & 15 & 113.51 \\
NGC1672 & 1.67$\pm$0.10 & 37.55$\pm$0.12 & 37.03 & 2.32 & 40.03 & 41.40 & 1051 & 570 & 152 & 89.80 \\
NGC4321 & 1.95$\pm$0.15 & 37.95$\pm$0.27 & 36.93 & 5.82 & 39.78 & 41.31 & 1385 & 382 & 111 & 85.17 \\
NGC1566 & 1.57$\pm$0.09 & 37.14$\pm$0.14 & 36.77 & 0.91 & 40.53 & 41.58 & 1655 & 1026 & 186 & 68.27 \\
NGC3627 & 1.57$\pm$0.09 & 37.43$\pm$0.23 & 36.84 & 1.78 & 40.10 & 41.56 & 1007 & 657 & 188 & 57.44 \\
NGC1433 & 1.96$\pm$0.16 & 37.32$\pm$0.14 & 36.20 & 1.37 & 39.41 & 40.69 & 1258 & 407 & 28 & 81.76 \\
NGC1365 & 1.74$\pm$0.11 & 37.82$\pm$0.20 & 37.04 & 4.31 & 40.76 & 41.44 & 854 & 347 & 119 & 108.51 \\
\bottomrule
\end{tabular}

    \tablefoot{The table reports the galaxy name (col\,1), the LF slope (col\,2) and \Lmin\ (col\,3), the LF completeness limit (col\,4), the number of equivalent O7~V stars corresponding to \Lmin\ (col\,5), the luminosity of the brightest \HII\ region (col\,6), the integrated \HII\ regions luminosity (col\,7), the total number of \HII\ regions (col\,8), the number of \HII\ regions used to fit the LF slope (col\,9), the number of ``density bounded'' \HII\ regions with $L_\mathrm{H\alpha} > 10^{38.6}$~\ergs\ (col\,10), and the physical resolution of the \Ha\ maps used to detect \HII\ regions in parsecs (col\,11).
    \label{tab:Table3}}
\end{table*}

The slope of the LF is commonly constrained by fitting the bright end of the LF or, more specifically, \HII\ regions with a luminosity above a given \Lmin. This is the luminosity below which the LF starts to flatten -- it is often referred to as the ``turnover point'' and, depending on the sensitivity of the observations, can arise due to incompleteness of the data. \Lmin\ is an important parameter for determining the LF slope and in many studies it has been, quite arbitrarily, fixed to the luminosity where the histogram of the LF peaks or, alternatively, where the bin count number steadily decreases \citep[e.g.][]{Cook2016, Azimlu2011}. As described in \cite{Clauset2009}, by adopting a maximum likelihood estimation (MLE) method, in combination with the Kolmogorov--Smirnov (KS) statistic, it is possible to obtain both \Lmin\ and $\alpha$ of a given LF in a statistically robust way (see Sec.~\ref{subsec:complcrow} for further discussion). 

In this paper, we adopt this method to fit an empirical LF: it does not depend on any binning scheme and, as discussed in \cite{Clauset2009}, provides a statistically robust way to fit data following heavily tailed distributions. \cite{Whitmore2014} compared the performance of this method to classical ones based on histograms by fitting the LF of star clusters observed with \textit{HST} and found an overall good agreement, with MLE fits giving steeper slopes for steeper LFs.
To perform the LF fits, we make use of the Python package \powerlaw\ \citep{Alstott2014} which follows the prescription given in \cite{Clauset2009} and models the probability distribution function (PDF) $p(L)$ connected to the empirical LF using a power-law of the form:

\begin{equation}
    p(L) = (\alpha-1) \, L_\mathrm{min}^{\alpha-1} \, L^{-\alpha} \quad \text{with} \quad L \ge L_\mathrm{min}~.
\end{equation}

The algorithm performs an MLE fit by recursively fixing \Lmin\ equal to each empirical data point. For a given \Lmin, the algorithm tries a range of different slopes and selects as the best-fitting slope the one maximizing the likelihood estimator. At this point we have a set of models that includes the best-fitting model for each fixed \Lmin. The KS statistic is then used to compute the maximum distance between this set of models and the empirical cumulative distribution function. The final best-fitting model is chosen as the one that minimizes the KS statistic.    

We limit the search of \Lmin\ to an interval of $\pm1\sigma$ around the median of the distribution (i.e.\ where, visually, the LFs start to flatten) and the LF slope to the interval of $\alpha = 1{-}3$.
The error on \Lmin\ is taken to be the standard deviation of the \Lmin\ values found by rerunning the fitting procedure on $1000$ mock LFs extracted via bootstrapping from a given LF, after making sure that the error converges to a stable value as the number of bootstraps approaches 1000.
%rather than taking the formal error provided by the \powerlaw\ package. 
For the error on $\alpha$, we consider the trend of the likelihood estimator as a function of $\alpha$ for the models with the best \Lmin. This trend can be well represented by a Gaussian that peaks at the best $\alpha$ and we compute the error on $\alpha$ as the standard deviation of the best-fitting Gaussian function modeling the data \citep[similar to what has been done by][]{Whitmore2014}. We do not run any test to check for the existence of an upper cutoff in the LFs.

The best-fitting models of our LF are shown in Fig.~\ref{Figure:global_LF} and reported in Table~\ref{tab:Table3}. The slopes we find are in the interval of $\alpha = 1.5{-}2$ with a mean value of $1.73$ and a $\sigma$ of $0.15$, in agreement with what has been discussed in the literature for Sa--Sc spiral galaxies \citep[see e.g.][]{Kennicutt1989, Elmegreen1999, Whitmore2014}.
We note that for our sample, the minimum number of \HII\ regions used to fit the LF is $300$ (i.e.\ for \object{NGC7496}) and that our fits can be considered robust against biases due to low number statistics (see Sec.~\ref{subsec:complcrow} for further discussion). In Fig.~\ref{Figure:global_LF}, we also mark the completeness limit of our LFs that will be discussed in Sec.~\ref{subsec:complcrow}.
In Appendix~\ref{appendix3}, we discuss how the results would change if \Ha\ luminosities were not corrected for dust extinction (e.g.\ in narrow-band imaging). We find that neglecting the dust extinction correction causes a slight steepening of the LFs that, within their errors, remain compatible with the slopes we report in this section.

Before describing further the details of our data analysis, it is worth discussing the overall properties of our LFs. As can be seen in Fig.~\ref{Figure:global_LF}, our \HII\ regions span a luminosity range of $\log(L_\mathrm{H\alpha} \ [\mathrm{erg~s^{-1}}]) \sim 35{-}40$. \HII\ regions with luminosities less than a few times $10^{37}$~\ergs\ are typically ionized by single O- or B-type stars, whereas at higher luminosities and up to about $10^{39}$~\ergs\ we enter the regime where \HII\ regions are ionized by stellar associations or clusters containing multiple OB stars. In this regime, the \Ha\ luminosity of an \HII\ region is expected to be roughly proportional to its number of ionizing stars. \HII\ regions with luminosities higher than a few $10^{39}$~\ergs\ are referred to as giant or supergiant \HII\ regions (e.g.\ 30~Doradus in the LMC) and are typical of late-type galaxies \citep[Sc and Irr Hubble types; see e.g.][]{Kennicutt1989,Elmegreen1996}.

The ionizing luminosities corresponding to \Lmin\ in each galaxy of our sample are reported in Table~\ref{tab:Table3} in terms of the number of equivalent O7~V type stars, assuming that the number of ionizing photons per second for an O7~V star along the zero-age main sequence (ZAMS) is $\log (Q_{0}^\mathrm{O7\,V} \ [\mathrm{photons~s^{-1}}])= 49.05$ following \cite{Vacca1994}. The number of equivalent O7~V stars is used here as a first order indicator of the number of ionizing stars in an \HII\ region and at \Lmin\ it is in the range of $0.05{-}6$ in our sample. This testifies that our LF fits, extending ${\sim}2{-}3$ orders of magnitude above \Lmin, are mainly probing \HII\ regions ionized by star clusters and associations, and only for a few targets (i.e.\ \object{NGC5068}, \object{IC3352} with smaller distances) are we probing the regime of \HII\ regions ionized by single massive stars.

Different studies in the literature argue that the LF steepens at luminosities higher than $\log(L_\mathrm{H\alpha} \ [\mathrm{erg~s^{-1}}]) = 38.6\pm0.1$ (i.e.\ the so-called LF cutoff or break) and is better reproduced by a double power-law \citep[i.e.\ type~II LF, reported in individual galaxy studies; e.g.][]{Kennicutt1989, Rand1992, Rozas1996, Beckman2000, Thilker2000, Gutierrez2011, Lee2011}.
\cite{Beckman2000} also observed a local sharp peak (adopting their nomenclature, we refer to this feature as ``glitch'') in the LF of their galaxies at $\log(L_\mathrm{H\alpha} \ [\mathrm{erg~s^{-1}}]) = 38.6$ and suggested that it marks the transition of \HII\ regions from ionization bounding (the central star/\linebreak[0]{}star cluster only ionizes the gas within the \HII\ region) at low luminosities to density bounding (a~large amount of Lyman continuum photons from the central ionizing source escapes the \HII\ region and ionizes the diffuse surrounding medium) at higher luminosities.

The number of putative density bounded \HII\ regions in our galaxies (i.e.\ $L_\mathrm{H\alpha} \geq 10^{38.6}$~\ergs) varies from a few up to about $350$ and is reported in Table~\ref{tab:Table3}. Neither the number of these regions nor their fraction with respect to the total number of \HII\ regions correlates with galaxy morphological type (two targets with Sa morphology have a few tens of these regions while Sb and Sc galaxies show comparable numbers and scatter) or the physical resolution of the observations. This suggests that in our sample the detection of this kind of \HII\ region is not purely an effect of blending due to limited spatial resolution \citep[as suggested by the study of][]{Lee2011}. 

\begin{figure*}[ht]
    \centering
    \includegraphics[trim=1cm 1cm 1cm 1cm, clip, width=0.7\textwidth]{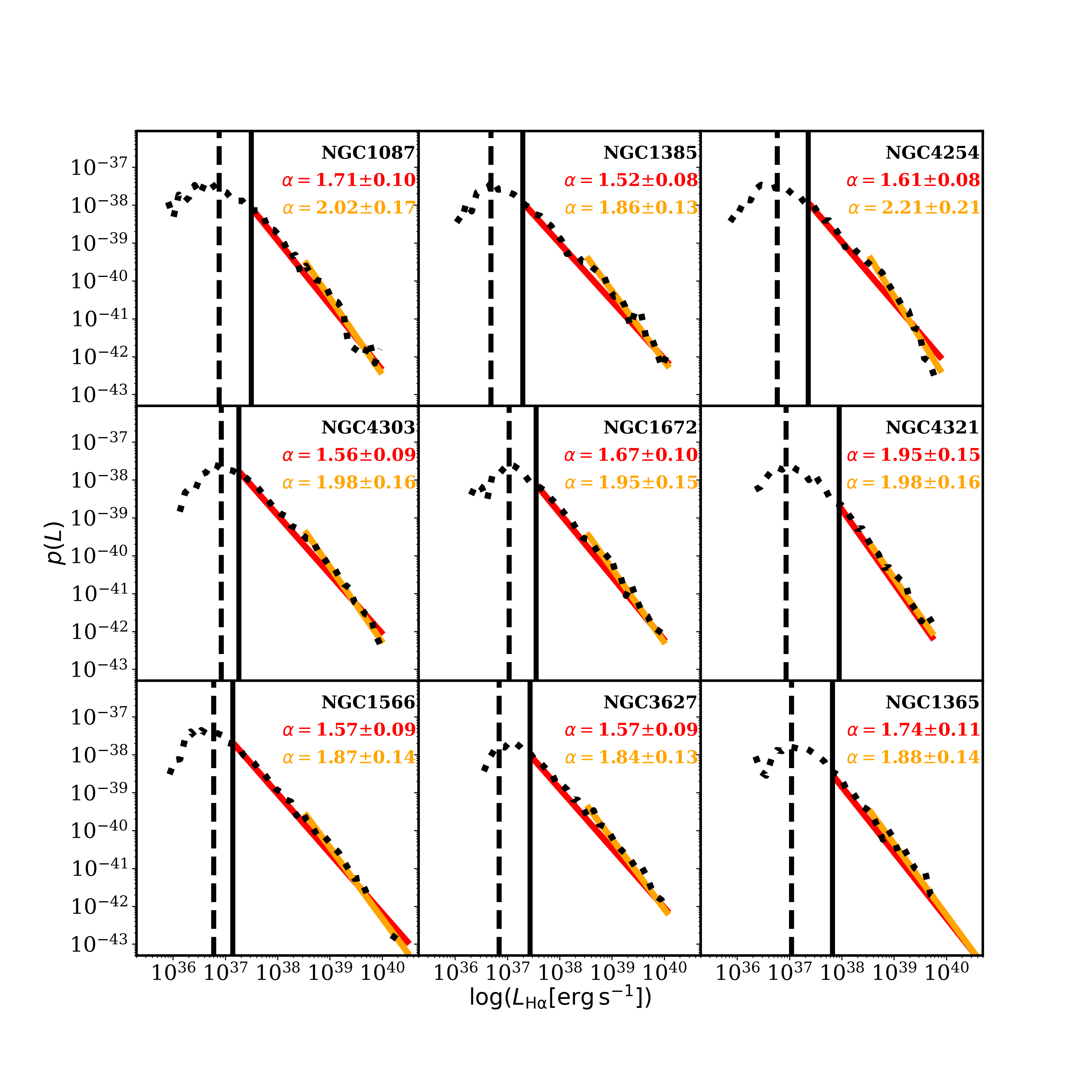}
    \caption{Best-fitting model of the upper end of the \HII\ region LF for the sub-sample of nine galaxies with more than $100$ density bounded \HII\ regions. Galaxies are ordered by increasing stellar mass from top left to bottom right and their names are indicated within each panel. The dotted black line indicates the empirical PDF, plotted as $p(L)$, versus the logarithm of the \HII\ regions \Ha\ luminosity expressed in \ergs. The vertical dashed line indicates the estimated completeness limit. The vertical solid black line and the solid red line indicate the best-fitting \Lmin\ and $\alpha$ of the global LF as shown in Fig.~\ref{Figure:global_LF}. The orange solid line marks the best-fitting model for the regions with $L_\mathrm{H\alpha} \geq 10^{38.5}$~\ergs. The LF slopes of the two best-fitting models are indicated in the upper right corner of each panel following the same color coding as the models.
    \label{Figure:double_PL_LF}}
\end{figure*}

For the nine galaxies in which we detect more than $100$ density bounded \HII\ regions, we perform an additional fit of the LF only at $L_\mathrm{H\alpha} \geq 10^{38.5}$~\ergs\ and show the results in Fig.~\ref{Figure:double_PL_LF}. We find that for these galaxies the slope of the upper end is indeed steeper with respect to the slope measured globally; only for two of the nine targets (i.e.\ \object{NGC1365} and \object{NGC4321}) the variations are within the errors. 
However, looking at the LFs of our sample in Fig.~\ref{Figure:global_LF}, the steepening at $\log(L_\mathrm{H\alpha} \ [\mathrm{erg~s^{-1}}]) > 38.6\pm0.1$ and the glitch at $\log(L_\mathrm{H\alpha} \ [\mathrm{erg~s^{-1}}]) = 38.6\pm0.1$ are subtle features, contrary to what has been shown by \cite{Beckman2000}. 
Assessing which LFs are better described by a double power-law is behind the scope of the current paper. Overall, a single power-law fit gives a good representation of the empirical data in our sample and is it thus adopted as our fiducial LF model. In the following, we will take the advantage of this simpler parametrization of the LF to directly compare the LF slopes of different galaxies and of \HII\ region sub-samples within individual galaxies.

\subsection{LF variations among galaxies}\label{sec:LF variations between galaxies}

 \begin{figure*}[ht]
    \centering
    \includegraphics[width=\textwidth]{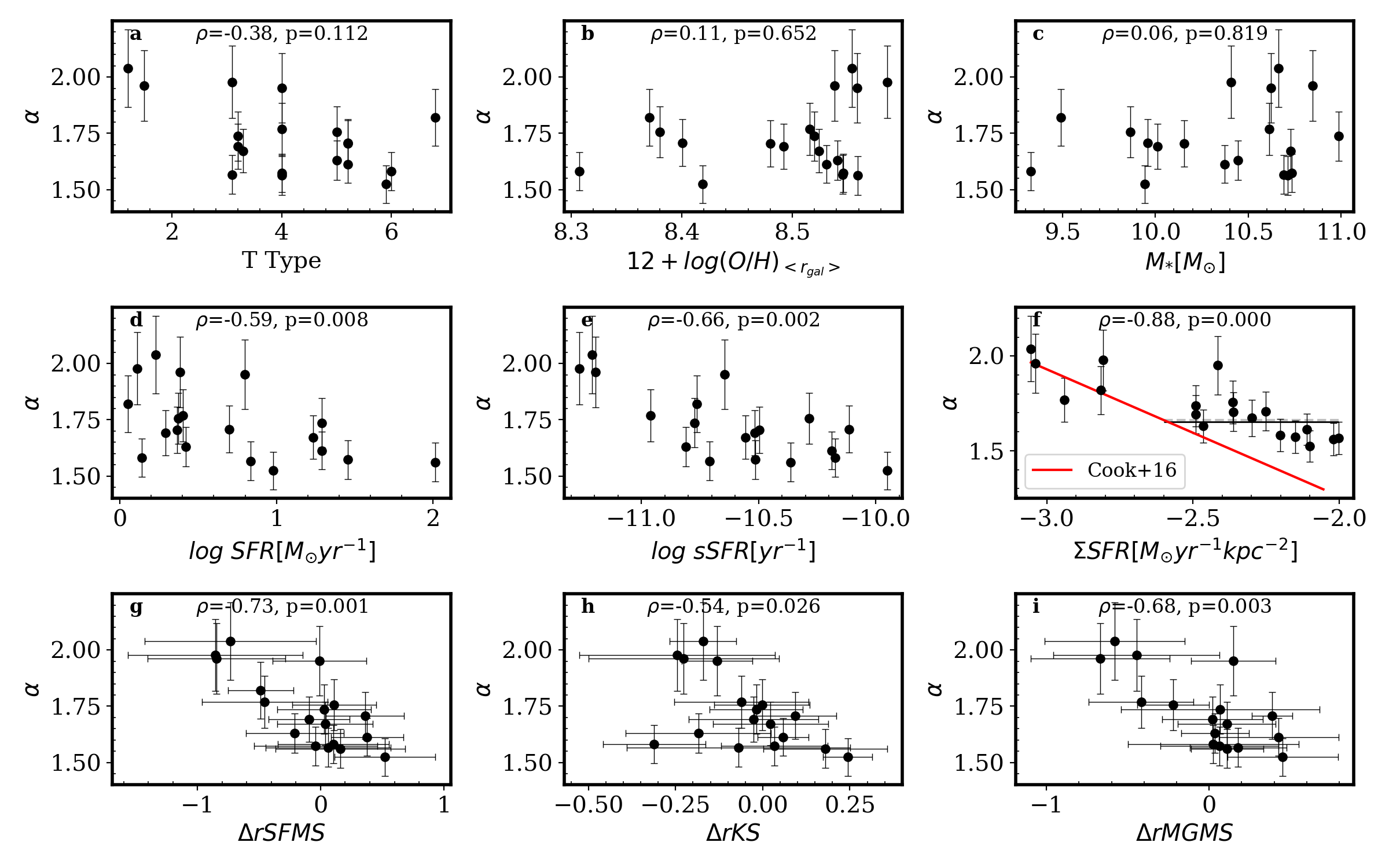}
    \caption{Trends of the LF slope $\alpha$ with global galaxy properties. From top left to bottom right we show $\alpha$ as a function of the galaxy morphological T~class, the Scal metallicity at the median galactocentric radius covered by our observations $12+\log(\mathrm{O/H})_{\langle r_\mathrm{gal}\rangle}$, the total stellar mass $M_{*}$, the total star formation rate SFR, the specific star formation rate sSFR, the SFR surface density \Ssfr, the offset from the resolved star formation main sequence \DrSFMS, the offset from the resolved Kennicutt--Schmidt relation \DrKS, and the offset from the resolved molecular gas main sequence \DrMGMS. For each panel we report the Spearman correlation coefficient $\rho$ and $p$-value of the plotted quantities. In the middle right panel the solid red line shows the best-fitting relation found by \cite{Cook2016} while the dashed gray and  solid back horizontal lines respectively mark the $\alpha$ of the \cite{Cook2016} line at \Ssfr$=-2.6$ and the mean $\alpha$ of the points at \Ssfr$>-2.6$ (see Sec.~\ref{sec:The SFR and the cold gas surface density} for further details).
    \label{fig:LF_SLOPE_corr}}
\end{figure*}

To look at what may drive galaxy-to-galaxy variations of the LF slope in our sample, we verified that $\alpha$ does not have any dependence on galaxy inclination reported in Table~\ref{tab:Table1} and then estimate a few global galaxy parameters, excluding the areas defined as galaxy centers (see Sec.~\ref{sec:Environmental and foreground stars masks}), to be in line with the \HII\ regions probed by our LFs. More specifically, we estimate:
\begin{itemize}
    \item The gas-phase metallicity at a representative radius, taken to be the median galactocentric radius of the \HII\ regions detected in each galaxy. The metallicity at this radius is calculated from a linear fit to the radial metallicity gradient (shown in Fig.~\ref{fig:met_radial_gradients} of Appendix~\ref{appendix2}).
    \item The total stellar mass ($M_{*}$) extracted by integrating the PHANGS--MUSE stellar mass maps produced by the DAP. 
    \item The total star formation rate (SFR) extracted by integrating the PHANGS--MUSE \Ha\ maps (not corrected for dust extinction) and applying the calibration reported in \cite{Kennicutt2012} taken from the work of \cite{Hao2011} and \cite{Murphy2011}, who adopted a \cite{Kroupa2003} IMF, with a Salpeter slope ($\alpha_{*} = -2.35$) from~$1$ to~$100~M_{\odot}$ and $\alpha_{*} = -1.3$ for $0.1{-}1~M_{\odot}$, and solar metallicity. 
    \item The specific SFR ($\mathrm{sSFR} = \mathrm{SFR}/M_{*}$) and the SFR surface density (\Ssfr) obtained by dividing the SFR by $M_{*}$ and the area covered by the MUSE FoV, respectively.
\end{itemize}

Taking advantage of the study on the resolved star formation scaling relations at ${\sim}100$~pc scales carried out by \cite{Pessa2021}, we also consider the offset $\Delta$ of individual galaxies with respect to the trend of the full sample for the resolved star formation main sequence (\rSFMS; \SMstar\ vs.\ \Ssfr), the Kennicutt--Schmidt relation (\rKS; \SHtwo\ vs.\ \Ssfr), and the molecular gas main sequence (\rMGMS; \SMstar\ vs.\ \SHtwo). 
The offsets \DrSFMS, \DrKS, and \DrMGMS\ have been estimated taking as a reference the modeled trends for the star formation scaling relations of the whole sample and fitting the same trends for each galaxy. The slope of the individual galaxy trends, for each scaling relation, was set to be the same as the one of the entire sample, and the offset of each galaxy from the global relation was estimated as the difference between the normalization factors of the linear fits \citep[see][for additional details]{Pessa2021}. It should be noted that these measurements are not available for \object{NGC0628} due to the quality of the MUSE data while, due the lack of significant \mbox{CO(2--1)} emission, \object{IC5332} does not have measurements of \DrKS\ and \DrMGMS. The \DrSFMS, \DrKS, and \DrMGMS\ offsets are interesting quantities because, by construction of the star formation scaling relations, they measure the relative variations of the sSFR (\Ssfr/\SMstar), the gas depletion time ($t_\mathrm{dep} =$ \SHtwo/\Ssfr), and the molecular gas fraction (\SHtwo/\SMstar) within our sample. We verified that estimating such offsets by excluding galaxy centers has no significant effect on our findings.

The trends between the LF slope $\alpha$ and the aforementioned parameters, with the addition of the morphological T~type, are shown in Fig.~\ref{fig:LF_SLOPE_corr} together with their Spearman correlation coefficient $\rho$ and $p$-value, indicating the probability that the two sets of data are uncorrelated. We summarize the properties for which we look for a correlation in Table~\ref{tab:Table_LFsubsamples}.
We define a correlation to be negligible when $\lvert\rho\rvert = [0{-}0.2]$, weak when $\lvert\rho\rvert = [0.2{-}0.4]$, moderate when $\lvert\rho\rvert = [0.4{-}0.6]$, strong when $\lvert\rho\rvert = [0.6{-}0.8]$, and very strong when $\lvert\rho\rvert = [0.8{-}1]$; using the $p$-value to evaluate the probability that, despite showing a correlation, two variables may be uncorrelated.
It should be noted that only a handful of studies so far have looked at the correlation between $\alpha$ and global galaxy properties: \citet{Kennicutt1989, Elmegreen1999, Youngblood1999, vanZee2000}, and \citet{Thilker2002} investigated nebular LFs as in this paper, while \citet{Liu2013} identified \HII\ regions via Pa$\alpha$, and \cite{Cook2016} studied the \textit{GALEX} far-ultraviolet (FUV) LFs of \HII\ regions. While the sample of \citet{Cook2016} includes a few hundred galaxies, the other studies are based on samples ranging from~$10$ to~$35$ galaxies, similar to our study. 
In this section and in Sec.~\ref{sec:The SFR and the cold gas surface density}, we compare our results to those studies which, like in our case, applied a uniform analysis methodology on galaxy samples. It should be noted that using different tracers means probing different source ages and, as reported by \cite{Oey1998}, older \HII\, regions tend to have steeper LF slopes, mainly due to the short main-sequence lifetimes of the more massive stars constituting the brighter \HII\, regions. This is the reason why e.g. FUV observations, probing \HII\, regions with ages less than $100$~Myr, are expected to deliver steeper LF compared to \Ha\, observations, typically probing \HII\, regions younger than 10~Myr, and our comparison remains qualitative.

We find a weak correlation between $\alpha$ and the galaxy morphology (Fig.~\ref{fig:LF_SLOPE_corr}a), and a negligible correlation with the gas-phase metallicity (Fig.~\ref{fig:LF_SLOPE_corr}b) and the total stellar mass of our galaxies (Fig.~\ref{fig:LF_SLOPE_corr}c).
\cite{Kennicutt1989} and \cite{Elmegreen1999} found that the LF slope flattens for later-type galaxies where the relative number of giant \HII\ regions is noticeably higher. On the other hand, \citet{Cook2016} found no trend between $\alpha$ and morphology when studying a sample more representative for irregulars, arguing that the trends reported in the literature were relying on few data points in this regime.
Due to the selection of the PHANGS--MUSE sample we are only probing typical star-forming galaxies with morphologies in the range Sa--Sb (corresponding roughly to T~types between~1 and~6). As noted also by \citet{Kennicutt1989} the scatter between the $\alpha$ of spirals of the same type in their sample is comparable to the magnitude of the trend they report to exist between $\alpha$ and morphology. The lack of a correlation between $\alpha$ and morphology in our sample is thus fully compatible with the results in the literature.  
The lack of a correlation with the gas-phase metallicity and the total stellar mass of our galaxies, despite the differences extracting these quantities with respect to previous studies, can also be explained by sample selection effects. Our sample mainly comprises galaxies with high metallicities (i.e.\ $12+\log(\mathrm{O/H}) > 8.3$) and stellar masses (i.e.\ $\log M_{*} \, [M_{\odot}] > 9$). \cite{Cook2016} found no trend between $\alpha$ and gas-phase metallicity in the range $7.2 < 12+\log(\mathrm{O/H}) < 9.2$ and only a moderate trend between $\alpha$ and $M_{*}$ in the range $6 < \log M_{*} \, [M_{\odot}] < 11$. The parameter space occupied by our galaxies is compatible with what has been shown by \cite{Cook2016} and we do not expect to find strong correlations. 

On the other hand, SFR-related quantities show a better correlations with $\alpha$ (Fig.~\ref{fig:LF_SLOPE_corr}d-e-f). SFR, sSFR, and \Ssfr\ have a moderate, strong, and very strong correlation with $\alpha$, respectively. The presence of a tighter correlation with \Ssfr\ is in line with the findings of \cite{Cook2016}, whose best-fitting relation is also shown in Fig.~\ref{fig:LF_SLOPE_corr}f. The negative Spearman correlation coefficient indicates that the quantities anti-correlate with $\alpha$, meaning that the LF has a shallower slope (i.e.\ higher relative number of bright \HII\ regions) for galaxies with higher SFR, sSFR, and \Ssfr.

Looking at where our galaxies lie in the plane of the three classical star formation scaling relations (Fig.~\ref{fig:LF_SLOPE_corr}g-h-i), we find that $\alpha$ anti-correlates strongly with \DrSFMS\ and \DrMGMS, and moderately with \DrKS. The weaker correlation in the latter case can be somehow explained by the fact that, between the three, the \rKS\ relation is the one that shows the lowest galaxy-to-galaxy scatter in our sample \citep{Pessa2021}. While the trend with \DrSFMS\ confirms the trend we find with sSFR, \DrMGMS\ adds information on the molecular gas fraction and indicates that galaxies with a lower fraction of molecular gas tend to have steeper LFs.

In this paper we choose to not perform any fit for the strongest correlations shown in Fig.~\ref{fig:LF_SLOPE_corr} because there is no strong evidence for a particular functional form from the data (e.g. considering the scatter with respect to the errors) nor is there one expected from theoretical studies. We note that future work could explore local correlations (e.g. using \HII\, region samples from different areas of the same galaxy) which would give more insights about this matter but this is behind the scope of the current paper.

\begin{table*}
    \caption{Global properties of the PHANGS--MUSE galaxies}
    \centering
    \noindent\adjustbox{max width=\textwidth}{
    \begin{tabular}{ccccccccccc}
\toprule
Galaxy & T~type & Slope~$\alpha$ & $12+\log(\mathrm{O/H})_{\langle r_\mathrm{gal} \rangle}$ & $\log M_{*} $ & $\log \mathrm{SFR}$ & $\log \mathrm{sSFR}$ & $\log \Sigma_\mathrm{SFR}$ & $\Delta \mathrm{rSFMS}$ & $\Delta \mathrm{rKS}$ & $\Delta \mathrm{rMGMS}$ \\
 &  &  &  & $M_{\odot}$ & $M_{\odot}~\mathrm{yr}^{-1}$ & $\mathrm{yr}^{-1}$ & $M_{\odot}~\mathrm{yr}^{-1}~\mathrm{kpc}^{-2}$ &  &  &  \\
\toprule
NGC5068 & 6.0 & 1.58$\pm$0.09 & 8.31 & 9.33 & 0.14 & -10.17 & -2.20 & 0.10 & -0.31 & 0.03 \\
IC5332 & 6.8 & 1.82$\pm$0.13 & 8.37 & 9.49 & 0.05 & -10.76 & -2.82 & -0.49 & -- & -- \\
NGC1087 & 5.2 & 1.71$\pm$0.10 & 8.40 & 9.96 & 0.70 & -10.11 & -2.25 & 0.36 & 0.09 & 0.39 \\
NGC1385 & 5.9 & 1.52$\pm$0.08 & 8.42 & 9.94 & 0.98 & -9.95 & -2.10 & 0.52 & 0.24 & 0.45 \\
NGC7496 & 3.2 & 1.69$\pm$0.10 & 8.49 & 10.01 & 0.29 & -10.52 & -2.49 & -0.09 & -0.03 & 0.02 \\
NGC2835 & 5.0 & 1.76$\pm$0.11 & 8.38 & 9.87 & 0.37 & -10.28 & -2.36 & 0.11 & -0.00 & -0.22 \\
NGC0628 & 5.2 & 1.71$\pm$0.10 & 8.48 & 10.15 & 0.37 & -10.50 & -2.36 & -- & -- & -- \\
NGC3351 & 3.1 & 1.98$\pm$0.16 & 8.59 & 10.41 & 0.11 & -11.26 & -2.81 & -0.86 & -0.25 & -0.45 \\
NGC4254 & 5.2 & 1.61$\pm$0.08 & 8.53 & 10.37 & 1.29 & -10.19 & -2.11 & 0.38 & 0.06 & 0.43 \\
NGC4303 & 4.0 & 1.56$\pm$0.09 & 8.56 & 10.71 & 2.02 & -10.36 & -2.02 & 0.16 & 0.18 & 0.11 \\
NGC4535 & 5.0 & 1.63$\pm$0.09 & 8.54 & 10.44 & 0.42 & -10.81 & -2.47 & -0.21 & -0.18 & 0.04 \\
NGC1300 & 4.0 & 1.77$\pm$0.12 & 8.52 & 10.61 & 0.40 & -10.96 & -2.94 & -0.45 & -0.06 & -0.42 \\
NGC1512 & 1.2 & 2.04$\pm$0.17 & 8.55 & 10.66 & 0.23 & -11.21 & -3.05 & -0.73 & -0.17 & -0.58 \\
NGC1672 & 3.3 & 1.67$\pm$0.10 & 8.52 & 10.73 & 1.23 & -10.56 & -2.30 & 0.04 & 0.02 & 0.11 \\
NGC4321 & 4.0 & 1.95$\pm$0.15 & 8.56 & 10.62 & 0.80 & -10.64 & -2.42 & -0.01 & -0.13 & 0.15 \\
NGC1566 & 4.0 & 1.57$\pm$0.09 & 8.55 & 10.73 & 1.46 & -10.51 & -2.15 & -0.04 & 0.03 & 0.06 \\
NGC3627 & 3.1 & 1.57$\pm$0.09 & 8.55 & 10.69 & 0.84 & -10.71 & -2.00 & 0.06 & -0.07 & 0.18 \\
NGC1433 & 1.5 & 1.96$\pm$0.16 & 8.54 & 10.85 & 0.38 & -11.20 & -3.04 & -0.85 & -0.23 & -0.67 \\
NGC1365 & 3.2 & 1.74$\pm$0.11 & 8.52 & 10.99 & 1.29 & -10.77 & -2.49 & 0.03 & -0.02 & 0.07 \\
\bottomrule
\end{tabular}
}
    \tablefoot{The table reports the galaxy name (col\,1), morphological T~type taken from HyperLEDA \citep{Makarov2014} (col\,2), the LF slope (col\,3), the metallicity at the median \HII\ regions galactocentric radius (col\,4), the stellar mass (col\,5), the star formation rate (col\,6), the specific star formation rate (col\,7), the star formation rate density (col\,8), the offset from the resolved star formation main sequence (col\,9), the offset from the resolved Kennicutt--Schmidt relation (col\,10), and the offset from resolved the molecular gas main sequence (col\,11). All the properties reported in col\,3{-}11 have been extracted using the PHANGS-MUSE data.
    \label{tab:Table_LFslope_trends}}
\end{table*}

\subsection{LF variations within galaxies}\label{sec:LF variations within galaxies}

After having investigated LF slope variations across our sample, in this section, we focus our attention on possible variations within a given galaxy by looking at the LF of \HII\ region sub-samples.
In the following, we use galactic environment (spiral arm and inter-arm areas), galactocentric radius, and ionization parameter to draw, for each galaxy, \HII\ region sub-samples and fit their LFs while keeping \Lmin\ fixed to the value found in Sec.~\ref{sec:HII regions luminosity function} for each parent sample. All the details of the LFs fits performed in this section are reported in Table~\ref{tab:Table_LFsubsamples}, while the actual fits are shown in Appendix~\ref{appendix2}.

\subsubsection{Spiral arm vs.\ inter-arm areas}\label{sec:Environment}

Differences in the LF slope between spiral arm and inter-arm areas have been reported in the past \citep{Rand1992,Banfi1993,Thilker2000}, even though not always significant, with steeper LFs in inter-arm areas. However, this does not appear to be a universal property of spiral galaxies, as numerous cases are known where such variations have not been found \citep{Knapen1998,Rozas1996,Azimlu2011,Gutierrez2011}. This has to do with the fact that a simple spatial classification of \HII\ regions as spiral or inter-arm regions does not guarantee to probe the conditions under which star formation happens in the two environments. Part of the \HII\ region population in the inter-arm areas, for example, can consist of aged \HII\ regions previously formed within spiral arms and/or as part of a recent star formation burst, depending on the star formation history (SFH) of a galaxy.

\begin{figure}
\begin{center}
    \includegraphics[width=0.5\textwidth]{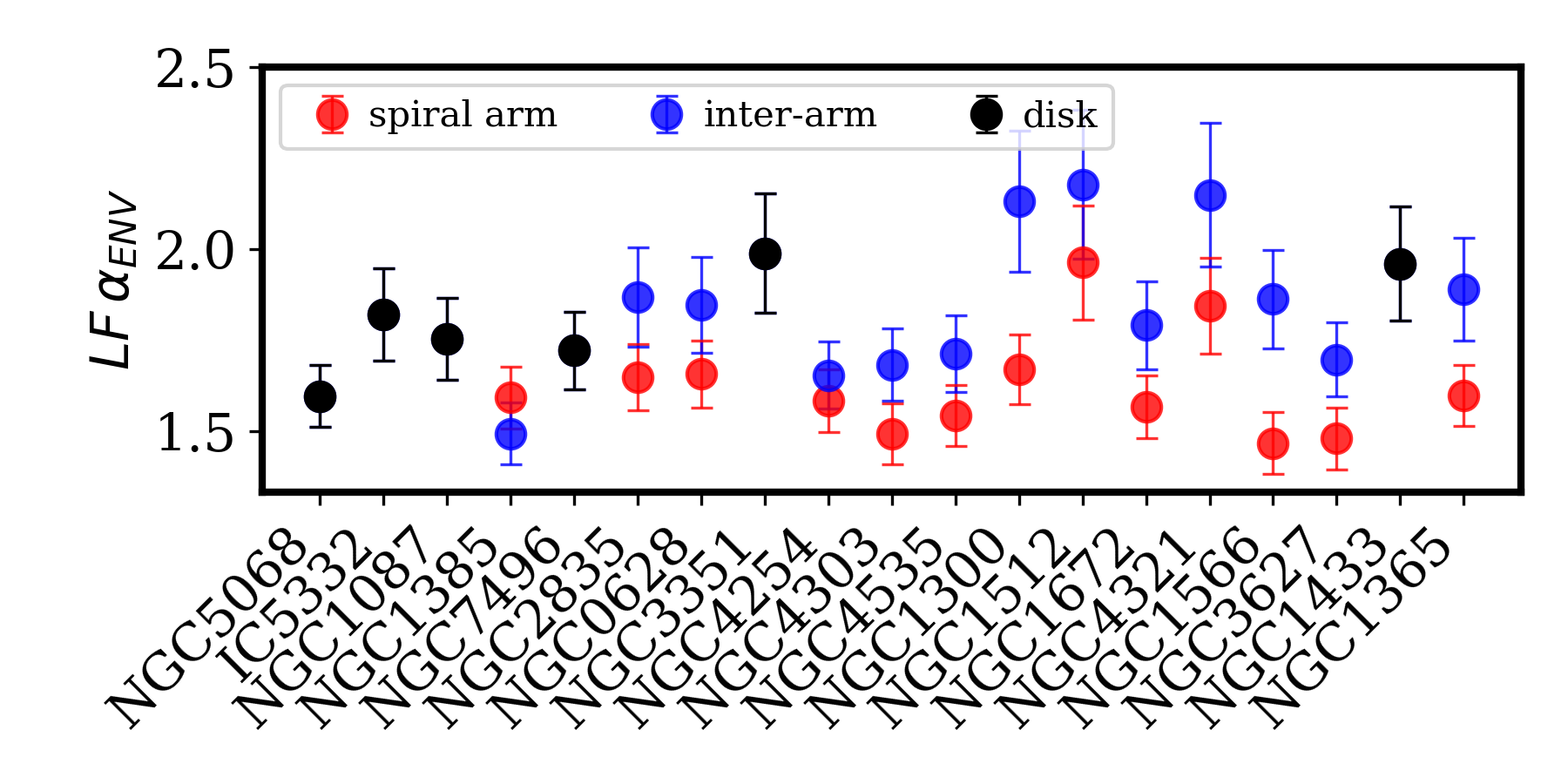}
    \caption{Slope of the LF for \HII\ regions located in spiral arms (red points) and inter-arm (blue points) areas. Galaxies are ordered by increasing stellar mass from left to right and their names are indicated along the abscissa. Galaxies with no evident spiral arms are marked by black points and their LF slope refers to the disk area (excluding bars).
    \label{Figure:LF_arm_interarm_smallfig}}
\end{center}
\end{figure}

Figure~\ref{Figure:LF_arm_interarm_smallfig} summarizes the result of fitting the LFs of the \HII\ regions belonging to spiral arm and inter-arm areas (see Sec.~\ref{sec:Environmental and foreground stars masks} for our definition of the environments and Fig.~\ref{Figure:LF_arm_interarm} for the actual LFs fits). Six of our 19~galaxies do not show evident spiral arms, for these galaxies we report the LF slope for the \HII\ regions across the entire galaxy disk but excluding bar regions (i.e.\ this is the reason for slight variations with respect to the slopes reported in Table~\ref{tab:Table3}).

We find that the LFs in spiral arm areas are either comparable to or shallower than in inter-arm areas. We find that six out of the 13~galaxies with spiral arms, preferentially at higher stellar masses, show variations between spiral arm and inter-arm LF $\alpha$ that go beyond the uncertainties. By comparing the LF shapes we see that the fraction of bright to faint \HII\ regions tends to be higher in spiral arms and lower in inter-arms areas (especially for \object{NGC1566} and \object{NGC1300}), however, in general, the two LFs span quite similar ranges of luminosities and have similar turnover points. The interpretation of these results is further discussed in Sec.~\ref{sec:The strength of spiral arms}.

\subsubsection{Inner vs.\ outer disk}\label{sec:Radial Trends}

There exists evidence that the conditions under which SF happens in the outer disk of spiral galaxies are different from the inner disk, for example, as suggested by an observed steepening of the GMC mass function \citep[e.g.][]{Rosolowsky2007,Colombo2014,Rice2016,Faesi2018,Schruba2019}. In line with this, radial variations of the LF are expected and have been reported in the literature suggesting that the LF is steeper in the outer disk \citep[i.e.\ extending beyond \Rtwentyfive; see e.g.][]{Lelievre2000} where \HII\ regions tend to be smaller and fainter \citep[see, e.g.\ ][]{Ferguson1998,Helmboldt2005}. 

To assess the presence of radial trends (which may also be connected with parameters varying radially such as the gas-phase metallicity), we study the LF at small and large galactocentric radii. To mitigate effects due to different sub-sample sizes, for each galaxy we select the median \HII\ region galactocentric radius as a demarcation between the inner and the outer star-forming disk and extract two \HII\ region sub-samples of equal sizes.
It should be noted that the area covered by our MUSE observations is largely contained within ${\sim}1$\Rtwentyfive\ of our targets, as can be seen looking at the $r_\mathrm{max}$ values reported in Table~\ref{tab:Table1}. What we define as the outer star-forming disk is still covering the main (i.e.\ molecular gas dominated) star-forming disk of our spiral galaxies and should not be confused with what is conventionally called outer (i.e.\ atomic gas dominated) disk in the literature (typically extending far beyond \Rtwentyfive). 

\begin{figure}
\begin{center}
    \includegraphics[width=0.5\textwidth]{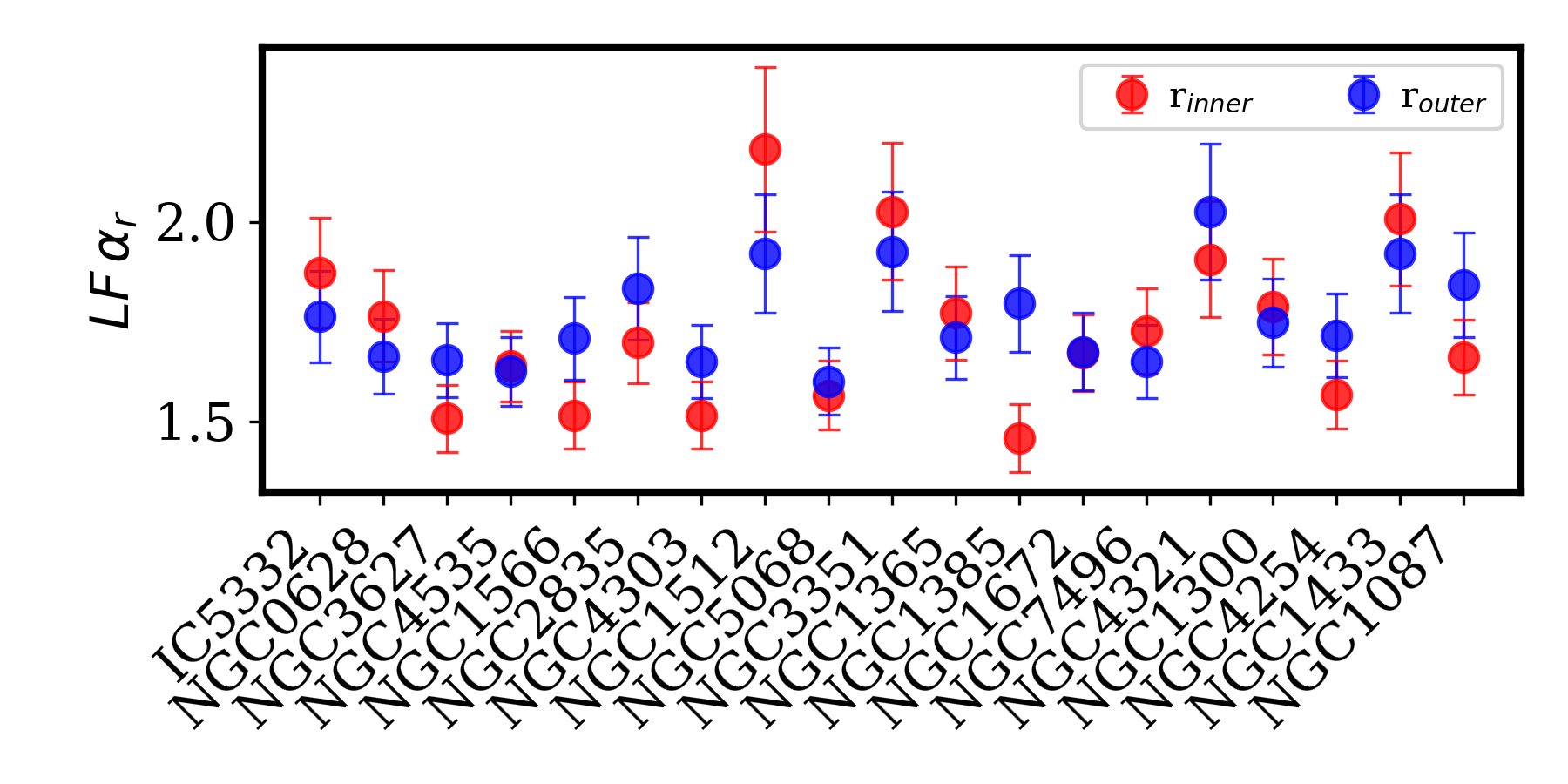}
    \caption{Slope of the LF for \HII\ regions located in the inner (red points) and outer (blue points) star-forming disk. Galaxies are ordered by increasing value of their \HII\ regions median galactocentric radius from left to right and their names are indicated along the abscissa.
    \label{Figure:LF_radial_smallfig}}
\end{center}
\end{figure}

Fig.~\ref{Figure:LF_radial_smallfig} summarizes the results of our fits of the LF slope in the inner and outer star-forming disk (shown in Fig.~\ref{Figure:LF_radial}). The median \HII\ region galactocentric radii, used to draw the \HII\ region sub-samples, varies between ${\sim}0.2$\Rtwentyfive\ and $0.5$\Rtwentyfive\ across our galaxies. 
Most of our galaxies do not show significant variations between the LF slope of \HII\ regions located at small and large galactocentric radii. Our results do not change if we adopt a fixed value of $0.3$\Rtwentyfive\ to draw \HII\ region sub-samples (not shown here). Only two out of 19~galaxies (i.e.\  \object{NGC1385} and \object{NGC1566}) show a significant variation, indicating a steeper LF in the outer disk. Our results indicate that within the star-forming disk of spiral galaxies there are no significant radial trends for the LF slope. This suggests that the radial metallicity trends in our galaxies (no more than $0.2$dex variations; see Fig.~\ref{fig:met_radial_gradients}) do not affect the LF slope.

\subsubsection{High vs.\ low ionization parameter}\label{sec:Age Trends}

Lastly, we investigate variations of the \HII\ region LF slope connected with the ionization parameter~$q$ of the \HII\ regions. As for the radial bins, for each galaxy we use the median \HII\ region $q$ to draw two sub-samples of \HII\ regions of equal size. The median values that we find are in the range $\log q = 6.4{-}7$. We visually inspected the spatial location of the two sub-samples of \HII\ regions for each galaxy and they both appear similarly distributed across the galaxy disks. 

\begin{figure}
\begin{center}
    \includegraphics[width=0.5\textwidth]{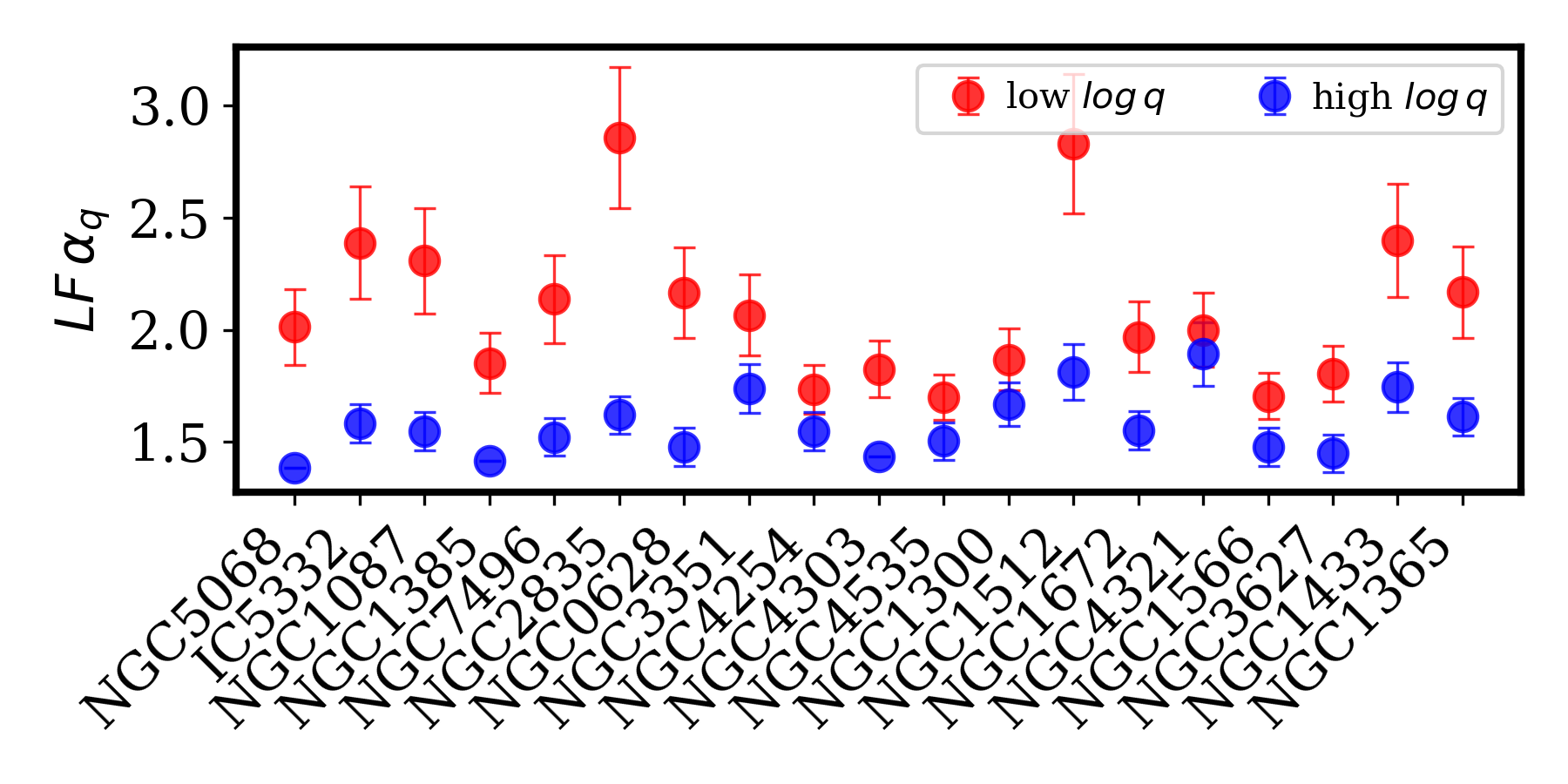}
    \caption{Slope of the LF for \HII\ regions with high (blue) and low (red) ionization parameter. Galaxies are ordered by increasing stellar mass from left to right and their names are indicated along the abscissa.
    \label{Figure:LF_Q_smallfig}}
\end{center}
\end{figure}

Fig.~\ref{Figure:LF_Q_smallfig} summarizes the results of the LF fit for the population of \HII\ regions with high and low ionization parameter (shown in Fig.~\ref{Figure:LF_Q}). It is evident that for 15 out of 19~galaxies the population of \HII\ regions with high $q$ has, within the uncertainties, a shallower LF compared to the \HII\ regions with low $q$. Also, looking at the shapes of the LFs we see that for a number of targets (e.g.\ \object{IC5332}, \object{NGC0628}, \object{NGC2835}, \object{NGC3627}, and \object{NGC5068}) the \HII\ regions with high/low $q$ constitute the bulk of the bright/\linebreak[0]{}faint \HII\ region populations. The interpretation of these results is further discussed in Sec.~\ref{sec:The aging effect and the ionization parameter}.  

\begin{table*}
    \caption{LF properties of \HII\ regions sub-samples in the PHANGS--MUSE galaxies.}
    \centering
    \resizebox{1\textwidth}{!}{
    \begin{tabular}{ccccccccccccc}
\toprule
Galaxy & $\alpha_{a}$ & $n_{a}$ & $\alpha_{i}$ & $n_{i}$ & $\alpha_{r<}$ & $n_{r<}$ & $\alpha_{r>}$ & $n_{r>}$ & $\alpha_{q<}$ & $n_{q<}$ & $\alpha_{q>}$ & $n_{q>}$ \\
\toprule
NGC5068 & -- & -- & -- & -- & 1.57$\pm$0.09 & 607 & 1.60$\pm$0.08 & 598 & 2.01$\pm$0.17 & 525 & 1.39$\pm$0.00 & 587 \\
IC5332 & -- & -- & -- & -- & 1.87$\pm$0.14 & 199 & 1.76$\pm$0.11 & 158 & 2.39$\pm$0.25 & 134 & 1.58$\pm$0.09 & 189 \\
NGC1087 & -- & -- & -- & -- & 1.66$\pm$0.09 & 303 & 1.84$\pm$0.13 & 135 & 2.31$\pm$0.23 & 170 & 1.55$\pm$0.08 & 268 \\
NGC1385 & 1.59$\pm$0.08 & 197 & 1.49$\pm$0.08 & 359 & 1.46$\pm$0.08 & 389 & 1.80$\pm$0.12 & 167 & 1.85$\pm$0.13 & 224 & 1.42$\pm$0.00 & 331 \\
NGC7496 & -- & -- & -- & -- & 1.73$\pm$0.11 & 174 & 1.65$\pm$0.09 & 127 & 2.14$\pm$0.20 & 136 & 1.52$\pm$0.08 & 164 \\
NGC2835 & 1.65$\pm$0.09 & 186 & 1.87$\pm$0.14 & 229 & 1.70$\pm$0.10 & 238 & 1.83$\pm$0.13 & 194 & 2.86$\pm$0.31 & 119 & 1.62$\pm$0.08 & 311 \\
NGC0628 & 1.66$\pm$0.09 & 883 & 1.85$\pm$0.13 & 389 & 1.76$\pm$0.11 & 572 & 1.66$\pm$0.09 & 700 & 2.17$\pm$0.20 & 510 & 1.48$\pm$0.09 & 633 \\
NGC3351 & -- & -- & -- & -- & 2.03$\pm$0.17 & 204 & 1.93$\pm$0.15 & 165 & 2.07$\pm$0.18 & 158 & 1.74$\pm$0.11 & 152 \\
NGC4254 & 1.58$\pm$0.09 & 792 & 1.65$\pm$0.09 & 637 & 1.57$\pm$0.09 & 921 & 1.72$\pm$0.10 & 508 & 1.74$\pm$0.11 & 583 & 1.55$\pm$0.08 & 846 \\
NGC4303 & 1.49$\pm$0.08 & 864 & 1.68$\pm$0.10 & 668 & 1.52$\pm$0.08 & 937 & 1.65$\pm$0.09 & 630 & 1.83$\pm$0.13 & 731 & 1.44$\pm$0.00 & 823 \\
NGC4535 & 1.54$\pm$0.08 & 341 & 1.71$\pm$0.10 & 545 & 1.64$\pm$0.09 & 455 & 1.62$\pm$0.09 & 479 & 1.70$\pm$0.10 & 459 & 1.50$\pm$0.08 & 393 \\
NGC1300 & 1.67$\pm$0.10 & 397 & 2.13$\pm$0.19 & 184 & 1.79$\pm$0.12 & 344 & 1.75$\pm$0.11 & 289 & 1.87$\pm$0.14 & 337 & 1.67$\pm$0.10 & 289 \\
NGC1512 & 1.96$\pm$0.16 & 125 & 2.18$\pm$0.20 & 77 & 2.18$\pm$0.21 & 108 & 1.92$\pm$0.15 & 101 & 2.83$\pm$0.31 & 78 & 1.81$\pm$0.13 & 128 \\
NGC1672 & 1.57$\pm$0.09 & 145 & 1.79$\pm$0.12 & 265 & 1.67$\pm$0.10 & 307 & 1.67$\pm$0.10 & 263 & 1.97$\pm$0.16 & 228 & 1.55$\pm$0.08 & 337 \\
NGC4321 & 1.84$\pm$0.13 & 217 & 2.15$\pm$0.20 & 134 & 1.91$\pm$0.14 & 233 & 2.02$\pm$0.17 & 150 & 2.00$\pm$0.17 & 213 & 1.89$\pm$0.14 & 168 \\
NGC1566 & 1.47$\pm$0.09 & 586 & 1.86$\pm$0.14 & 407 & 1.52$\pm$0.08 & 644 & 1.71$\pm$0.10 & 383 & 1.70$\pm$0.10 & 506 & 1.48$\pm$0.09 & 513 \\
NGC3627 & 1.48$\pm$0.09 & 320 & 1.70$\pm$0.10 & 282 & 1.51$\pm$0.08 & 350 & 1.65$\pm$0.09 & 307 & 1.80$\pm$0.12 & 259 & 1.45$\pm$0.08 & 371 \\
NGC1433 & -- & -- & -- & -- & 2.01$\pm$0.17 & 203 & 1.92$\pm$0.15 & 204 & 2.40$\pm$0.25 & 182 & 1.74$\pm$0.11 & 217 \\
NGC1365 & 1.60$\pm$0.08 & 111 & 1.89$\pm$0.14 & 182 & 1.77$\pm$0.12 & 146 & 1.71$\pm$0.10 & 200 & 2.17$\pm$0.20 & 119 & 1.61$\pm$0.08 & 226 \\
\bottomrule
\end{tabular}

    }
    \tablefoot{The table reports the galaxy name (col\,1) and the LF slope and the number of \HII\ regions involved in the LF fit for the \HII\ regions sub-samples in spiral arms areas (col\,2--3), in inter-arms areas (col\,4--5), in the inner star-forming disk (col\,6--7), in the outer star-forming disk (col\,8--9), with low ionization parameter (col\,10--11), and with high ionization parameter (col\,12--13).
    \label{tab:Table_LFsubsamples}}
\end{table*}

\section{Completeness, blending, and selection effects}\label{subsec:complcrow}

\begin{figure}[ht]
    \centering
    \includegraphics[width=0.5\textwidth]{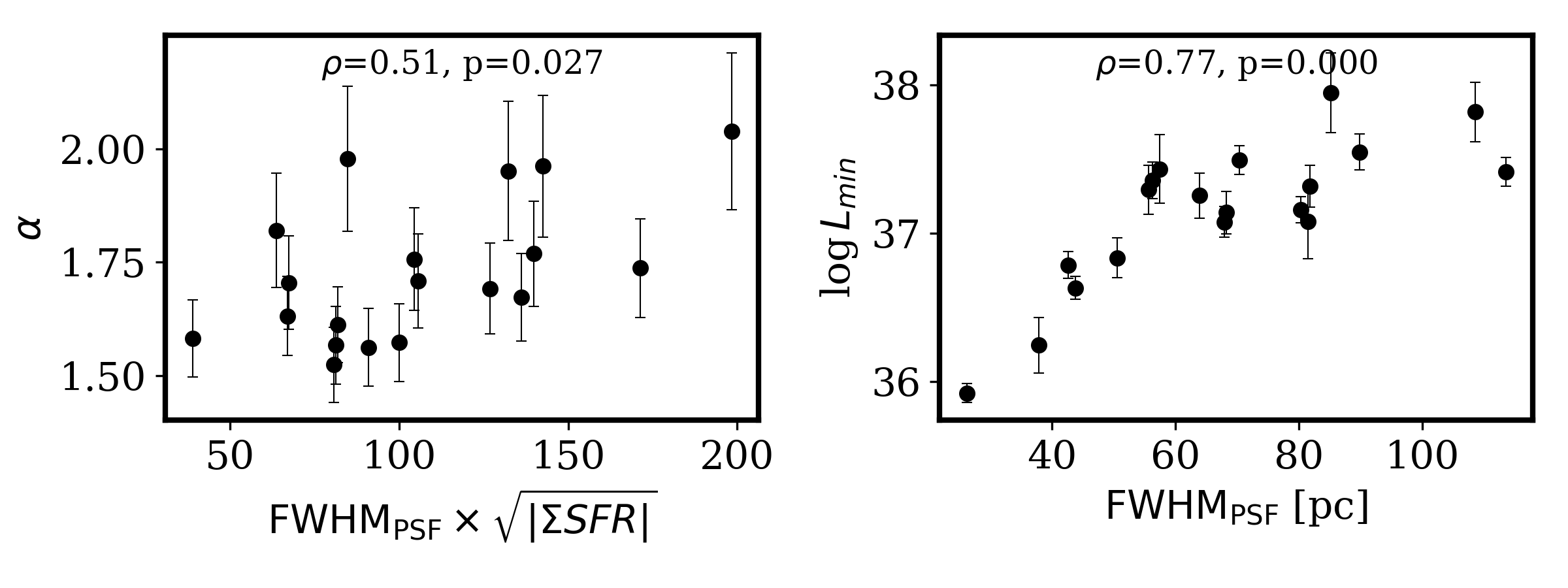}
    \caption{Left panel. LF slope plotted as a function of the $\mathrm{FWHM}_\mathrm{PSF}\times \sqrt{|\Ssfr|}$, used as a proxy to quantify the effect of blending. Right panel. LF \Lmin\ plotted as a function of the observations spatial resolution. For both panels we report the Spearman correlation coefficient $\rho$ and $p$-value of the plotted quantities.
    \label{fig:LF_validation}}
\end{figure}

Observational studies of LFs suffer from two main limitations. On the one hand, faint \HII\ regions are more difficult to detect against the diffuse \Ha\ background. On the other hand, the brighter \HII\ regions, often located in crowded areas such as spiral arms, are more subject to blending effects. This can artificially lower the number of detected faint regions and increase the luminosity of the brightest regions, respectively, with repercussions on the slope of the LF.  

The first limitation is related to the sensitivity and the completeness limit of the observations. We estimate the completeness limit of our LFs (shown in Fig.~\ref{Figure:global_LF}) in an empirical way by looking at the distribution of the \Ha\ flux outside the ionized nebula footprints in the \Ha\ emission line maps, which we refer to as the \Ha\ diffuse emission for simplicity. Given the PSF of the MUSE observations for a given galaxy, we estimate the completeness limit as the luminosity of a mock circular \HII\ region with the size of the PSF and a uniform \SHa\ equal to the 90th percentile (i.e.\ slightly larger than $1\sigma$ of the distribution) of the \Ha\ diffuse emission distribution. In this way, we take into account the bright tail of the diffuse \Ha\ background against which \HII\ regions can remain undetected. It is worth noting that tests injecting mock \HII\ regions are a good way forward to assess more robustly the completeness limit in the future.

As shown in Fig.~\ref{Figure:global_LF}, the completeness limit estimated in this way lies below the best-fitting \Lmin\ (i.e.\ the LF turnover luminosity) suggesting that the presence of a turnover point is not a completeness problem but rather an intrinsic feature of the LFs of our galaxies \citep[see][and references therein for a discussion]{Youngblood1999}.
Models have shown that a turnover point is to be expected when considering a constant SFR over time, while different SFH (e.g.\ a star formation burst in the past) can affect the shape of the LF in a more complex way \citep[see e.g.\ M31's double-peaked LF is attributed to a recent starburst;][]{Azimlu2011}. We note that, by taking into account the bright tail of the diffuse \Ha\ emission, we estimate a quite conservative completeness limit especially for the targets with stronger diffuse background (i.e.\ the galaxies for which the completeness limit is located close to the peak of the LF in Fig.~\ref{Figure:global_LF}). 
However, it should be noted that higher sensitivity and resolution are expected to move the turnover point to lower luminosities and, as shown in the right panel of Fig.~\ref{fig:LF_validation}, \Lmin\ is indeed driven by resolution of the observations in our survey. This constitutes a validation of our fitting approach and ensures that, at a given spatial resolution, the LFs $\alpha$ is extracted from a well-sampled luminosity regime.

On the other hand, the issue of blending is related to the spatial resolution of the observations and the number density (or filling factor) of star-forming regions. Intuitively, observations with coarser spatial resolution or a higher filling factor will cause nearby \HII\ regions to blend and be detected as a single \HII\ region, leading to a flattening of the LF. 
To assess the effect of blending to a first order, for each target, we multiply the $\mathrm{FWHM}_\mathrm{PSF}$, that takes into account the spatial resolution of the data, by $\sqrt{|\Ssfr|}$, that gives us a handle on the mean separation between \HII\ regions.
We use this parameter as a proxy for blending and check its relation with $\alpha$; as shown in the left panel of Fig.~\ref{fig:LF_validation}, we find a moderate correlation. 
Different studies \citep{Kennicutt1989,Bastian2007,Cook2016} showed that spatial resolution does not significantly change the slope of the LF at resolutions finer than few hundred parsecs like in the case of our data (i.e. the coarser physical resolution in our sample is $\sim110$~pc for \object{NGC1512} and \object{NGC1365}).
This indicates that, although present, the effect of blending is not the main driver of the correlations we report in Sec.~\ref{sec:LF variations between galaxies}. 
It is worth noting that our checks for the blending effect are either indirect or made a posteriori. In the future, progressively degrading the MUSE data cubes at a common resolution and analyzing how the LF parameters vary will give us a more comprehensive understanding of this matter.

In Appendix~\ref{appendix3}, we also discuss how the selection criteria, we use to select our \HII\ regions (see Sec.~\ref{sec:The final HII region catalogs}), have no significant effect on the measurements of the LF slope presented in Sec.~\ref{sec:HII regions luminosity function}. 

\section{Discussion: what sets the LF slope?}\label{sec:Discussion}

\subsection{The SFR surface density}\label{sec:The SFR and the cold gas surface density}

In Sec.~\ref{sec:LF variations between galaxies}, we find that in general $\alpha$ correlates better with the global star formation properties of our galaxies and especially with \Ssfr. We do not find any clear trends within the T-type, $\mathrm{[O/H]}$, and $M_{*}$ ranges probed by our sample. This indicates that the properties of star-forming regions are more closely connected to the global star formation properties in our sample. 

\begin{figure}
    \centering
    \includegraphics[width=0.5\textwidth]{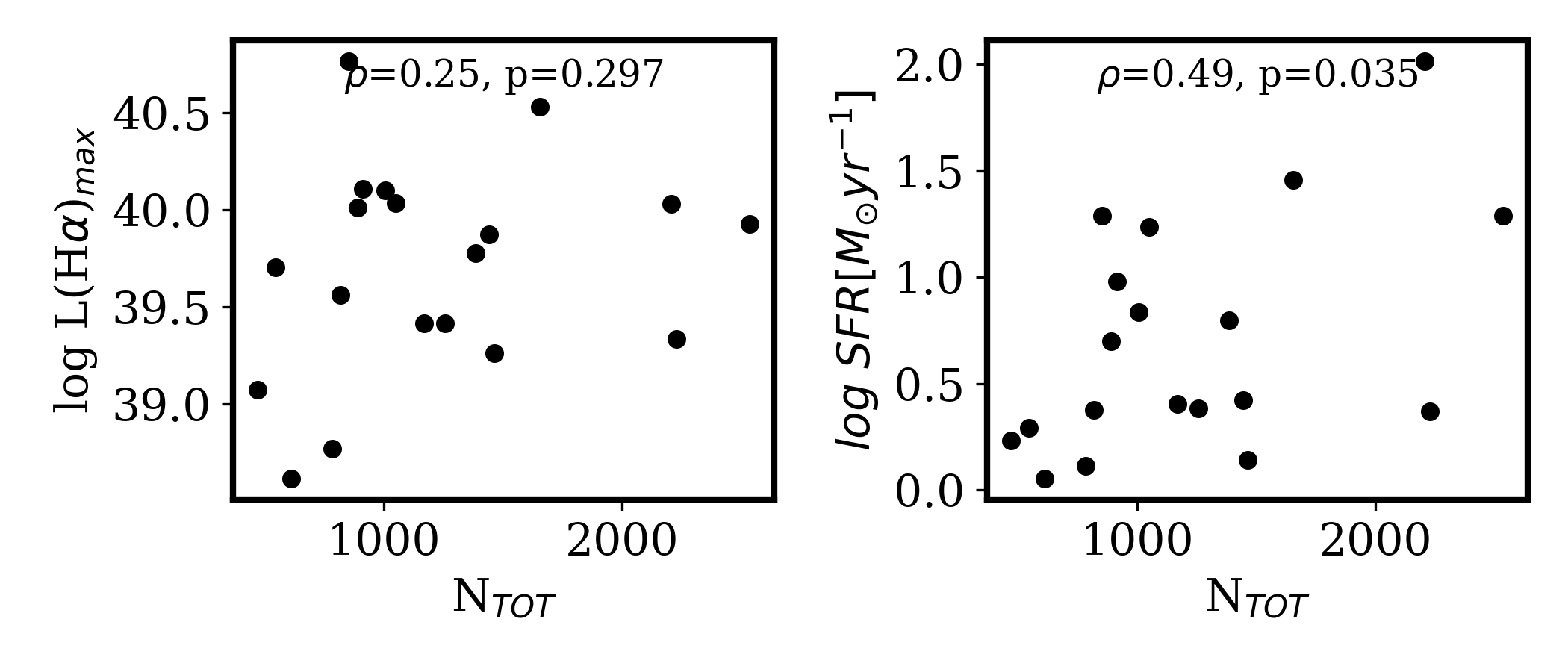}
    \caption{Maximum luminosity among the detected \HII\ regions (left panel) and the total SFR (right panel) as a function of the total number of detected \HII\ regions for the galaxies of our sample. For each panel we report the Spearman correlation coefficient~$\rho$ between the plotted quantities.
    \label{fig:SOS-effect}}
\end{figure}

One caveat to keep in mind is that the correlation between the LF and the SFR-related properties may be driven by the so-called size-of-sample effect \citep{Larsen2002,Weidner2004,Bastian2008,Cook2012,Whitmore2014}. This is a statistical effect (i.e. stochastic sampling) due to the fact that galaxies with higher SFRs tend to have more \HII\ regions and, in turn, the chances of observing a higher number of bright \HII\ regions increase in these galaxies. This is expected to drive a tight correlation between the total number of detected \HII\ regions (and subsequently the SFR) and the luminosity of the brightest star-forming region. Intuitively, the resulting effect is a flattening of the LF for galaxies with higher SFRs.
To check for this effect in our sample, in Fig.~\ref{fig:SOS-effect}, we plot the number of \HII\ regions (or similarly the number of ionized nebulae) against the maximum \HII\ region luminosity $L(\mathrm{H}\alpha)_\mathrm{max}$ and the global SFR. We find only a weak and a moderate trend, respectively.
This indicates that, thanks to the resolution of our observations and the nature of our sample, the flatter slopes we observe for the more star-forming galaxies are not simply due to a statistical effect. In addition, \cite{Cook2016} tested the size-of-sample effect using simulated LFs with $\alpha = -2$ and showed that the stochastic scatter of the LF slope starts to increase symmetrically when the number of detected \HII\ regions drops below~$100$. This implies that low number statistics increases the scatter in the measured LF slope but does not drive systematic changes of $\alpha$ in a specific direction. Moreover, our LFs are built with \HII\ region samples containing at least ${\sim}500$ objects, and the LF fits presented in Sec.~\ref{sec:HII regions luminosity function} rely on at least $200$ objects. For this reason, we expect the stochastic scatter of $\alpha$ to be lower in our sample compared to previous studies that often relied on smaller \HII\ region samples.

In agreement with \cite{Cook2016}, who studied the LF in a sample of $258$~nearby galaxies using \textit{GALEX} FUV data, we find that the LF slope correlates best with~\Ssfr. In Fig.~\ref{fig:LF_SLOPE_corr}f, we show the best-fitting relation found by \cite{Cook2016} for a qualitative comparison. We note that our points mostly lie above this relation. A direct comparison is, however, not straightforward because of the different type of observations. In fact, FUV and \Ha\ LFs probe different ages of \HII\ regions, ${\lesssim}100$~Myr and ${\lesssim}10$~Myr, respectively \citep[see e.g.][]{Haydon2020}. \cite{Oey1998A} predicted that older \HII\ region populations (e.g.\ derived from FUV) should have steeper slopes than younger populations (e.g.\ derived from~\Ha). Overall, the average slope of the nebular LF \citep[$-2\pm0.5$;][]{Kennicutt1989,Elmegreen1999}, including our study ($-1.73\pm0.15$), is compatible with what has been found in the FUV \citep[$-1.76\pm0.3$;][]{Cook2016}. 

The observed trend indicates that a shallower LF and, thus, the formation of higher relative number of massive, luminous \HII\ regions is connected with a higher~\Ssfr. We know that the star formation process is fueled by cold gas. We thus expect the SFR-related properties we measure to be connected with the way how cold gas reservoir fuels star formation, which in turn is connected with the sSFR, the $t_\mathrm{dep}$, and the molecular gas fraction. 

As discussed in Sec.~\ref{sec:LF variations between galaxies}, the offsets from the resolved star formation scaling relations can be used as a proxy for these parameters.  We find that the stronger correlation is between the LF slope and \DrSFMS\ (i.e.\ a measurement of the sSFR) indicating that, when the mass fraction between young and old stars is higher, galaxies tend to form a higher relative number of luminous \HII\ regions, in agreement with what we find when estimating the global sSFR directly from the MUSE data (Fig.~\ref{fig:LF_SLOPE_corr}e). A~slightly weaker but still significant trend is found between the LF slope and \DrMGMS\ (i.e.\ a measurement of the molecular gas fraction) suggesting that, when in a galaxy the fraction of cold gas that will eventually be converted into stars is higher, the relative number of bright and massive \HII\ regions produced by ongoing star formation is also higher.

\cite{Kruijssen2012} formulated a theoretical framework (invoked also by \citealp{Cook2016} to explain their results), in which higher SFR and cold gas densities can result in a higher pc-scale star formation efficiency that, in turn, enhances the cluster formation efficiency, defined as the ratio between the cluster formation rate (CFR) and the SFR ($\Gamma \equiv \mathrm{CFR}/\mathrm{SFR}$). This means that, at higher SFR and gas densities, more stars form in (gravitationally bound) star clusters.
Observational evidence has indeed been found that $\Gamma$ correlates well with the SFR surface density \citep[see e.g.]{Goddard2010,Adamo2011,Cook2012,Adamo2015,Johnson2016,Adamo2020a,Adamo2020b}.
The trend that we find in Fig.~\ref{fig:LF_SLOPE_corr}e between $\alpha$ and \Ssfr\ goes in the direction predicted by this scenario. 
It is worth noting that, with the available PHANGS--HST data \citep{Lee2021}, it will be possible to directly measure $\Gamma$ and compare it to our measured \HII\ region LF slopes.

Quantitatively, the slope of the \HII\ region LF might be expected to be similar to the slope of the GMC mass function \citep[historically estimated to lie between $1.6$ and $2.0$, e.g.][although more recent works find a wider range of slopes, see e.g.\ \citealt{Colombo2014,Rosolowsky2021}]{Kennicutt2012}. If star formation within GMCs is self-similar \citep{Efremov1998} and the star formation efficiency is independent of the GMC mass at the observable, high-mass end of the mass range, then the slope of the \HII\ region LF is also expected to be $\alpha=1.6{-}2.0$. Steeper slopes may then be the result of incomplete sampling of the stellar initial mass function (IMF), which occurs if the stellar populations born within the parent GMCs have a low mass (${\la}10^4~M_\odot$; \citealt{Krumholz2015}). Simple analytical models, in which the maximum GMC and stellar population mass scales are regulated by gravitational instability and stellar feedback, predict that the maximum mass increases approximately with the total gas surface density as $\Sigma_\textrm{gas}^{1.4}$, and therefore approximately linearly with \Ssfr\ \citep{Kruijssen2014b,Reinacampos2017}. This linear relation is indeed observed by \citet{Johnson2017}, who found that a maximum mass of $10^4~\textrm{M}_\odot$ is reached at $\log{(\Sigma_\textrm{SFR}\ [\textrm{M}_\odot~\textrm{yr}^{-1}~\textrm{kpc}^{-2}])} \approx -2.6$. Above this SFR surface density, we expect the \HII\ region LF to be close to the slope of the GMC mass function (which in environments of high gas and SFR surface densities is about $1.6$, e.g.\ \citealt{Colombo2014,Hughes2016}). Toward lower SFR surface densities, we expect the \HII\ region LF to steepen. The same behavior, in which $\alpha$ declines with \Ssfr\ for $\log{(\Sigma_\textrm{SFR}\ [\textrm{M}_\odot~\textrm{yr}^{-1}~\textrm{kpc}^{-2}])} \la -2.6$ and stays approximately constant (or declines less steeply) for $\log{(\Sigma_\textrm{SFR}\ [\textrm{M}_\odot~\textrm{yr}^{-1}~\textrm{kpc}^{-2}])} \ga -2.6$ was also found by \citet{Cook2016} and is illustrated in (Fig.~\ref{fig:LF_SLOPE_corr}f). Here, as a reference to illustrate this flattening, we marked with horizontal lines the $\alpha$ corresponding to the \cite{Cook2016} best-fitting line at $\Ssfr \approx 2.6$ and the mean $\alpha$ of our samples at $\Ssfr>2.6$. Both values are remarkably close to $\alpha=1.6$, the slope of the GMC mass function in the high gas and \Ssfr regime. Based on the discussion in this section, we expect this behavior to arise from a combination of increased stellar clustering at birth toward higher gas densities (and therefore pressures; \citealt{Kruijssen2012}), as well as of decreased IMF sampling toward low SFR surface densities. Higher-resolution observations may be necessary to assess how blending of adjacent \HII\ regions may affect these trends.

\subsection{The strength of spiral arms}\label{sec:The strength of spiral arms}

In Sec.~\ref{sec:Environment}, we showed that a flattening of the LF slope in spiral arm areas compared to inter-arm areas is evident for only about half of the galaxies with spiral arms in our sample (i.e.\ six out of~13 galaxies, namely \object{NGC1300}, \object{NGC1365}, \object{NGC1566}, \object{NGC1672}, \object{NGC3627}, and \object{NGC4303}). 
The Monte Carlo simulations by \cite{Oey1998} suggested that such variations can be expected when the \HII\ regions populating the spiral arm areas trace a current burst of coeval star formation while the inter-arm population is an aged version thereof (i.e.\ aging effect). However, the same authors noted that, at any given location in a galaxy disk, the timescale between the passage of spiral density waves is of the order of $40$~Myr \citep[see e.g.][]{Rand1993}, much longer than the observed lifetime of \HII\ regions \citep[in the interval $5{-}10$~Myr;][]{Chevance2020R}. This implies that \HII\ regions detected in inter-arm areas are, potentially, tracing both a population of aging \HII\ regions and recent star formation that is genuinely happening in inter-arm areas. 

Large-scale dynamical processes shaping the distribution of cold gas that ultimately fuels the star formation can also influence the variation (or not) in the LF slopes between spiral arm and inter-arm areas. We might expect this to depend on the nature of the dynamical perturbation present in the disk (material spiral arms vs.\ density waves) and its strength.  \cite{Rand1992}, for example, found a significant difference between the LF slopes of spiral arm and inter-arm \HII\ regions in the grand-design interacting galaxy M51 and attributed it to a shallower GMCs MF in the spiral arms. In NGC~6814, on the other hand, \cite{Knapen1993} argued that the spiral arms are not strong enough to build up the large cloud masses needed to produce the number of giant \HII\ regions that make the spiral arm LF shallower than the inter-arm LF.

Strong spiral arms are not only capable of building high gas densities and growing high mass molecular clouds, they can also impact the likelihood of in situ inter-arm star formation, thus influencing the LFs of \HII\ regions.  Indeed, the presence of spiral arms imposes a strong organization on the gas distribution, which acts to concentrate star formation into the spiral arms.  Comparing a high resolution simulation of an M51-like galaxy with a comparable isolated galaxy, \cite{Tress2020,Tress2021} concluded that strong spiral arms gather molecular gas and clouds without dramatically affecting their properties, yielding similar overall star formation rates. 

In our case, we might thus expect that variations in cloud mass functions and \HII\ region LF slopes will be weak in systems with more uniformly distributed \HII\ regions across the disk (weaker spirals) and stronger in systems with strong spiral arms. In line with this, we find that the galaxies showing a significant environmental change in $\alpha$ are those in which the distribution of the \HII\ regions traces quite clearly the spiral arms and fewer \HII\ regions are detected in between spiral arms (i.e.\ less ongoing star formation in the inter-arm areas).  %(We note that all these galaxies have strong bars or grand design spiral arms.)  
The galaxies with more uniformly distributed \HII\ regions across the disk (e.g.\ \object{NGC0628}, \object{NGC4254}, and \object{NGC4321}), on the other hand, exhibit comparable $\alpha$ in the two environments.

We quantify this further using the azimuthal contrast in CO brightness across our targets, which serves as a proxy for spiral arm strength \citep{Meidt2021}.  For our targets, we measure the CO contrasts in a set of 150-pc wide radial bins, calculating the ratio between the brightness of the $84$th percentile of the CO distribution in a given ring and a reference level defined to capture the level of low brightness inter-arm emission in \cite{Meidt2021}.  We then take the average of the CO contrasts measured across the full area of interest.

\begin{figure}
    \centering
    \includegraphics[width=0.5\textwidth]{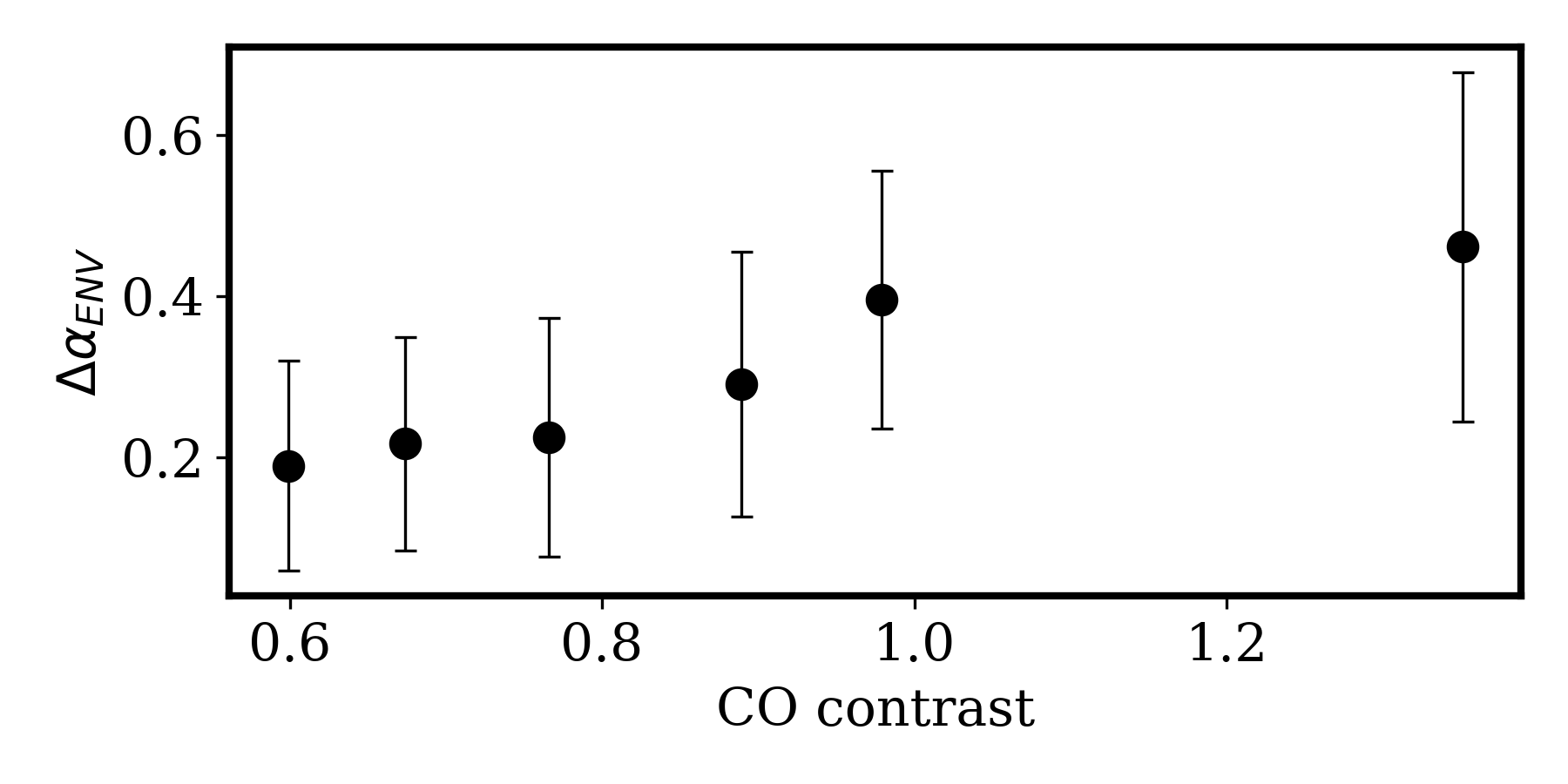}
    \caption{Change in the LF slope between inter-arm and spiral arm environments $\Delta\alpha_{ENV}$ (derived as $\alpha_{i}-\alpha_{a}$ as reported in Table~\ref{tab:Table_LFsubsamples}) as a function of the \mbox{CO(2--1)} contrast between the $84$th percentile and the reference level defined in \cite{Meidt2021} for six galaxies in our sample.
    \label{fig:LF_DeltaSlope_ENV}}
\end{figure}

As shown in Fig.~\ref{fig:LF_DeltaSlope_ENV}, for the six galaxies showing environmental changes of the LF slope, the magnitude of such variations appears to correlate with the spiral arm contrast.  The remaining galaxies with evidently weaker spiral arms, more uniformly distributed \HII\ regions, and no significant variation in LF slope from arm to inter-arm,  do not exhibit a similar trend.

Overall, these findings suggest that environmental variations in $\alpha$ are sensitive to how spiral arms shape the distribution of the more massive molecular clouds that serve as the birthplace of the brightest \HII\ regions.  A study of environmental variations in GMC mass functions in our sample \citep[e.g.][Hughes et al.\ in preparation]{Rosolowsky2021} will be an important tool for recognizing how much of the change in LF slope from arm to inter-arm is due to the evolution in the distribution of progenitor clouds and how much is due to, i.e. aging, as considered in the next section. The morphology of inter-arm gas (in the form of spurs/\linebreak[0]{}feathers and non-spiral clump features) may also yield important clues on the nature of inter-arm star formation and the change in \HII\  region LF slope.

Given the resolution of our observations, we cannot fully exclude that \HII\ region blending, which affects spiral arms to a greater extent, has an effect on our findings. An additional factor that we are not considering and that can play a role in this context is the SFH of the galaxies. \cite{Feinstein1997} predicted that a past starburst can manifest as an additional peak below the LF turnover luminosity and result in a double-peaked LF as the one observed for M31 by \cite{Azimlu2011}. We would need observations that are at least one order of magnitude deeper in terms of $L_\mathrm{H\alpha}$ to search for such features and, for this reason, we cannot deduce or fold in information on the SFH in our study of the LF.

\subsection{The aging effect and the ionization parameter}\label{sec:The aging effect and the ionization parameter} 

\begin{figure}[ht]
    \centering
    \includegraphics[width=0.5\textwidth]{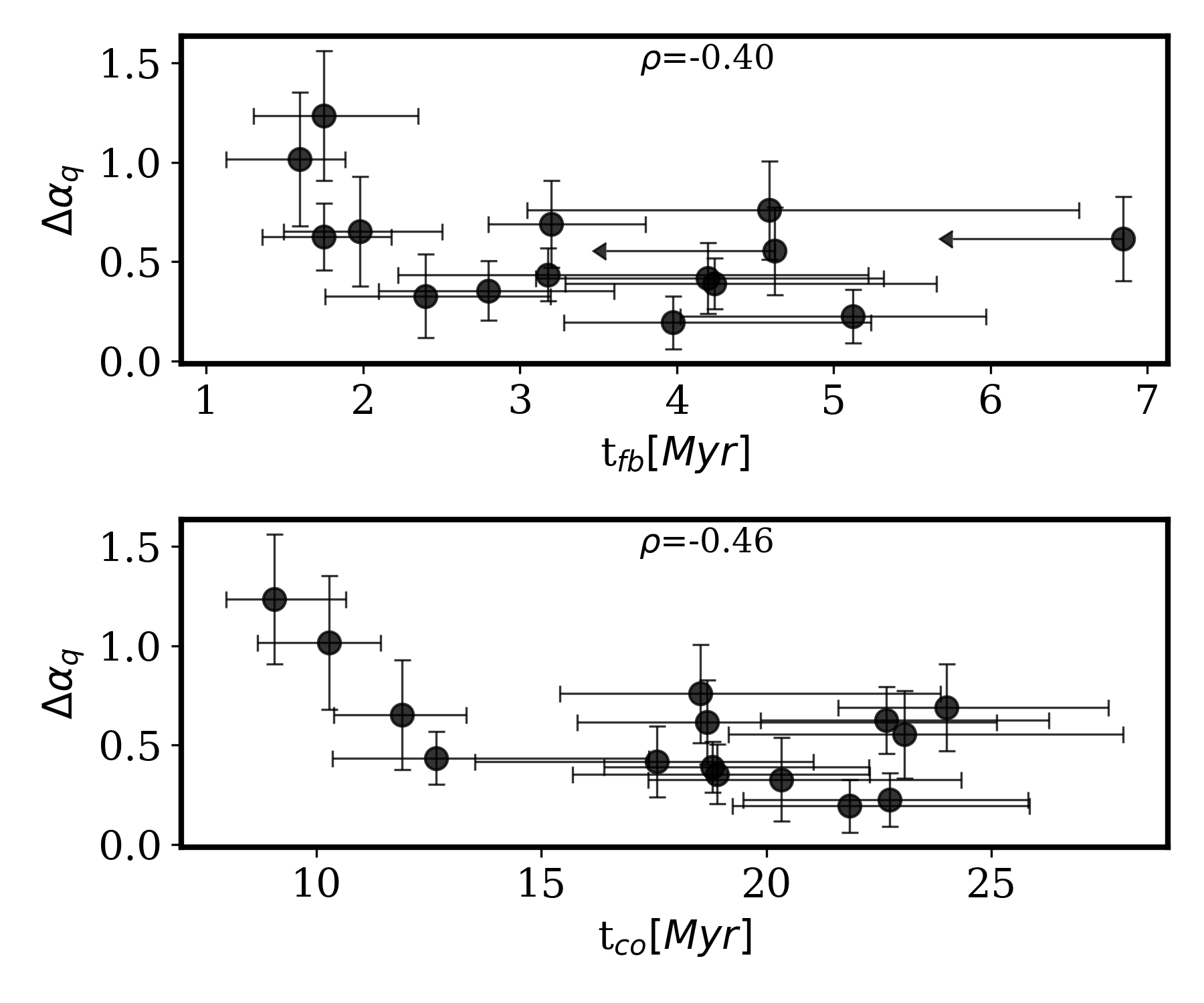}
    \caption{Change in LF slope between \HII\ regions with low and high ionization parameter (derived as $\alpha_{q<}-\alpha_{q>}$ as reported in Table~\ref{tab:Table_LFsubsamples}) as a function of the feedback timescales (upper panel) and the GMC lifetime (lower panel). For each panel we report the Spearman correlation coefficient~$\rho$ between the plotted quantities.
    \label{fig:LF_DeltaSlope_Q}}
\end{figure}

According to \citet[eq.~(10)]{Charlot2001}, by assuming that \HII\ regions are well approximated by Str{\"o}mgren spheres, the ionization parameter changes with time as $U(t) \propto (Q(t) n_\mathrm{H})^{1/3}$, where $Q$ is the rate of ionizing photons produced by the central star/\linebreak[0]{}cluster and $n_\mathrm{H}$ is the hydrogen density. Considering the typical evolution of an \HII\ region, we can expect the ionization parameter to drop as time passes: as the central ionizing star/\linebreak[0]{}cluster evolves its ionizing flux drops while the gas density decreases due to stellar feedback dispersing the \HII\ region's gas \citep[see e.g.][]{Rahner2017, Rahner2019, Pellegrini2020}.

The LF slope can, in principle, be used to test this expectation by using the ionization parameter to draw sub-samples of \HII\ regions in the way we did in Sec.~\ref{sec:Age Trends}. If the ionization parameter~$q$ is related to the evolutionary stage of an \HII\ region, it would separate less and more evolved \HII\ regions and, ultimately, we would expect the aging effect to drive changes in the LF slope \citep{Oey1998}. This is the same effect discussed for environmental variations in the previous section. What we find in Sec.~\ref{sec:Age Trends} is indeed a steeper LF for the population of \HII\ regions with lower~$q$ that, for a number of galaxies, constitutes the bulk of the faint \HII\ regions. This corroborates the expectations of \cite{Charlot2001} from an empirical point of view, indicating that the ionization parameter can be used as a good proxy for the evolutionary stage of \HII\ regions.

However, we know that \HII\ regions are not always well represented by Str{\"o}mgren spheres, e.g.\ they can be asymmetric or even broken shells \citep[see e.g.][]{Pellegrini2012}. Also, assuming equal physical conditions for an \HII\ region, the same ionizing flux (and thus ionization parameter) can be produced by a single very young star or a cluster of slightly older stars. Another parameter that is relevant for the interpretation of our results is the metallicity, and its long-debated degeneracy with age.
In fact, an increase in stellar metallicity is also expected to lower the ionization parameter: as stellar atmospheres of O~stars become cooler \citep{Massey2005} and the mechanical energy due to photon scattering increases, dispersing gas more efficiently \citep{Dopita2006}.

Interestingly, using PHANGS--MUSE data, \cite{Kreckel2019} showed that \HII\ regions with higher \SIII/\SII line ratios (and thus higher ionization parameter) also tend to have higher \Ha-to-FUV flux ratios, indicating younger star cluster ages.
We note that the PHANGS--HST data set will be an ideal test-bed to actually connect our \HII\ regions (and their ionization parameters) to the ages of their ionizing star(s)/\linebreak[0]{}cluster and shed a light on this topic.
The \SIII/\SII line ratio we use to compute the ionization parameter has been found to have only a weak dependence on the gas-phase metallicity for \HII\ regions located in spiral galaxies. \citep{Dors2005,Garnett1997,Kennicutt1996,Dors2011}. However, the topic is still debated as other works presented contrasting results (see e.g.\ \citealp{Bresolin1999} for disk galaxies and \citealp{Dopita2014} for Luminous Infrared Galaxies) and, above all, in our sample a clear correlation has been found between \SIII/\SII and the gas-phase metallicity \citep[see fig.~4 in][]{Kreckel2019}. We confirm this by looking at the distribution of the \HII\ region metallicities for the sub-samples with high and low~$q$.
However, if the gas-phase metallicity plays a role in setting the slope of the LF, we would see it by defining \HII\ region sub-samples based on their metallicity and looking for variations in~$\alpha$. We conducted this test and do not find such variations. We thus conclude that the variations in the LF slope we observe in Fig.~\ref{Figure:LF_Q} are mainly driven by aging of the \HII\ regions rather than their metal content.

By comparing the spatial distribution of the warm ionized gas (i.e.\ \HII\ regions) and the cold molecular gas (i.e.\ GMCs), it is now possible to infer e.g.\ the lifetime of GMCs $t_\mathrm{CO}$ and the timescale over which stellar feedback acts $t_\mathrm{fb}$ \cite[see e.g.][]{Kruijssen2019,Chevance2020,Kim2021}.
Assuming that the aging effect is the main driver behind the LF slope variations we observe between the \HII\ region sub-samples with high and low~$q$, we can reasonably expect the magnitude of such variations $\Delta \alpha_{q}$ to correlate with the timescales that regulate the star formation process (e.g.\ the lifetime of GMCs serving as the stellar nurseries, the timescales of stellar feedback) in a given galaxy.
Following the ``uncertainty principle for star formation'' method of \citet{Kruijssen2018} and expanding the results of \cite{Chevance2020}, Kim et al. (in preparation) derived $t_\mathrm{CO}$ and $t_\mathrm{fb}$ for most of the galaxies in our sample.
In that work, \HII\ regions are traced via the \Ha\ emission in the PHANGS--H$\alpha$ data (Razza et al. in preparation), while GMCs are traced via the \mbox{CO(2--1)} emission in the PHANGS--ALMA data \citep{Leroy2021}.
It should be noted that for \object{IC5332} there is no significant \mbox{CO(2--1)} emission detected to apply this method and the $t_\mathrm{CO}$ and $t_\mathrm{fb}$ measurements for \object{NGC0628} and \object{NGC3627} are taken from \cite{Chevance2020}.
As shown in Fig.~\ref{fig:LF_DeltaSlope_Q}, we find a weak-to-moderate/\linebreak[0]{}moderate anti-correlation between $\Delta \alpha_{q}$ and both $t_\mathrm{fb}$ and $t_\mathrm{CO}$. 
These trends suggest that when stellar feedback occurs on shorter timescales and the GMC lifetime is shorter (i.e.\ they get dispersed faster due to stellar feedback) the difference between the LF slope of \HII\ regions with high and low~$q$ increases, as it would be expected if those populations are intrinsically different in terms of their age.

\section{Conclusions}\label{sec:conclusions}

In this paper, we have studied the nebular luminosity functions (LFs; i.e.\ built using the \Ha\ luminosity of \HII\ regions) in the star-forming disks of 19~nearby galaxies making up the PHANGS--MUSE sample. Thanks to the exquisite spatial resolution (mean $\mathrm{PSF}_\mathrm{FWHM} = 67$~pc) and sensitivity of the data, we were able to build a catalog of about $31\,400$ ionized nebulae from which we extracted an unprecedented sample of about $23\,000$ \HII\ regions. With MUSE covering a large part of the optical spectrum we were able to derive global properties (e.g.\ $M_{*}$, SFR, sSFR) for our galaxies and characterize our \HII\ regions in terms of their dust attenuation (via the \Ha/\Hb\ Balmer decrement), gas-phase metallicity $\mathrm{O/H}$, and ionization parameter~$q$.
The average number of \HII\ regions detected per galaxy is about $1200$, marking a significant improvement with respect to previous spectroscopic studies at a comparable spatial resolution.

We fit the LFs via maximum likelihood estimation \citep[MLE;][]{Alstott2014}. This method does not require us to bin the data, is especially suited for heavily tailed distributions, and overcomes typical pitfalls of more commonly used methods (e.g.\ histograms combined with linear regression). We find an LF slope $\alpha = -1.73\pm0.15$ in agreement with what has been found in the literature for Sa--Sc type galaxies \cite[see e.g.][]{Kennicutt1989,Elmegreen1999,Whitmore2014}.   
Given the sensitivity of the MUSE observations, our LFs are mainly probing massive star formation happening in star clusters and associations. We report evidence that for a number of galaxies their LFs steepen at luminosities above $\log(L_\mathrm{H\alpha} \ [\mathrm{erg~s^{-1}}]) = 38.6$ (i.e.\ type~II LF). However, this appears as a subtle feature compared to what has been shown in past studies \citep[e.g.\ ][]{Beckman2000}. In general, a power-law with a single slope offers a good representation of the LFs and allows us to analyze uniformly the LFs across our sample.

We find that the LF slope remains largely unchanged for the inner and outer parts of the star-forming galaxy disks (Sec.~\ref{sec:Radial Trends}). For the galaxies with \HII\ regions more uniformly distributed across the star-forming disk, we do not find any significant difference between the spiral arm and inter-arm LFs slope. On the other hand, we find environmental (i.e. arm vs.\ inter-arm) LF variations for galaxies showing \HII\ regions preferentially located along spiral arms. We attribute these variations to the spiral arms increasing the molecular clouds arm--inter-arm mass contrast and find suggestive evidence that they are more pronounced in galaxies with spiral arms driving stronger dynamical perturbations. 

In line with the results of \cite{Cook2016}, who studied the LF of star-forming regions as traced by \textit{GALEX} FUV imaging, we find that galaxies with a higher \Ssfr\ have a flatter LF, meaning that their relative number of bright \HII\ regions is higher compared to galaxies with lower~\Ssfr. 
This potentially connects to fundamental changes in the physics regulating star formation in galaxy disks.
The trend we observe between $\alpha$ and \Ssfr\ is not driven by low-number statistics and brings further evidence to the prediction that the clustering of young stars is enhanced at high gas surface densities and star formation efficiencies \citep{Kruijssen2012}. We propose that $\alpha$ may increase even further at $\log{(\Sigma_\textrm{SFR}\ [M_\odot~\textrm{yr}^{-1}~\textrm{kpc}^{-2}])} \la -2.6$, where observations and models find that the maximum masses of the new-born stellar populations should become so low \citep[e.g.][]{Johnson2017,Reinacampos2017} that they are affected by stochastic sampling of the stellar IMF \citep[e.g.][]{Krumholz2015}.

Finally, we find that, within each of our galaxies, \HII\ regions with high ionization parameter have a shallower LF compared to \HII\ regions with low ionization parameter.
Such variations are compatible with an aging effect suggesting that \HII\ regions with high~$q$ are the youngest star-forming regions while the ones with low~$q$ are more evolved. The lack of $\alpha$ trends/\linebreak[0]{}variations related to the \HII\ regions' gas-phase metallicity persuades us that the main parameter regulating changes in $q$ is age. We also bring some tentative evidence suggesting that $\alpha$ variations between these \HII\ region populations are more evident when stellar feedback acts on shorter timescales and GMCs lifetimes are shorter.

\section{Future prospects}\label{sec:future prospects}

We note that this work can be expanded in a number of promising directions in the near future, especially thanks to the availability of \textit{HST} and ALMA data for the entire PHANGS--MUSE sample.
The PHANGS--ALMA data will allow us to characterize GMCs and study their mass function (MF). Comparing GMC MF slopes with the \HII\ regions LF slopes reported in this paper can bring new insights into if/how physical properties of the molecular gas influence the physics of star formation. 
On the other side, the PHANGS--HST data can be used to estimate the ages and metallicities of the star(s)/\linebreak[0]{}cluster ionizing the \HII\ regions and help disentangling what is the main driver behind changes in the \HII\ regions' ionization parameter. Furthermore, by providing a direct estimate of the clustered star formation efficiency (defined as the ratio between the cluster formation rate and the SFR; $\Gamma \equiv \mathrm{CFR}/\mathrm{SFR}$) it will be possible to compare $\Gamma$ to the \HII\ region LF slopes measured in this paper and test whether this is the mechanism behind the flattening of the LF that we observe with increasing~\Ssfr.

\begin{acknowledgements}
This work was carried out as part of the PHANGS collaboration.
FS, ES, RME, TS, and TGW acknowledge funding from the European Research Council (ERC) under the European Union’s Horizon 2020 research and innovation programme (grant agreement No. 694343).
ATB would like to acknowledge funding from the European Research Council (ERC) under the European Union’s Horizon 2020 research and innovation programme (grant agreement No.726384/Empire). JMDK and MC gratefully acknowledge funding from the Deutsche Forschungsgemeinschaft (DFG, German Research Foundation) through an Emmy Noether Research Group (grant number KR4801/1-1), as well as from the European Research Council (ERC) under the European Union's Horizon 2020 research and innovation programme via the ERC Starting Grant MUSTANG (grant agreement number 714907). JMDK, MC, and JK gratefully acknowledge funding from the DFG through the DFG Sachbeihilfe (grant number KR4801/2-1).
ER acknowledges the support of the Natural Sciences and Engineering Research Council of Canada (NSERC), funding reference number RGPIN-2017-03987. KK gratefully acknowledges funding from the German Research Foundation (DFG) in the form of an Emmy Noether Research Group (grant number KR4598/2-1, PI Kreckel). EJW acknowledges support from the Deutsche Forschungsgemeinschaft (DFG, German Research Foundation) – Project-ID 138713538 – SFB 881 (“The Milky Way System”, subproject P2).
Based on observations collected at the European Southern Observatory under ESO programmes 1100.B-0651, 095.C-0473, and 094.C-0623 (PHANGS-MUSE; PI Schinnerer), as well as 094.B-0321 (MAGNUM; PI Marconi), 099.B-0242, 0100.B-0116, 098.B-0551 (MAD; PI Carollo) and 097.B-0640 (TIMER; PI Gadotti).
\end{acknowledgements}

\bibliographystyle{aa}
\bibliography{biblio.bib}

\begin{thebibliography}{133}
\expandafter\ifx\csname natexlab\endcsname\relax\def\natexlab#1{#1}\fi

\bibitem[{{Adamo} {et~al.}(2020{\natexlab{a}}){Adamo}, {Hollyhead}, {Messa},
  {Ryon}, {Bajaj}, {Runnholm}, {Aalto}, {Calzetti}, {Gallagher}, {Hayes},
  {Kruijssen}, {K{\"o}nig}, {Larsen}, {Melinder}, {Sabbi}, {Smith}, \&
  {{\"O}stlin}}]{Adamo2020b}
{Adamo}, A., {Hollyhead}, K., {Messa}, M., {et~al.} 2020{\natexlab{a}}, \mnras,
  499, 3267

\bibitem[{{Adamo} {et~al.}(2015){Adamo}, {Kruijssen}, {Bastian}, {Silva-Villa},
  \& {Ryon}}]{Adamo2015}
{Adamo}, A., {Kruijssen}, J.~M.~D., {Bastian}, N., {Silva-Villa}, E., \&
  {Ryon}, J. 2015, \mnras, 452, 246

\bibitem[{{Adamo} {et~al.}(2011){Adamo}, {{\"O}stlin}, \&
  {Zackrisson}}]{Adamo2011}
{Adamo}, A., {{\"O}stlin}, G., \& {Zackrisson}, E. 2011, \mnras, 417, 1904

\bibitem[{{Adamo} {et~al.}(2020{\natexlab{b}}){Adamo}, {Zeidler}, {Kruijssen},
  {Chevance}, {Gieles}, {Calzetti}, {Charbonnel}, {Zinnecker}, \&
  {Krause}}]{Adamo2020a}
{Adamo}, A., {Zeidler}, P., {Kruijssen}, J.~M.~D., {et~al.} 2020{\natexlab{b}},
  \ssr, 216, 69

\bibitem[{{Alstott} {et~al.}(2014){Alstott}, {Bullmore}, \&
  {Plenz}}]{Alstott2014}
{Alstott}, J., {Bullmore}, E., \& {Plenz}, D. 2014, PLoS ONE, 9, e85777

\bibitem[{{Anand} {et~al.}(2021){Anand}, {Lee}, {Van Dyk}, {Leroy},
  {Rosolowsky}, {Schinnerer}, {Larson}, {Kourkchi}, {Kreckel}, {Scheuermann},
  {Rizzi}, {Thilker}, {Tully}, {Bigiel}, {Blanc}, {Boquien}, {Chandar}, {Dale},
  {Emsellem}, {Deger}, {Glover}, {Grasha}, {Groves}, {Klessen}, {Kruijssen},
  {Querejeta}, {S{\'a}nchez-Bl{\'a}zquez}, {Schruba}, {Turner}, {Ubeda},
  {Williams}, \& {Whitmore}}]{Anand2021}
{Anand}, G.~S., {Lee}, J.~C., {Van Dyk}, S.~D., {et~al.} 2021, \mnras, 501,
  3621

\bibitem[{{Azimlu} {et~al.}(2011){Azimlu}, {Marciniak}, \&
  {Barmby}}]{Azimlu2011}
{Azimlu}, M., {Marciniak}, R., \& {Barmby}, P. 2011, \aj, 142, 139

\bibitem[{{Bacon} {et~al.}(2010){Bacon}, {Accardo}, {Adjali}, {Anwand},
  {Bauer}, {Biswas}, {Blaizot}, {Boudon}, {Brau-Nogue}, {Brinchmann},
  {Caillier}, {Capoani}, {Carollo}, {Contini}, {Couderc}, {Daguis{\'e}},
  {Deiries}, {Delabre}, {Dreizler}, {Dubois}, {Dupieux}, {Dupuy}, {Emsellem},
  {Fechner}, {Fleischmann}, {Fran{\c{c}}ois}, {Gallou}, {Gharsa}, {Glindemann},
  {Gojak}, {Guiderdoni}, {Hansali}, {Hahn}, {Jarno}, {Kelz}, {Koehler},
  {Kosmalski}, {Laurent}, {Le Floch}, {Lilly}, {Lizon}, {Loupias}, {Manescau},
  {Monstein}, {Nicklas}, {Olaya}, {Pares}, {Pasquini}, {P{\'e}contal-Rousset},
  {Pell{\'o}}, {Petit}, {Popow}, {Reiss}, {Remillieux}, {Renault}, {Roth},
  {Rupprecht}, {Serre}, {Schaye}, {Soucail}, {Steinmetz}, {Streicher}, {Stuik},
  {Valentin}, {Vernet}, {Weilbacher}, {Wisotzki}, \& {Yerle}}]{Bacon2010}
{Bacon}, R., {Accardo}, M., {Adjali}, L., {et~al.} 2010, in Society of
  Photo-Optical Instrumentation Engineers (SPIE) Conference Series, Vol. 7735,
  Ground-based and Airborne Instrumentation for Astronomy III, ed. I.~S.
  {McLean}, S.~K. {Ramsay}, \& H.~{Takami}, 773508

\bibitem[{{Bacon} {et~al.}(2017){Bacon}, {Conseil}, {Mary}, {Brinchmann},
  {Shepherd}, {Akhlaghi}, {Weilbacher}, {Piqueras}, {Wisotzki}, {Lagattuta},
  {Epinat}, {Guerou}, {Inami}, {Cantalupo}, {Courbot}, {Contini}, {Richard},
  {Maseda}, {Bouwens}, {Bouch{\'e}}, {Kollatschny}, {Schaye}, {Marino},
  {Pello}, {Herenz}, {Guiderdoni}, \& {Carollo}}]{Bacon2017}
{Bacon}, R., {Conseil}, S., {Mary}, D., {et~al.} 2017, \aap, 608, A1

\bibitem[{{Baldwin} {et~al.}(1981){Baldwin}, {Phillips}, \&
  {Terlevich}}]{Baldwin1981}
{Baldwin}, J.~A., {Phillips}, M.~M., \& {Terlevich}, R. 1981, \pasp, 93, 5

\bibitem[{{Banfi} {et~al.}(1993){Banfi}, {Rampazzo}, {Chincarini}, \&
  {Henry}}]{Banfi1993}
{Banfi}, M., {Rampazzo}, R., {Chincarini}, G., \& {Henry}, R.~B.~C. 1993, \aap,
  280, 373

\bibitem[{{Barnes} {et~al.}(2020){Barnes}, {Longmore}, {Dale}, {Krumholz},
  {Kruijssen}, \& {Bigiel}}]{Barnes2020}
{Barnes}, A.~T., {Longmore}, S.~N., {Dale}, J.~E., {et~al.} 2020, \mnras, 498,
  4906

\bibitem[{{Bastian}(2008)}]{Bastian2008}
{Bastian}, N. 2008, \mnras, 390, 759

\bibitem[{{Bastian} {et~al.}(2007){Bastian}, {Ercolano}, {Gieles},
  {Rosolowsky}, {Scheepmaker}, {Gutermuth}, \& {Efremov}}]{Bastian2007}
{Bastian}, N., {Ercolano}, B., {Gieles}, M., {et~al.} 2007, \mnras, 379, 1302

\bibitem[{{Bastian} {et~al.}(2009){Bastian}, {Gieles}, {Ercolano}, \&
  {Gutermuth}}]{Bastian2009}
{Bastian}, N., {Gieles}, M., {Ercolano}, B., \& {Gutermuth}, R. 2009, \mnras,
  392, 868

\bibitem[{{Beckman} {et~al.}(2000){Beckman}, {Rozas}, {Zurita}, {Watson}, \&
  {Knapen}}]{Beckman2000}
{Beckman}, J.~E., {Rozas}, M., {Zurita}, A., {Watson}, R.~A., \& {Knapen},
  J.~H. 2000, \aj, 119, 2728

\bibitem[{{Belfiore} {et~al.}(2021){Belfiore}, {Santoro}, \&
  {Groves}}]{Belfiore2021}
{Belfiore}, F., {Santoro}, F., \& {Groves}, B. 2021, \aap, submitted

\bibitem[{{Bittner} {et~al.}(2019){Bittner}, {Falc{\'o}n-Barroso}, {Nedelchev},
  {Dorta}, {Gadotti}, {Sarzi}, {Molaeinezhad}, {Iodice}, {Rosado-Belza}, {de
  Lorenzo-C{\'a}ceres}, {Fragkoudi}, {Gal{\'a}n-de Anta}, {Husemann},
  {M{\'e}ndez-Abreu}, {Neumann}, {Pinna}, {Querejeta},
  {S{\'a}nchez-Bl{\'a}zquez}, \& {Seidel}}]{Bittner2019}
{Bittner}, A., {Falc{\'o}n-Barroso}, J., {Nedelchev}, B., {et~al.} 2019, \aap,
  628, A117

\bibitem[{{Blanc} {et~al.}(2009){Blanc}, {Heiderman}, {Gebhardt}, {Evans}, \&
  {Adams}}]{Blanc2009}
{Blanc}, G.~A., {Heiderman}, A., {Gebhardt}, K., {Evans}, Neal~J., I., \&
  {Adams}, J. 2009, \apj, 704, 842

\bibitem[{{Bresolin} {et~al.}(1999){Bresolin}, {Kennicutt}, \&
  {Garnett}}]{Bresolin1999}
{Bresolin}, F., {Kennicutt}, Robert~C., J., \& {Garnett}, D.~R. 1999, \apj,
  510, 104

\bibitem[{{Cappellari}(2017)}]{Cappellari2017}
{Cappellari}, M. 2017, \mnras, 466, 798

\bibitem[{{Cepa} \& {Beckman}(1990)}]{Cepa1990}
{Cepa}, J. \& {Beckman}, J.~E. 1990, \aaps, 83, 211

\bibitem[{{Chabrier}(2003)}]{Chabrier2003}
{Chabrier}, G. 2003, \pasp, 115, 763

\bibitem[{{Charlot} \& {Longhetti}(2001)}]{Charlot2001}
{Charlot}, S. \& {Longhetti}, M. 2001, \mnras, 323, 887

\bibitem[{{Chevance} {et~al.}(2020{\natexlab{a}}){Chevance}, {Kruijssen},
  {Hygate}, {Schruba}, {Longmore}, {Groves}, {Henshaw}, {Herrera}, {Hughes},
  {Jeffreson}, {Lang}, {Leroy}, {Meidt}, {Pety}, {Razza}, {Rosolowsky},
  {Schinnerer}, {Bigiel}, {Blanc}, {Emsellem}, {Faesi}, {Glover}, {Haydon},
  {Ho}, {Kreckel}, {Lee}, {Liu}, {Querejeta}, {Saito}, {Sun}, {Usero}, \&
  {Utomo}}]{Chevance2020}
{Chevance}, M., {Kruijssen}, J.~M.~D., {Hygate}, A. P.~S., {et~al.}
  2020{\natexlab{a}}, \mnras, 493, 2872

\bibitem[{{Chevance} {et~al.}(2020{\natexlab{b}}){Chevance}, {Kruijssen},
  {Vazquez-Semadeni}, {Nakamura}, {Klessen}, {Ballesteros-Paredes}, {Inutsuka},
  {Adamo}, \& {Hennebelle}}]{Chevance2020R}
{Chevance}, M., {Kruijssen}, J.~M.~D., {Vazquez-Semadeni}, E., {et~al.}
  2020{\natexlab{b}}, \ssr, 216, 50

\bibitem[{{Clauset} {et~al.}(2009){Clauset}, {Shalizi}, \&
  {Newman}}]{Clauset2009}
{Clauset}, A., {Shalizi}, C.~R., \& {Newman}, M.~E.~J. 2009, SIAM Review, 51,
  661

\bibitem[{{Colombo} {et~al.}(2014){Colombo}, {Hughes}, {Schinnerer}, {Meidt},
  {Leroy}, {Pety}, {Dobbs}, {Garc{\'\i}a-Burillo}, {Dumas}, {Thompson},
  {Schuster}, \& {Kramer}}]{Colombo2014}
{Colombo}, D., {Hughes}, A., {Schinnerer}, E., {et~al.} 2014, \apj, 784, 3

\bibitem[{{Cook} {et~al.}(2016){Cook}, {Dale}, {Lee}, {Thilker}, {Calzetti}, \&
  {Kennicutt}}]{Cook2016}
{Cook}, D.~O., {Dale}, D.~A., {Lee}, J.~C., {et~al.} 2016, \mnras, 462, 3766

\bibitem[{{Cook} {et~al.}(2012){Cook}, {Seth}, {Dale}, {Johnson}, {Weisz},
  {Fouesneau}, {Olsen}, {Engelbracht}, \& {Dalcanton}}]{Cook2012}
{Cook}, D.~O., {Seth}, A.~C., {Dale}, D.~A., {et~al.} 2012, \apj, 751, 100

\bibitem[{{Diaz} {et~al.}(1991){Diaz}, {Terlevich}, {Vilchez}, {Pagel}, \&
  {Edmunds}}]{Diaz1991}
{Diaz}, A.~I., {Terlevich}, E., {Vilchez}, J.~M., {Pagel}, B. E.~J., \&
  {Edmunds}, M.~G. 1991, \mnras, 253, 245

\bibitem[{{Dopita} {et~al.}(2006){Dopita}, {Fischera}, {Sutherland}, {Kewley},
  {Leitherer}, {Tuffs}, {Popescu}, {van Breugel}, \& {Groves}}]{Dopita2006}
{Dopita}, M.~A., {Fischera}, J., {Sutherland}, R.~S., {et~al.} 2006, \apjs,
  167, 177

\bibitem[{{Dopita} {et~al.}(2014){Dopita}, {Rich}, {Vogt}, {Kewley}, {Ho},
  {Basurah}, {Ali}, \& {Amer}}]{Dopita2014}
{Dopita}, M.~A., {Rich}, J., {Vogt}, F. P.~A., {et~al.} 2014, \apss, 350, 741

\bibitem[{{Dors} \& {Copetti}(2005)}]{Dors2005}
{Dors}, O.~L., J. \& {Copetti}, M.~V.~F. 2005, \aap, 437, 837

\bibitem[{{Dors} {et~al.}(2011){Dors}, {Krabbe}, {H{\"a}gele}, \&
  {P{\'e}rez-Montero}}]{Dors2011}
{Dors}, O.~L., J., {Krabbe}, A., {H{\"a}gele}, G.~F., \& {P{\'e}rez-Montero},
  E. 2011, \mnras, 415, 3616

\bibitem[{{Efremov} \& {Elmegreen}(1998)}]{Efremov1998}
{Efremov}, Y.~N. \& {Elmegreen}, B.~G. 1998, \mnras, 299, 588

\bibitem[{{Elmegreen} {et~al.}(1996){Elmegreen}, {Elmegreen}, {Salzer}, \&
  {Mann}}]{Elmegreen1996}
{Elmegreen}, B.~G., {Elmegreen}, D.~M., {Salzer}, J.~J., \& {Mann}, H. 1996,
  \apj, 467, 579

\bibitem[{{Elmegreen} \& {Falgarone}(1996)}]{Elmegreen1996b}
{Elmegreen}, B.~G. \& {Falgarone}, E. 1996, \apj, 471, 816

\bibitem[{{Elmegreen} \& {Salzer}(1999)}]{Elmegreen1999}
{Elmegreen}, D.~M. \& {Salzer}, J.~J. 1999, \aj, 117, 764

\bibitem[{{Emsellem} {et~al.}(2021){Emsellem}, {Schinnerer}, {Santoro},
  {Belfiore}, {Pessa}, {McElroy}, {Blanc}, {Congiu}, {Groves}, {Ho}, {Kreckel},
  {Razza}, {Sanchez-Blazquez}, {Egorov}, {Faesi}, {Klessen}, {Leroy}, {Meidt},
  {Querejeta}, {Rosolowsky}, {Scheuermann}, {Anand}, {Barnes},
  {Be{\v{s}}li{\'c}}, {Bigiel}, {Boquien}, {Cao}, {Chevance}, {Dale},
  {Eibensteiner}, {Glover}, {Grasha}, {Henshaw}, {Hughes}, {Koch}, {Kruijssen},
  {Lee}, {Liu}, {Pan}, {Pety}, {Saito}, {Sandstrom}, {Schruba}, {Sun},
  {Thilker}, {Usero}, {Watkins}, \& {Williams}}]{Emsellem2021}
{Emsellem}, E., {Schinnerer}, E., {Santoro}, F., {et~al.} 2021, arXiv e-prints,
  arXiv:2110.03708

\bibitem[{{Espinosa-Ponce} {et~al.}(2020){Espinosa-Ponce}, {S{\'a}nchez},
  {Morisset}, {Barrera-Ballesteros}, {Galbany}, {Garc{\'\i}a-Benito},
  {Lacerda}, \& {Mast}}]{Espinosa2020}
{Espinosa-Ponce}, C., {S{\'a}nchez}, S.~F., {Morisset}, C., {et~al.} 2020,
  \mnras, 494, 1622

\bibitem[{{Faesi} {et~al.}(2018){Faesi}, {Lada}, \& {Forbrich}}]{Faesi2018}
{Faesi}, C.~M., {Lada}, C.~J., \& {Forbrich}, J. 2018, \apj, 857, 19

\bibitem[{{Feinstein}(1997)}]{Feinstein1997}
{Feinstein}, C. 1997, \apjs, 112, 29

\bibitem[{{Ferguson} {et~al.}(1998){Ferguson}, {Wyse}, {Gallagher}, \&
  {Hunter}}]{Ferguson1998}
{Ferguson}, A. M.~N., {Wyse}, R. F.~G., {Gallagher}, J.~S., \& {Hunter}, D.~A.
  1998, \apjl, 506, L19

\bibitem[{{Garnett} {et~al.}(1997){Garnett}, {Shields}, {Skillman}, {Sagan}, \&
  {Dufour}}]{Garnett1997}
{Garnett}, D.~R., {Shields}, G.~A., {Skillman}, E.~D., {Sagan}, S.~P., \&
  {Dufour}, R.~J. 1997, \apj, 489, 63

\bibitem[{{Goddard} {et~al.}(2010){Goddard}, {Bastian}, \&
  {Kennicutt}}]{Goddard2010}
{Goddard}, Q.~E., {Bastian}, N., \& {Kennicutt}, R.~C. 2010, \mnras, 405, 857

\bibitem[{{Guti{\'e}rrez} {et~al.}(2011){Guti{\'e}rrez}, {Beckman}, \&
  {Buenrostro}}]{Gutierrez2011}
{Guti{\'e}rrez}, L., {Beckman}, J.~E., \& {Buenrostro}, V. 2011, \aj, 141, 113

\bibitem[{{Haffner} {et~al.}(2009){Haffner}, {Dettmar}, {Beckman}, {Wood},
  {Slavin}, {Giammanco}, {Madsen}, {Zurita}, \& {Reynolds}}]{Haffner2009}
{Haffner}, L.~M., {Dettmar}, R.~J., {Beckman}, J.~E., {et~al.} 2009, Reviews of
  Modern Physics, 81, 969

\bibitem[{{Hao} {et~al.}(2011){Hao}, {Kennicutt}, {Johnson}, {Calzetti},
  {Dale}, \& {Moustakas}}]{Hao2011}
{Hao}, C.-N., {Kennicutt}, R.~C., {Johnson}, B.~D., {et~al.} 2011, \apj, 741,
  124

\bibitem[{{Haydon} {et~al.}(2020){Haydon}, {Kruijssen}, {Chevance}, {Hygate},
  {Krumholz}, {Schruba}, \& {Longmore}}]{Haydon2020}
{Haydon}, D.~T., {Kruijssen}, J.~M.~D., {Chevance}, M., {et~al.} 2020, \mnras,
  498, 235

\bibitem[{{Helmboldt} {et~al.}(2005){Helmboldt}, {Walterbos}, {Bothun}, \&
  {O'Neil}}]{Helmboldt2005}
{Helmboldt}, J.~F., {Walterbos}, R.~A.~M., {Bothun}, G.~D., \& {O'Neil}, K.
  2005, \apj, 630, 824

\bibitem[{{Herrera-Endoqui} {et~al.}(2015){Herrera-Endoqui},
  {D{\'\i}az-Garc{\'\i}a}, {Laurikainen}, \& {Salo}}]{Herrera-Endoqui2015}
{Herrera-Endoqui}, M., {D{\'\i}az-Garc{\'\i}a}, S., {Laurikainen}, E., \&
  {Salo}, H. 2015, \aap, 582, A86

\bibitem[{{Ho} {et~al.}(2019){Ho}, {Kreckel}, {Meidt}, {Groves}, {Blanc},
  {Bigiel}, {Dale}, {Emsellem}, {Glover}, {Grasha}, {Kewley}, {Kruijssen},
  {Lang}, {McElroy}, {Kudritzki}, {Sanchez-Blazquez}, {Sandstrom}, {Santoro},
  {Schinnerer}, \& {Schruba}}]{Ho2019}
{Ho}, I.~T., {Kreckel}, K., {Meidt}, S.~E., {et~al.} 2019, \apjl, 885, L31

\bibitem[{{Hoare}(2005)}]{Hoare2005}
{Hoare}, M.~G. 2005, \apss, 295, 203

\bibitem[{{Hopkins}(2013)}]{Hopkins2013}
{Hopkins}, P.~F. 2013, \mnras, 430, 1653

\bibitem[{{Hughes} {et~al.}(2016){Hughes}, {Meidt}, {Colombo}, {Schruba},
  {Schinnerer}, {Leroy}, \& {Wong}}]{Hughes2016}
{Hughes}, A., {Meidt}, S., {Colombo}, D., {et~al.} 2016, in From Interstellar
  Clouds to Star-Forming Galaxies: Universal Processes?, ed. P.~{Jablonka},
  P.~{Andr{\'e}}, \& F.~{van der Tak}, Vol. 315, 30--37

\bibitem[{{Johnson} {et~al.}(2016){Johnson}, {Seth}, {Dalcanton}, {Beerman},
  {Fouesneau}, {Lewis}, {Weisz}, {Williams}, {Bell}, {Dolphin}, {Larsen},
  {Sandstrom}, \& {Skillman}}]{Johnson2016}
{Johnson}, L.~C., {Seth}, A.~C., {Dalcanton}, J.~J., {et~al.} 2016, \apj, 827,
  33

\bibitem[{{Johnson} {et~al.}(2017){Johnson}, {Seth}, {Dalcanton}, {Beerman},
  {Fouesneau}, {Weisz}, {Bell}, {Dolphin}, {Sandstrom}, \&
  {Williams}}]{Johnson2017}
{Johnson}, L.~C., {Seth}, A.~C., {Dalcanton}, J.~J., {et~al.} 2017, \apj, 839,
  78

\bibitem[{{Kauffmann} {et~al.}(2003){Kauffmann}, {Heckman}, {Tremonti},
  {Brinchmann}, {Charlot}, {White}, {Ridgway}, {Brinkmann}, {Fukugita}, {Hall},
  {Ivezi{\'c}}, {Richards}, \& {Schneider}}]{Kauffmann2003}
{Kauffmann}, G., {Heckman}, T.~M., {Tremonti}, C., {et~al.} 2003, \mnras, 346,
  1055

\bibitem[{{Kennicutt} {et~al.}(1989){Kennicutt}, {Edgar}, \&
  {Hodge}}]{Kennicutt1989}
{Kennicutt}, Robert~C., J., {Edgar}, B.~K., \& {Hodge}, P.~W. 1989, \apj, 337,
  761

\bibitem[{{Kennicutt} \& {Garnett}(1996)}]{Kennicutt1996}
{Kennicutt}, Robert~C., J. \& {Garnett}, D.~R. 1996, \apj, 456, 504

\bibitem[{{Kennicutt} {et~al.}(2008){Kennicutt}, {Lee}, {Funes}, {J.}, {Sakai},
  \& {Akiyama}}]{Kennicutt2008}
{Kennicutt}, Robert~C., J., {Lee}, J.~C., {Funes}, J.~G., {et~al.} 2008, \apjs,
  178, 247

\bibitem[{{Kennicutt} \& {Evans}(2012{\natexlab{a}})}]{Kennicutt2012R}
{Kennicutt}, R.~C. \& {Evans}, N.~J. 2012{\natexlab{a}}, \araa, 50, 531

\bibitem[{{Kennicutt} \& {Evans}(2012{\natexlab{b}})}]{Kennicutt2012}
{Kennicutt}, R.~C. \& {Evans}, N.~J. 2012{\natexlab{b}}, \araa, 50, 531

\bibitem[{{Kewley} \& {Dopita}(2002)}]{Kewley2002}
{Kewley}, L.~J. \& {Dopita}, M.~A. 2002, \apjs, 142, 35

\bibitem[{{Kewley} {et~al.}(2001){Kewley}, {Heisler}, {Dopita}, \&
  {Lumsden}}]{Kewley2001}
{Kewley}, L.~J., {Heisler}, C.~A., {Dopita}, M.~A., \& {Lumsden}, S. 2001,
  \apjs, 132, 37

\bibitem[{{Kim} {et~al.}(2021){Kim}, {Chevance}, {Kruijssen}, {Schruba},
  {Sandstrom}, {Barnes}, {Bigiel}, {Blanc}, {Cao}, {Dale}, {Faesi}, {Glover},
  {Grasha}, {Groves}, {Herrera}, {Klessen}, {Kreckel}, {Lee}, {Leroy}, {Pety},
  {Querejeta}, {Schinnerer}, {Sun}, {Usero}, {Ward}, \& {Williams}}]{Kim2021}
{Kim}, J., {Chevance}, M., {Kruijssen}, J.~M.~D., {et~al.} 2021, \mnras, 504,
  487

\bibitem[{{Knapen}(1998)}]{Knapen1998}
{Knapen}, J.~H. 1998, \mnras, 297, 255

\bibitem[{{Knapen} {et~al.}(1993){Knapen}, {Arnth-Jensen}, {Cepa}, \&
  {Beckman}}]{Knapen1993}
{Knapen}, J.~H., {Arnth-Jensen}, N., {Cepa}, J., \& {Beckman}, J.~E. 1993, \aj,
  106, 56

\bibitem[{{Kreckel} {et~al.}(2020){Kreckel}, {Ho}, {Blanc}, {Glover}, {Groves},
  {Rosolowsky}, {Bigiel}, {Boqu{\'\i}en}, {Chevance}, {Dale}, {Deger},
  {Emsellem}, {Grasha}, {Kim}, {Klessen}, {Kruijssen}, {Lee}, {Leroy}, {Liu},
  {McElroy}, {Meidt}, {Pessa}, {Sanchez-Blazquez}, {Sandstrom}, {Santoro},
  {Scheuermann}, {Schinnerer}, {Schruba}, {Utomo}, {Watkins}, \&
  {Williams}}]{Kreckel2020}
{Kreckel}, K., {Ho}, I.~T., {Blanc}, G.~A., {et~al.} 2020, \mnras, 499, 193

\bibitem[{{Kreckel} {et~al.}(2019){Kreckel}, {Ho}, {Blanc}, {Groves},
  {Santoro}, {Schinnerer}, {Bigiel}, {Chevance}, {Congiu}, {Emsellem}, {Faesi},
  {Glover}, {Grasha}, {Kruijssen}, {Lang}, {Leroy}, {Meidt}, {McElroy}, {Pety},
  {Rosolowsky}, {Saito}, {Sandstrom}, {Sanchez-Blazquez}, \&
  {Schruba}}]{Kreckel2019}
{Kreckel}, K., {Ho}, I.~T., {Blanc}, G.~A., {et~al.} 2019, \apj, 887, 80

\bibitem[{{Kroupa} \& {Weidner}(2003)}]{Kroupa2003}
{Kroupa}, P. \& {Weidner}, C. 2003, \apj, 598, 1076

\bibitem[{{Kruijssen}(2012)}]{Kruijssen2012}
{Kruijssen}, J.~M.~D. 2012, \mnras, 426, 3008

\bibitem[{{Kruijssen} \& {Longmore}(2014)}]{Kruijssen2014b}
{Kruijssen}, J.~M.~D. \& {Longmore}, S.~N. 2014, \mnras, 439, 3239

\bibitem[{{Kruijssen} {et~al.}(2019){Kruijssen}, {Schruba}, {Chevance},
  {Longmore}, {Hygate}, {Haydon}, {McLeod}, {Dalcanton}, {Tacconi}, \& {van
  Dishoeck}}]{Kruijssen2019}
{Kruijssen}, J.~M.~D., {Schruba}, A., {Chevance}, M., {et~al.} 2019, \nat, 569,
  519

\bibitem[{{Kruijssen} {et~al.}(2018){Kruijssen}, {Schruba}, {Hygate}, {Hu},
  {Haydon}, \& {Longmore}}]{Kruijssen2018}
{Kruijssen}, J.~M.~D., {Schruba}, A., {Hygate}, A. P.~S., {et~al.} 2018,
  \mnras, 479, 1866

\bibitem[{{Krumholz}(2014)}]{Krumholz2014}
{Krumholz}, M.~R. 2014, \physrep, 539, 49

\bibitem[{{Krumholz} {et~al.}(2015){Krumholz}, {Adamo}, {Fumagalli}, {Wofford},
  {Calzetti}, {Lee}, {Whitmore}, {Bright}, {Grasha}, {Gouliermis}, {Kim},
  {Nair}, {Ryon}, {Smith}, {Thilker}, {Ubeda}, \& {Zackrisson}}]{Krumholz2015}
{Krumholz}, M.~R., {Adamo}, A., {Fumagalli}, M., {et~al.} 2015, \apj, 812, 147

\bibitem[{{Lang} {et~al.}(2020){Lang}, {Meidt}, {Rosolowsky}, {Nofech},
  {Schinnerer}, {Leroy}, {Emsellem}, {Pessa}, {Glover}, {Groves}, {Hughes},
  {Kruijssen}, {Querejeta}, {Schruba}, {Bigiel}, {Blanc}, {Chevance},
  {Colombo}, {Faesi}, {Henshaw}, {Herrera}, {Liu}, {Pety}, {Puschnig}, {Saito},
  {Sun}, \& {Usero}}]{Lang2020}
{Lang}, P., {Meidt}, S.~E., {Rosolowsky}, E., {et~al.} 2020, \apj, 897, 122

\bibitem[{{Larsen}(2002)}]{Larsen2002}
{Larsen}, S.~S. 2002, \aj, 124, 1393

\bibitem[{{Lawton} {et~al.}(2010){Lawton}, {Gordon}, {Babler}, {Block},
  {Bolatto}, {Bracker}, {Carlson}, {Engelbracht}, {Hora}, {Indebetouw},
  {Madden}, {Meade}, {Meixner}, {Misselt}, {Oey}, {Oliveira}, {Robitaille},
  {Sewilo}, {Shiao}, {Vijh}, \& {Whitney}}]{Lawton2010}
{Lawton}, B., {Gordon}, K.~D., {Babler}, B., {et~al.} 2010, \apj, 716, 453

\bibitem[{{Lee} {et~al.}(2021){Lee}, {Whitmore}, {Thilker}, {Deger}, {Larson},
  {Ubeda}, {Anand}, {Boquien}, {Chandar}, {Dale}, {Emsellem}, {Leroy},
  {Rosolowsky}, {Schinnerer}, {Schmidt}, {Turner}, {Van Dyk}, {White},
  {Barnes}, {Belfiore}, {Bigiel}, {Blanc}, {Cao}, {Chevance}, {Congiu},
  {Egorov}, {Glover}, {Grasha}, {Groves}, {Henshaw}, {Hughes}, {Klessen},
  {Koch}, {Kreckel}, {Kruijssen}, {Liu}, {Lopez}, {Mayker}, {Meidt}, {Murphy},
  {Pan}, {Pety}, {Querejeta}, {Razza}, {Saito}, {Sanchez-Blazquez}, {Santoro},
  {Sardone}, {Scheuermann}, {Schruba}, {Sun}, {Usero}, {Watkins}, \&
  {Williams}}]{Lee2021}
{Lee}, J.~C., {Whitmore}, B.~C., {Thilker}, D.~A., {et~al.} 2021, arXiv
  e-prints, arXiv:2101.02855

\bibitem[{{Lee} {et~al.}(2011){Lee}, {Hwang}, \& {Lee}}]{Lee2011}
{Lee}, J.~H., {Hwang}, N., \& {Lee}, M.~G. 2011, \apj, 735, 75

\bibitem[{{Leli{\`e}vre} \& {Roy}(2000)}]{Lelievre2000}
{Leli{\`e}vre}, M. \& {Roy}, J.-R. 2000, \aj, 120, 1306

\bibitem[{{Leroy} {et~al.}(2021){Leroy}, {Schinnerer}, {Hughes}, {Rosolowsky},
  {Pety}, {Schruba}, {Usero}, {Blanc}, {Chevance}, {Emsellem}, {Faesi},
  {Herrera}, {Liu}, {Meidt}, {Querejeta}, {Saito}, {Sandstrom}, {Sun},
  {Williams}, {Anand}, {Barnes}, {Behrens}, {Belfiore}, {Benincasa},
  {Be{\v{s}}li{\'c}}, {Bigiel}, {Bolatto}, {den Brok}, {Cao}, {Chandar},
  {Chastenet}, {Chiang}, {Congiu}, {Dale}, {Deger}, {Eibensteiner}, {Egorov},
  {Garc{\'\i}a-Rodr{\'\i}guez}, {Glover}, {Grasha}, {Henshaw}, {Ho}, {Kepley},
  {Kim}, {Klessen}, {Kreckel}, {Koch}, {Kruijssen}, {Larson}, {Lee}, {Lopez},
  {Machado}, {Mayker}, {McElroy}, {Murphy}, {Ostriker}, {Pan}, {Pessa},
  {Puschnig}, {Razza}, {S{\'a}nchez-Bl{\'a}zquez}, {Santoro}, {Sardone},
  {Scheuermann}, {Sliwa}, {Sormani}, {Stuber}, {Thilker}, {Turner}, {Utomo},
  {Watkins}, \& {Whitmore}}]{Leroy2021}
{Leroy}, A.~K., {Schinnerer}, E., {Hughes}, A., {et~al.} 2021, arXiv e-prints,
  arXiv:2104.07739

\bibitem[{{Liu} {et~al.}(2013){Liu}, {Calzetti}, {Kennicutt}, {Schinnerer},
  {Sofue}, {Komugi}, {Egusa}, \& {Scoville}}]{Liu2013}
{Liu}, G., {Calzetti}, D., {Kennicutt}, Robert~C., J., {et~al.} 2013, \apj,
  772, 27

\bibitem[{{Luridiana} {et~al.}(2015){Luridiana}, {Morisset}, \&
  {Shaw}}]{Luridiana2015}
{Luridiana}, V., {Morisset}, C., \& {Shaw}, R.~A. 2015, \aap, 573, A42

\bibitem[{{Ma{\'\i}z Apell{\'a}niz} \& {{\'U}beda}(2005)}]{2005ApJ...629..873M}
{Ma{\'\i}z Apell{\'a}niz}, J. \& {{\'U}beda}, L. 2005, \apj, 629, 873

\bibitem[{{Makarov} {et~al.}(2014){Makarov}, {Prugniel}, {Terekhova},
  {Courtois}, \& {Vauglin}}]{Makarov2014}
{Makarov}, D., {Prugniel}, P., {Terekhova}, N., {Courtois}, H., \& {Vauglin},
  I. 2014, \aap, 570, A13

\bibitem[{{Mascoop} {et~al.}(2021){Mascoop}, {Anderson}, {Wenger}, {Makai},
  {Armentrout}, {Balser}, \& {Bania}}]{Mascoop2021}
{Mascoop}, J.~L., {Anderson}, L.~D., {Wenger}, T.~V., {et~al.} 2021, \apj, 910,
  159

\bibitem[{{Massey} {et~al.}(2005){Massey}, {Puls}, {Pauldrach}, {Bresolin},
  {Kudritzki}, \& {Simon}}]{Massey2005}
{Massey}, P., {Puls}, J., {Pauldrach}, A.~W.~A., {et~al.} 2005, \apj, 627, 477

\bibitem[{{Meidt} {et~al.}(2021){Meidt}, {Leroy}, {Querejeta}, {Schinnerer},
  {Sun}, {van der Wel}, {Emsellem}, {Henshaw}, {Hughes}, {Kruijssen},
  {Rosolowsky}, {Schruba}, {Barnes}, {Bigiel}, {Blanc}, {Chevance}, {Cao},
  {Dale}, {Faesi}, {Glover}, {Grasha}, {Groves}, {Herrera}, {Klessen},
  {Kreckel}, {Liu}, {Pan}, {Pety}, {Saito}, {Usero}, {Watkins}, \&
  {Williams}}]{Meidt2021}
{Meidt}, S.~E., {Leroy}, A.~K., {Querejeta}, M., {et~al.} 2021, \apj, 913, 113

\bibitem[{{Murphy} {et~al.}(2011){Murphy}, {Condon}, {Schinnerer}, {Kennicutt},
  {Calzetti}, {Armus}, {Helou}, {Turner}, {Aniano}, {Beir{\~a}o}, {Bolatto},
  {Brandl}, {Croxall}, {Dale}, {Donovan Meyer}, {Draine}, {Engelbracht},
  {Hunt}, {Hao}, {Koda}, {Roussel}, {Skibba}, \& {Smith}}]{Murphy2011}
{Murphy}, E.~J., {Condon}, J.~J., {Schinnerer}, E., {et~al.} 2011, \apj, 737,
  67

\bibitem[{{O'Donnell}(1994)}]{ODonnell1994}
{O'Donnell}, J.~E. 1994, \apj, 422, 158

\bibitem[{{Oey}(1996)}]{Oey1996}
{Oey}, M.~S. 1996, \apj, 465, 231

\bibitem[{{Oey} \& {Clarke}(1998{\natexlab{a}})}]{Oey1998}
{Oey}, M.~S. \& {Clarke}, C.~J. 1998{\natexlab{a}}, \aj, 115, 1543

\bibitem[{{Oey} \& {Clarke}(1998{\natexlab{b}})}]{Oey1998A}
{Oey}, M.~S. \& {Clarke}, C.~J. 1998{\natexlab{b}}, \aj, 115, 1543

\bibitem[{{Oey} {et~al.}(2007){Oey}, {Meurer}, {Yelda}, {Furst},
  {Caballero-Nieves}, {Hanish}, {Levesque}, {Thilker}, {Walth},
  {Bland-Hawthorn}, {Dopita}, {Ferguson}, {Heckman}, {Doyle}, {Drinkwater},
  {Freeman}, {Kennicutt}, {Kilborn}, {Knezek}, {Koribalski}, {Meyer}, {Putman},
  {Ryan-Weber}, {Smith}, {Staveley-Smith}, {Webster}, {Werk}, \&
  {Zwaan}}]{Oey2007}
{Oey}, M.~S., {Meurer}, G.~R., {Yelda}, S., {et~al.} 2007, \apj, 661, 801

\bibitem[{{Pellegrini} {et~al.}(2010){Pellegrini}, {Baldwin}, \&
  {Ferland}}]{Pellegrini2010}
{Pellegrini}, E.~W., {Baldwin}, J.~A., \& {Ferland}, G.~J. 2010, \apjs, 191,
  160

\bibitem[{{Pellegrini} {et~al.}(2012){Pellegrini}, {Oey}, {Winkler}, {Points},
  {Smith}, {Jaskot}, \& {Zastrow}}]{Pellegrini2012}
{Pellegrini}, E.~W., {Oey}, M.~S., {Winkler}, P.~F., {et~al.} 2012, \apj, 755,
  40

\bibitem[{{Pellegrini} {et~al.}(2020){Pellegrini}, {Rahner}, {Reissl},
  {Glover}, {Klessen}, {Rousseau-Nepton}, \& {Herrera-Camus}}]{Pellegrini2020}
{Pellegrini}, E.~W., {Rahner}, D., {Reissl}, S., {et~al.} 2020, \mnras, 496,
  339

\bibitem[{{Pessa} {et~al.}(2021){Pessa}, {Schinnerer}, {Belfiore}, {Emsellem},
  {Leroy}, {Schruba}, {Kruijssen}, {Pan}, {Blanc}, {Sanchez-Blazquez},
  {Bigiel}, {Chevance}, {Congiu}, {Dale}, {Faesi}, {Glover}, {Grasha},
  {Groves}, {Ho}, {Jim{\'e}nez-Donaire}, {Klessen}, {Kreckel}, {Koch}, {Liu},
  {Meidt}, {Pety}, {Querejeta}, {Rosolowsky}, {Saito}, {Santoro}, {Sun},
  {Usero}, {Watkins}, \& {Williams}}]{Pessa2021}
{Pessa}, I., {Schinnerer}, E., {Belfiore}, F., {et~al.} 2021, \aap, 650, A134

\bibitem[{{Pietrinferni} {et~al.}(2004){Pietrinferni}, {Cassisi}, {Salaris}, \&
  {Castelli}}]{Pietrinferni2004}
{Pietrinferni}, A., {Cassisi}, S., {Salaris}, M., \& {Castelli}, F. 2004, \apj,
  612, 168

\bibitem[{{Pilyugin} \& {Grebel}(2016)}]{Pilyugin2016}
{Pilyugin}, L.~S. \& {Grebel}, E.~K. 2016, \mnras, 457, 3678

\bibitem[{{Querejeta} {et~al.}(2021){Querejeta}, {Schinnerer}, {Meidt}, {Sun},
  {Leroy}, {Emsellem}, {Klessen}, {Munoz-Mateos}, {Salo}, {Laurikainen},
  {Beslic}, {Blanc}, {Chevance}, {Dale}, {Eibensteiner}, {Faesi},
  {Garcia-Rodriguez}, {Glover}, {Grasha}, {Henshaw}, {Herrera}, {Hughes},
  {Kreckel}, {Kruijssen}, {Liu}, {Murphy}, {Pan}, {Pety}, {Razza},
  {Rosolowsky}, {Saito}, {Schruba}, {Usero}, {Watkins}, \&
  {Williams}}]{Querejeta2021}
{Querejeta}, M., {Schinnerer}, E., {Meidt}, S., {et~al.} 2021, arXiv e-prints,
  arXiv:2109.04491

\bibitem[{{Rahner} {et~al.}(2017){Rahner}, {Pellegrini}, {Glover}, \&
  {Klessen}}]{Rahner2017}
{Rahner}, D., {Pellegrini}, E.~W., {Glover}, S. C.~O., \& {Klessen}, R.~S.
  2017, \mnras, 470, 4453

\bibitem[{{Rahner} {et~al.}(2019){Rahner}, {Pellegrini}, {Glover}, \&
  {Klessen}}]{Rahner2019}
{Rahner}, D., {Pellegrini}, E.~W., {Glover}, S. C.~O., \& {Klessen}, R.~S.
  2019, \mnras, 483, 2547

\bibitem[{{Rand}(1992)}]{Rand1992}
{Rand}, R.~J. 1992, \aj, 103, 815

\bibitem[{{Rand}(1993)}]{Rand1993}
{Rand}, R.~J. 1993, \apj, 410, 68

\bibitem[{{Reina-Campos} \& {Kruijssen}(2017)}]{Reinacampos2017}
{Reina-Campos}, M. \& {Kruijssen}, J.~M.~D. 2017, \mnras, 469, 1282

\bibitem[{{Rice} {et~al.}(2016){Rice}, {Goodman}, {Bergin}, {Beaumont}, \&
  {Dame}}]{Rice2016}
{Rice}, T.~S., {Goodman}, A.~A., {Bergin}, E.~A., {Beaumont}, C., \& {Dame},
  T.~M. 2016, \apj, 822, 52

\bibitem[{{Rosolowsky} {et~al.}(2021){Rosolowsky}, {Hughes}, {Leroy}, {Sun},
  {Querejeta}, {Schruba}, {Usero}, {Herrera}, {Liu}, {Pety}, {Saito},
  {Be{\v{s}}li{\'c}}, {Bigiel}, {Blanc}, {Chevance}, {Dale}, {Deger}, {Faesi},
  {Glover}, {Henshaw}, {Klessen}, {Kruijssen}, {Larson}, {Lee}, {Meidt}, {Mok},
  {Schinnerer}, {Thilker}, \& {Williams}}]{Rosolowsky2021}
{Rosolowsky}, E., {Hughes}, A., {Leroy}, A.~K., {et~al.} 2021, \mnras, 502,
  1218

\bibitem[{{Rosolowsky} {et~al.}(2007){Rosolowsky}, {Keto}, {Matsushita}, \&
  {Willner}}]{Rosolowsky2007}
{Rosolowsky}, E., {Keto}, E., {Matsushita}, S., \& {Willner}, S.~P. 2007, \apj,
  661, 830

\bibitem[{{Rousseau-Nepton} {et~al.}(2018){Rousseau-Nepton}, {Robert},
  {Martin}, {Drissen}, \& {Martin}}]{Rousseau2018}
{Rousseau-Nepton}, L., {Robert}, C., {Martin}, R.~P., {Drissen}, L., \&
  {Martin}, T. 2018, \mnras, 477, 4152

\bibitem[{{Rozas} {et~al.}(1996){Rozas}, {Beckman}, \& {Knapen}}]{Rozas1996}
{Rozas}, M., {Beckman}, J.~E., \& {Knapen}, J.~H. 1996, \aap, 307, 735

\bibitem[{{S{\'a}nchez} {et~al.}(2012){S{\'a}nchez}, {Rosales-Ortega},
  {Marino}, {Iglesias-P{\'a}ramo}, {V{\'\i}lchez}, {Kennicutt}, {D{\'\i}az},
  {Mast}, {Monreal-Ibero}, {Garc{\'\i}a-Benito}, {Bland-Hawthorn}, {P{\'e}rez},
  {Gonz{\'a}lez Delgado}, {Husemann}, {L{\'o}pez-S{\'a}nchez}, {Cid Fernandes},
  {Kehrig}, {Walcher}, {Gil de Paz}, \& {Ellis}}]{Sanchez2012}
{S{\'a}nchez}, S.~F., {Rosales-Ortega}, F.~F., {Marino}, R.~A., {et~al.} 2012,
  \aap, 546, A2

\bibitem[{{Schlafly} \& {Finkbeiner}(2011)}]{Schlafly2011}
{Schlafly}, E.~F. \& {Finkbeiner}, D.~P. 2011, \apj, 737, 103

\bibitem[{{Schruba} {et~al.}(2019){Schruba}, {Kruijssen}, \&
  {Leroy}}]{Schruba2019}
{Schruba}, A., {Kruijssen}, J.~M.~D., \& {Leroy}, A.~K. 2019, \apj, 883, 2

\bibitem[{{Scoville} {et~al.}(2001){Scoville}, {Polletta}, {Ewald}, {Stolovy},
  {Thompson}, \& {Rieke}}]{Scoville2001}
{Scoville}, N.~Z., {Polletta}, M., {Ewald}, S., {et~al.} 2001, \aj, 122, 3017

\bibitem[{Thilker {et~al.}(2000)Thilker, Braun, \& Walterbos}]{Thilker2000}
Thilker, D.~A., Braun, R., \& Walterbos, R. A.~M. 2000, The Astronomical
  Journal, 120, 3070

\bibitem[{{Thilker} {et~al.}(2002){Thilker}, {Walterbos}, {Braun}, \&
  {Hoopes}}]{Thilker2002}
{Thilker}, D.~A., {Walterbos}, R. A.~M., {Braun}, R., \& {Hoopes}, C.~G. 2002,
  \aj, 124, 3118

\bibitem[{{Thilker} {et~al.}(2021){Thilker}, {Whitmore}, {Lee}, {Deger},
  {Chandar}, {Larson}, {Hannon}, {Ubeda}, {Dale}, {Glover}, {Grasha},
  {Klessen}, {Kruijssen}, {Rosolowsky}, {Schruba}, {White}, \&
  {Williams}}]{Thilker2021}
{Thilker}, D.~A., {Whitmore}, B.~C., {Lee}, J.~C., {et~al.} 2021, arXiv
  e-prints, arXiv:2106.13366

\bibitem[{{Tress} {et~al.}(2020){Tress}, {Smith}, {Sormani}, {Glover},
  {Klessen}, {Mac Low}, \& {Clark}}]{Tress2020}
{Tress}, R.~G., {Smith}, R.~J., {Sormani}, M.~C., {et~al.} 2020, \mnras, 492,
  2973

\bibitem[{{Tress} {et~al.}(2021){Tress}, {Sormani}, {Smith}, {Glover},
  {Klessen}, {Mac Low}, {Clark}, \& {Duarte-Cabral}}]{Tress2021}
{Tress}, R.~G., {Sormani}, M.~C., {Smith}, R.~J., {et~al.} 2021, \mnras, 505,
  5438

\bibitem[{{Vacca}(1994)}]{Vacca1994}
{Vacca}, W.~D. 1994, \apj, 421, 140

\bibitem[{{van Zee}(2000)}]{vanZee2000}
{van Zee}, L. 2000, \aj, 119, 2757

\bibitem[{{Vazdekis} {et~al.}(2016){Vazdekis}, {Koleva}, {Ricciardelli},
  {R{\"o}ck}, \& {Falc{\'o}n-Barroso}}]{Vazdekis2016}
{Vazdekis}, A., {Koleva}, M., {Ricciardelli}, E., {R{\"o}ck}, B., \&
  {Falc{\'o}n-Barroso}, J. 2016, \mnras, 463, 3409

\bibitem[{{Wei} {et~al.}(2020){Wei}, {Zou}, {Kong}, {Zhou}, {Hu}, {Lin}, {Mao},
  {Lin}, {Zhou}, {Liu}, {Ma}, {Ma}, {Zhong}, {Dang}, {Sun}, \& {Lin}}]{Wei2020}
{Wei}, P., {Zou}, H., {Kong}, X., {et~al.} 2020, \pasp, 132, 094101

\bibitem[{{Weidner} {et~al.}(2004){Weidner}, {Kroupa}, \&
  {Larsen}}]{Weidner2004}
{Weidner}, C., {Kroupa}, P., \& {Larsen}, S.~S. 2004, \mnras, 350, 1503

\bibitem[{{Whitmore} {et~al.}(2014){Whitmore}, {Chandar}, {Bowers}, {Larsen},
  {Lindsay}, {Ansari}, \& {Evans}}]{Whitmore2014}
{Whitmore}, B.~C., {Chandar}, R., {Bowers}, A.~S., {et~al.} 2014, \aj, 147, 78

\bibitem[{{Whitmore} {et~al.}(2021){Whitmore}, {Lee}, {Chandar}, {Thilker},
  {Hannon}, {Wei}, {Huerta}, {Bigiel}, {Boquien}, {Chevance}, {Dale}, {Deger},
  {Grasha}, {Klessen}, {Kruijssen}, {Larson}, {Mok}, {Rosolowsky},
  {Schinnerer}, {Schruba}, {Ubeda}, {Van Dyk}, {Watkins}, \&
  {Williams}}]{Whitmore2021}
{Whitmore}, B.~C., {Lee}, J.~C., {Chandar}, R., {et~al.} 2021, \mnras

\bibitem[{{Youngblood} \& {Hunter}(1999)}]{Youngblood1999}
{Youngblood}, A.~J. \& {Hunter}, D.~A. 1999, \apj, 519, 55

\bibitem[{{Zhang} {et~al.}(2017){Zhang}, {Yan}, {Bundy}, {Bershady}, {Haffner},
  {Walterbos}, {Maiolino}, {Tremonti}, {Thomas}, {Drory}, {Jones}, {Belfiore},
  {S{\'a}nchez}, {Diamond-Stanic}, {Bizyaev}, {Nitschelm}, {Andrews},
  {Brinkmann}, {Brownstein}, {Cheung}, {Li}, {Law}, {Roman Lopes}, {Oravetz},
  {Pan}, {Storchi Bergmann}, \& {Simmons}}]{Zhang2017}
{Zhang}, K., {Yan}, R., {Bundy}, K., {et~al.} 2017, \mnras, 466, 3217

\end{thebibliography}

\begin{appendix}

\section{Nebulae spatial masks and metallicity gradients}\label{appendix1}

\begin{figure*}
\begin{center}
\includegraphics[width=0.8\textwidth]{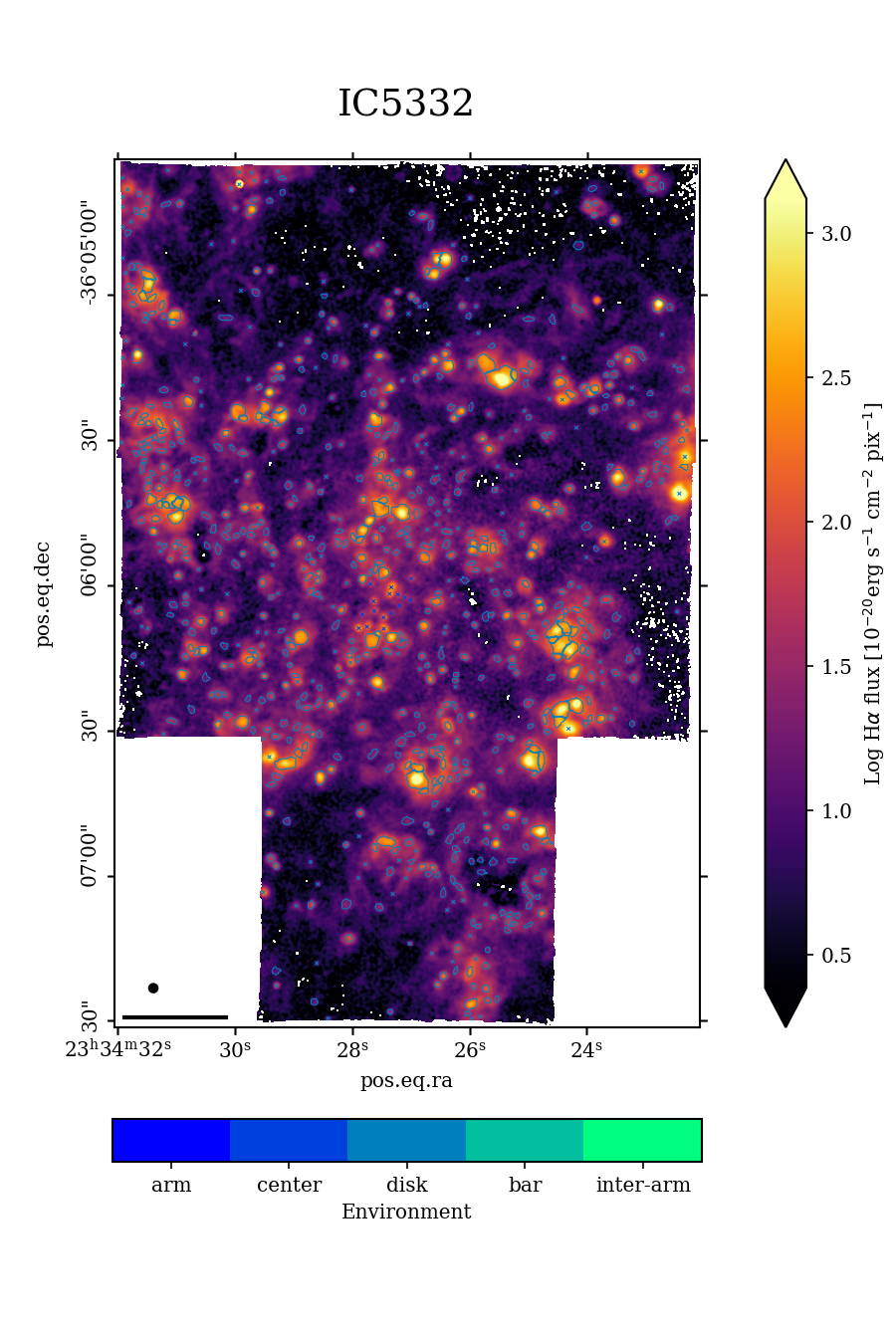}
 \caption{\HII\ regions and environments for \object{IC5332}. The figure shows the H$\alpha$ emission in the background, color coded according to the color scheme on the right, with overlaid the borders of the \HII\ regions in our catalog. The centers of the nebulae which have been discarded by our selection criteria are marked with crosses. In the lower left corner, the black circle indicates the PSF of the MUSE observations while the black line marks a physical scale corresponding to $1$~kpc. Both the \HII\ regions and the discarded nebulae are color coded according to our definition of environments as outlined by the color scheme at the bottom.
\label{fig:IC5332_regions}}
\end{center}
\end{figure*}

\begin{figure*}
\begin{center}
\includegraphics[width=1.0\textwidth]{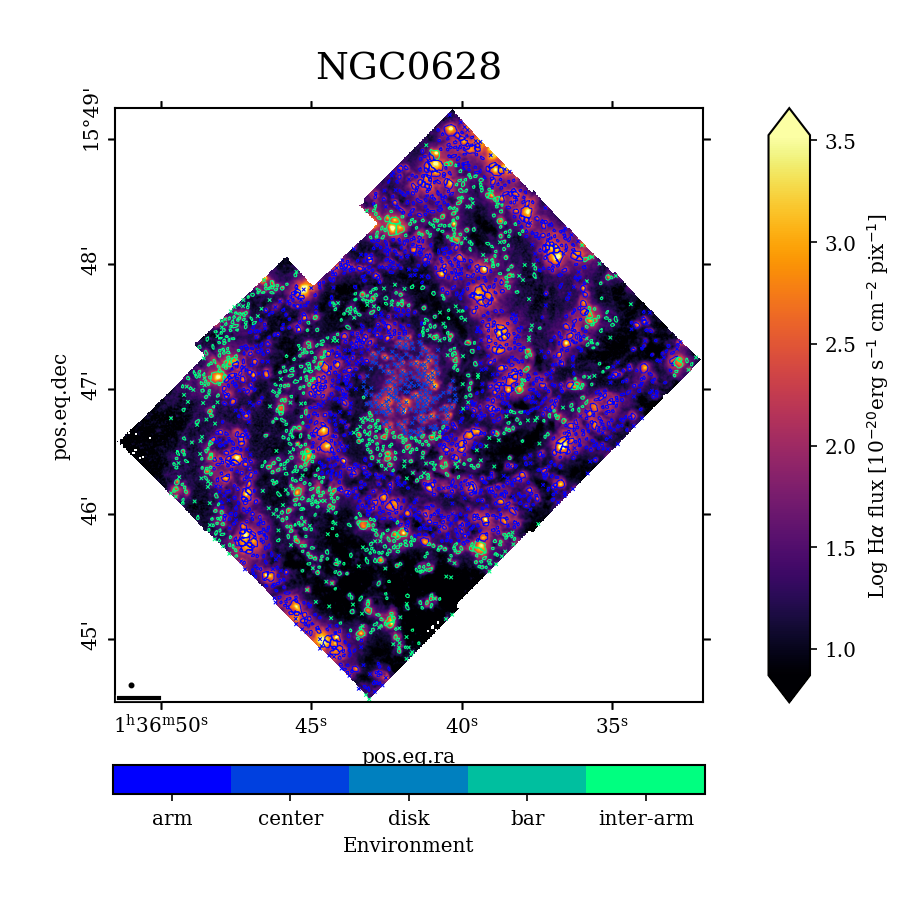}
\caption{\HII\ regions and environments for \object{NGC0628}. As in Fig.~\ref{fig:IC5332_regions}.
\label{NGC628_regions}}
\end{center}
\end{figure*}

\begin{figure*}
\begin{center}
\includegraphics[width=0.8\textwidth]{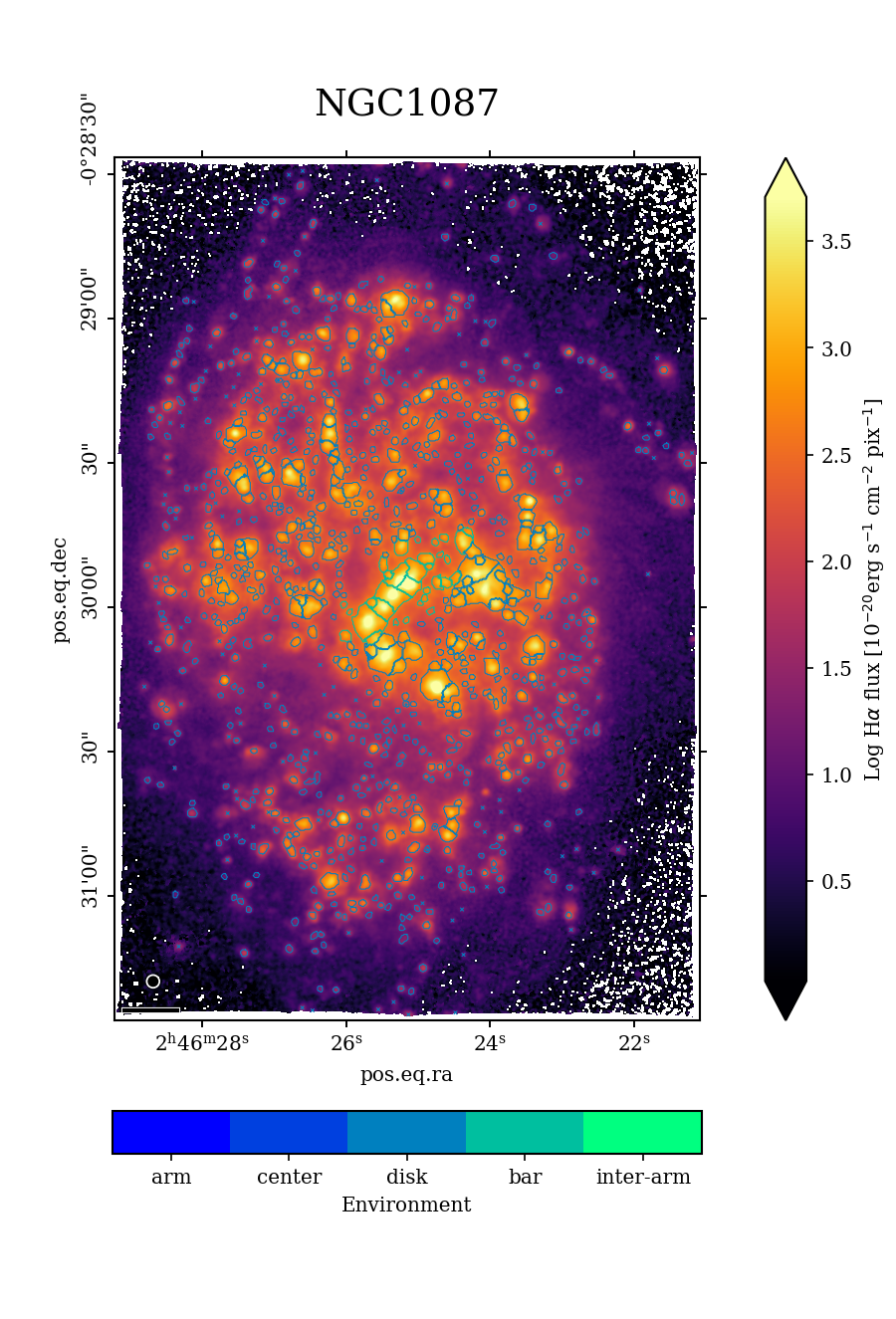}
\caption{\HII\ regions and environments for \object{NGC1087}. As in Fig.~\ref{fig:IC5332_regions}.
\label{NGC1087_regions}}
\end{center}
\end{figure*}

\begin{figure*}
\begin{center}
\includegraphics[angle=90,origin=c,width=0.75\textwidth]{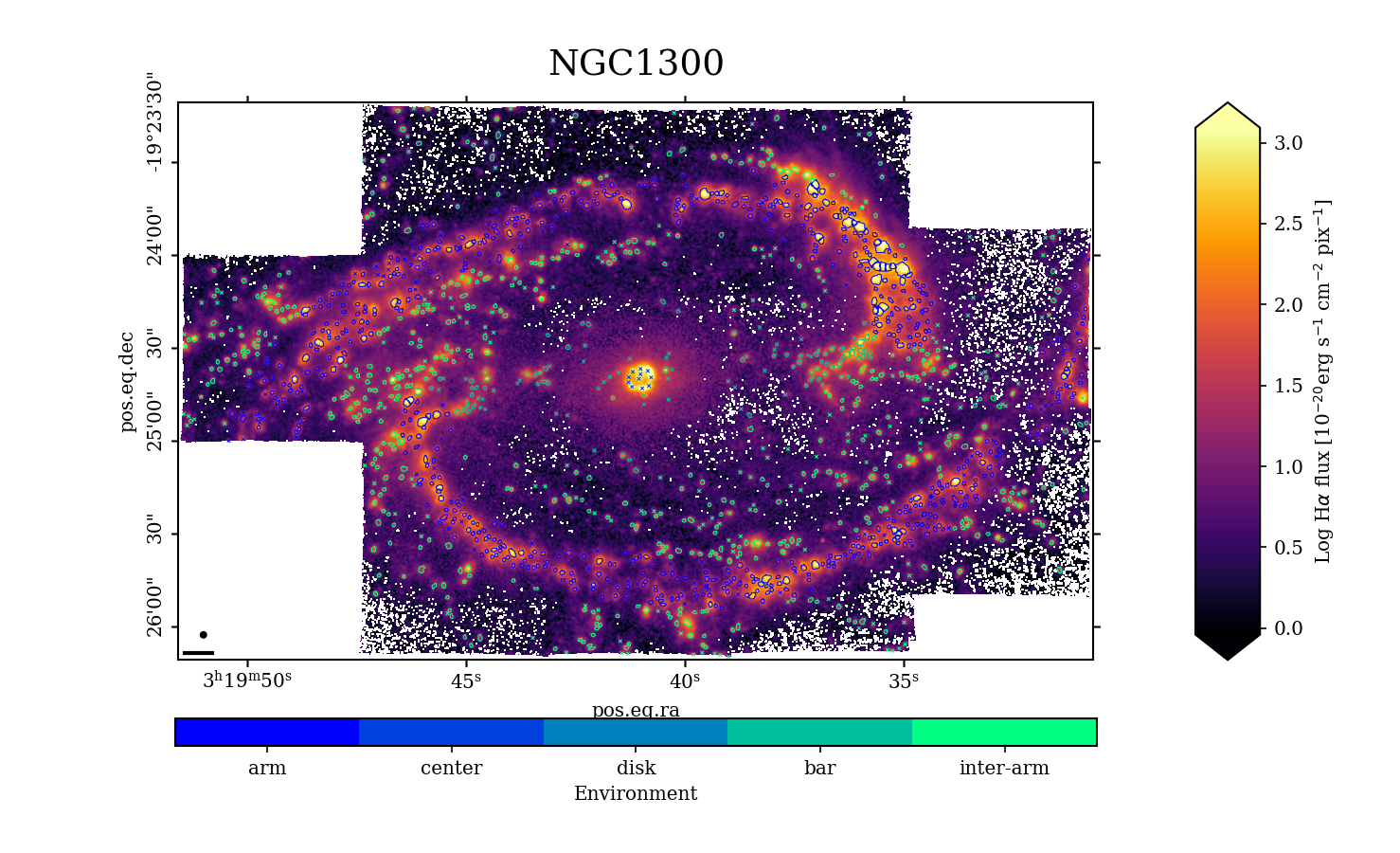}
\caption{\HII\ regions and environments for \object{NGC1300}. As in Fig.~\ref{fig:IC5332_regions}.
\label{NGC1300_regions} }
\end{center}
\end{figure*}

\begin{figure*}
\begin{center}
\includegraphics[angle=90,origin=c,width=0.75\textwidth]{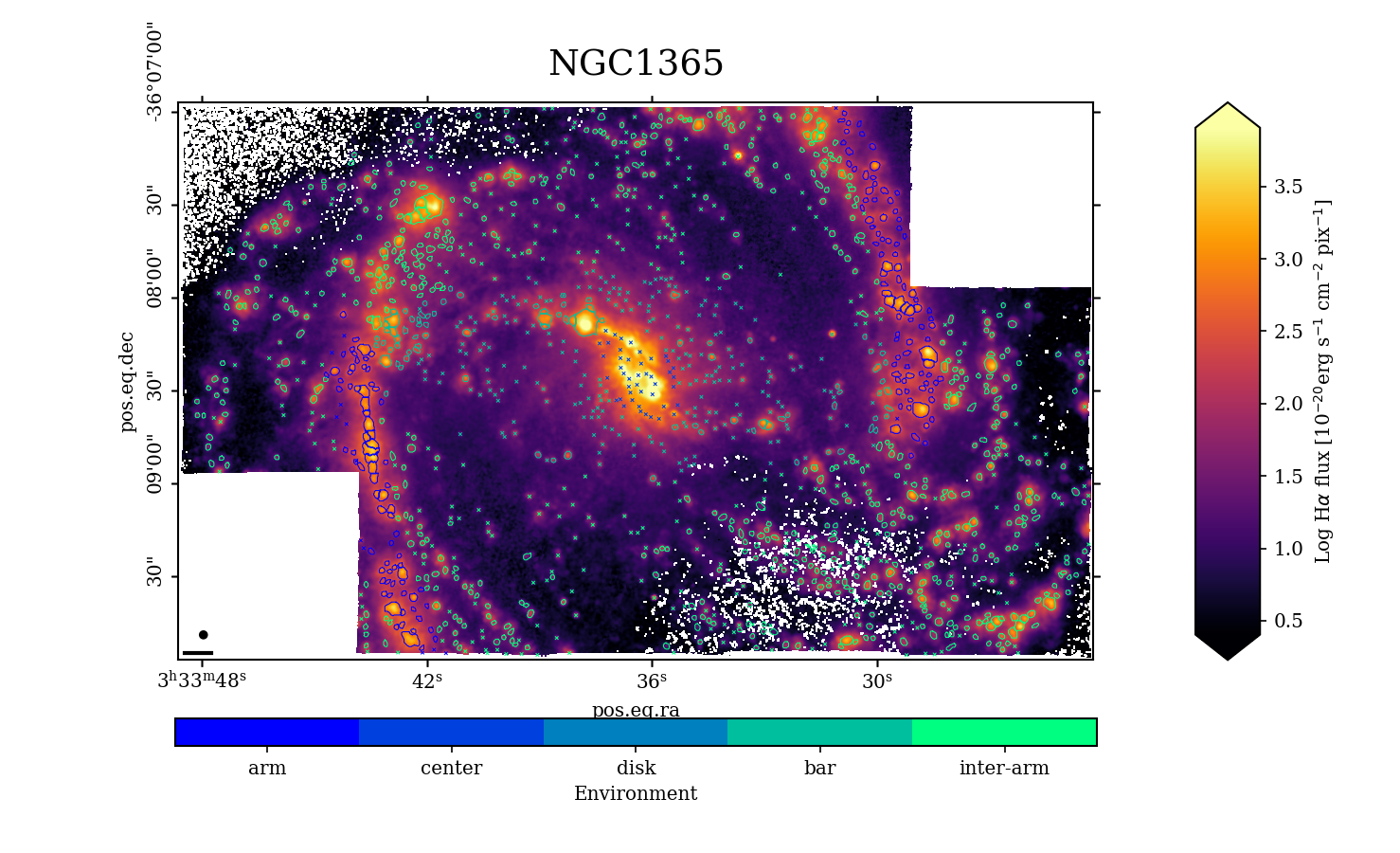}
\caption{\HII\ regions and environments for \object{NGC1365}. As in Fig.~\ref{fig:IC5332_regions}.
\label{NGC1365_regions} }
\end{center}
\end{figure*}

\begin{figure*}
\begin{center}
\includegraphics[width=0.8\textwidth]{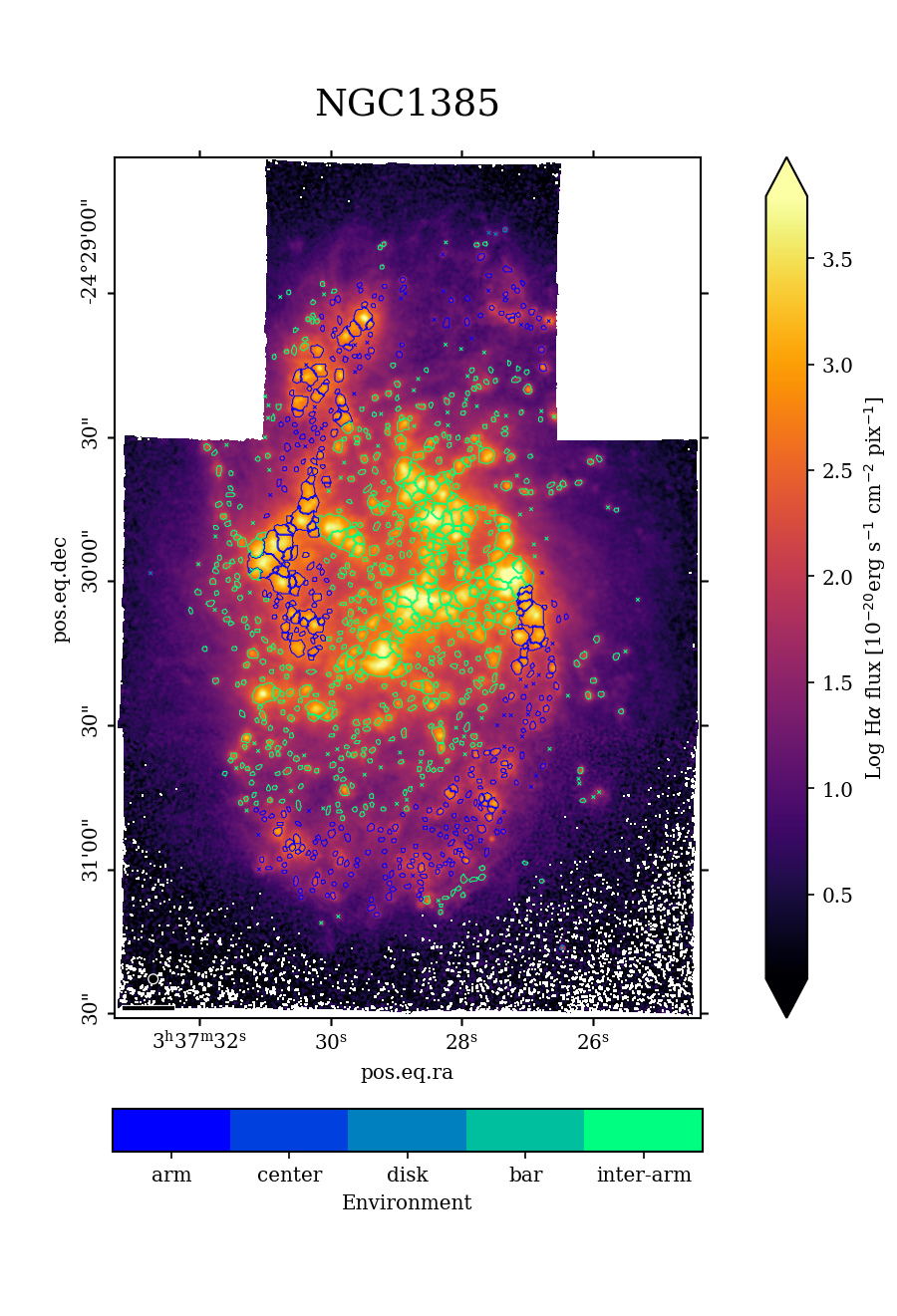}
\caption{\HII\ regions and environments for \object{NGC1385}. As in Fig.~\ref{fig:IC5332_regions}.
\label{NGC1385_regions} }
\end{center}
\end{figure*}

\begin{figure*}
\begin{center}
\includegraphics[angle=90,origin=c,width=0.8\textwidth]{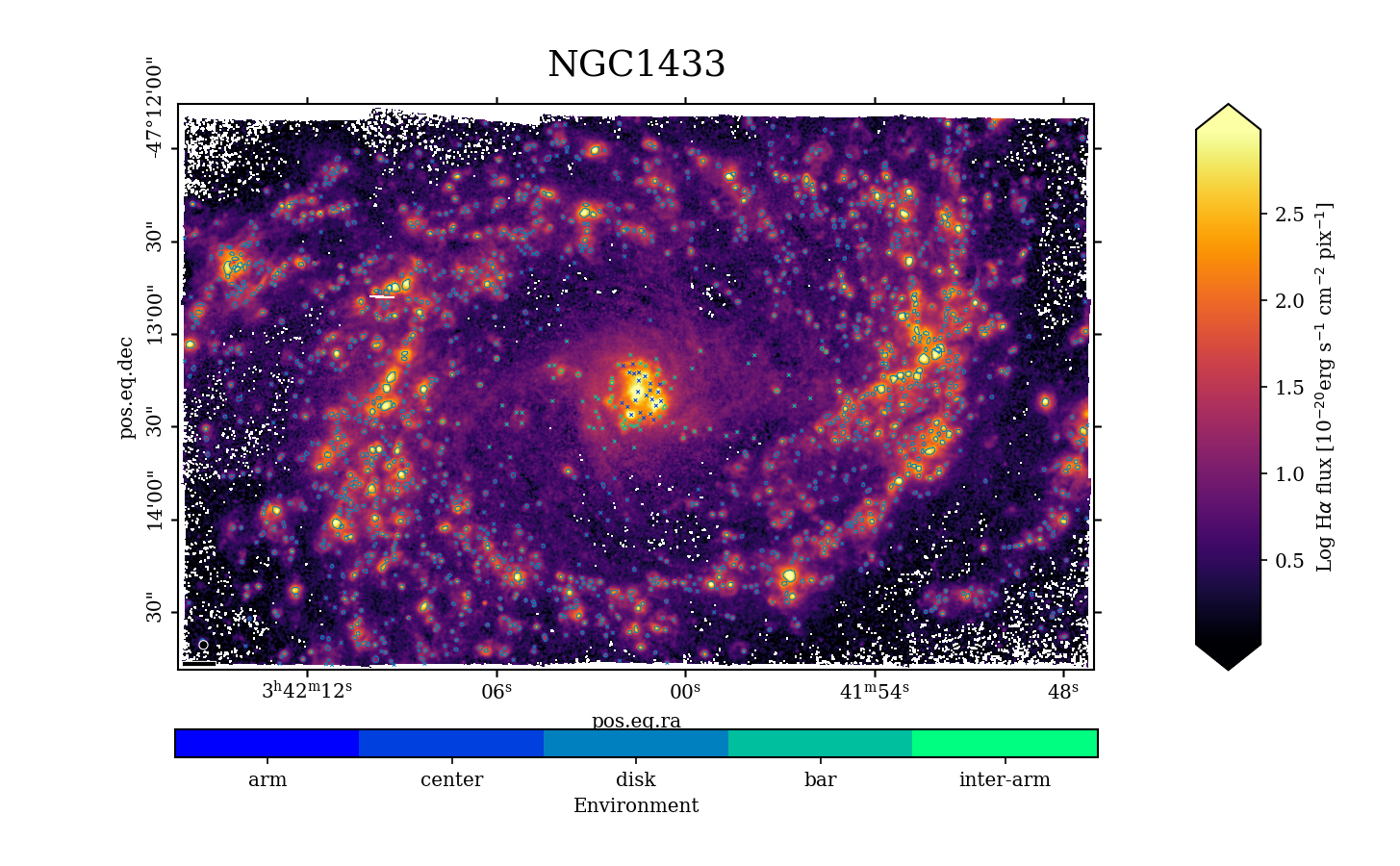}
\caption{\HII\ regions and environments for \object{NGC1433}. As in Fig.~\ref{fig:IC5332_regions}.
\label{NGC1433_regions} }
\end{center}
\end{figure*}

\begin{figure*}
\begin{center}
\includegraphics[width=1.0\textwidth]{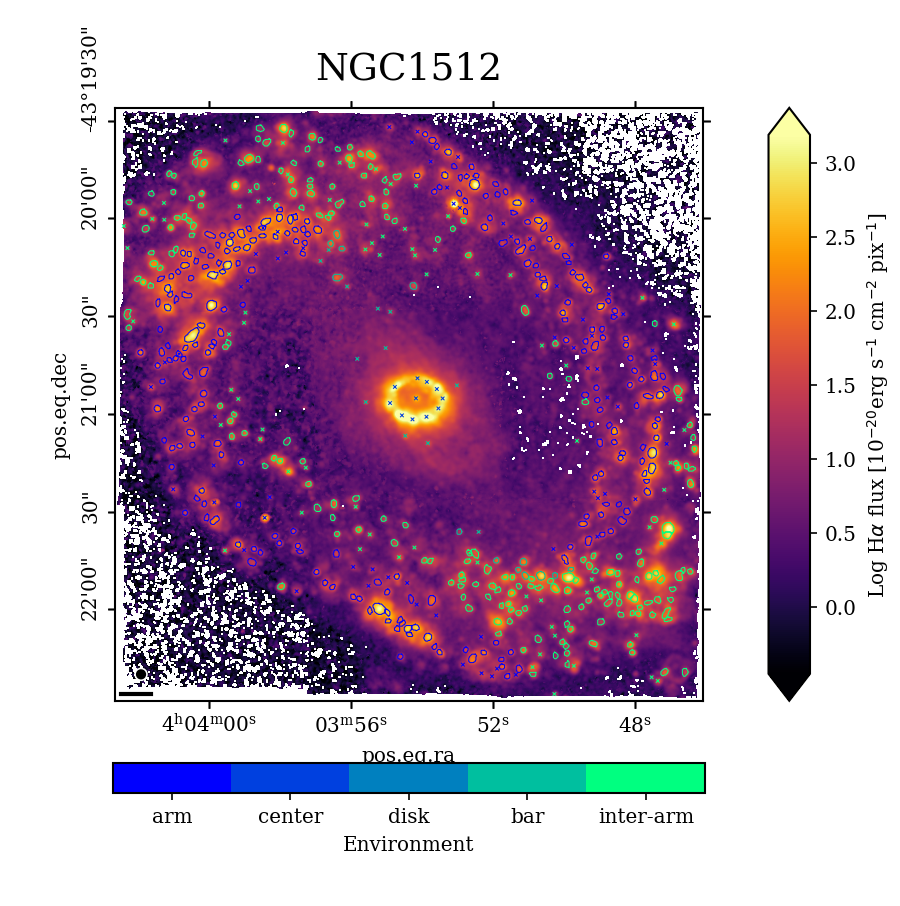}
\caption{\HII\ regions and environments for \object{NGC1512}. As in Fig.~\ref{fig:IC5332_regions}.
\label{NGC1512_regions} }
\end{center}
\end{figure*}

\begin{figure*}
\begin{center}
\includegraphics[width=0.9\textwidth]{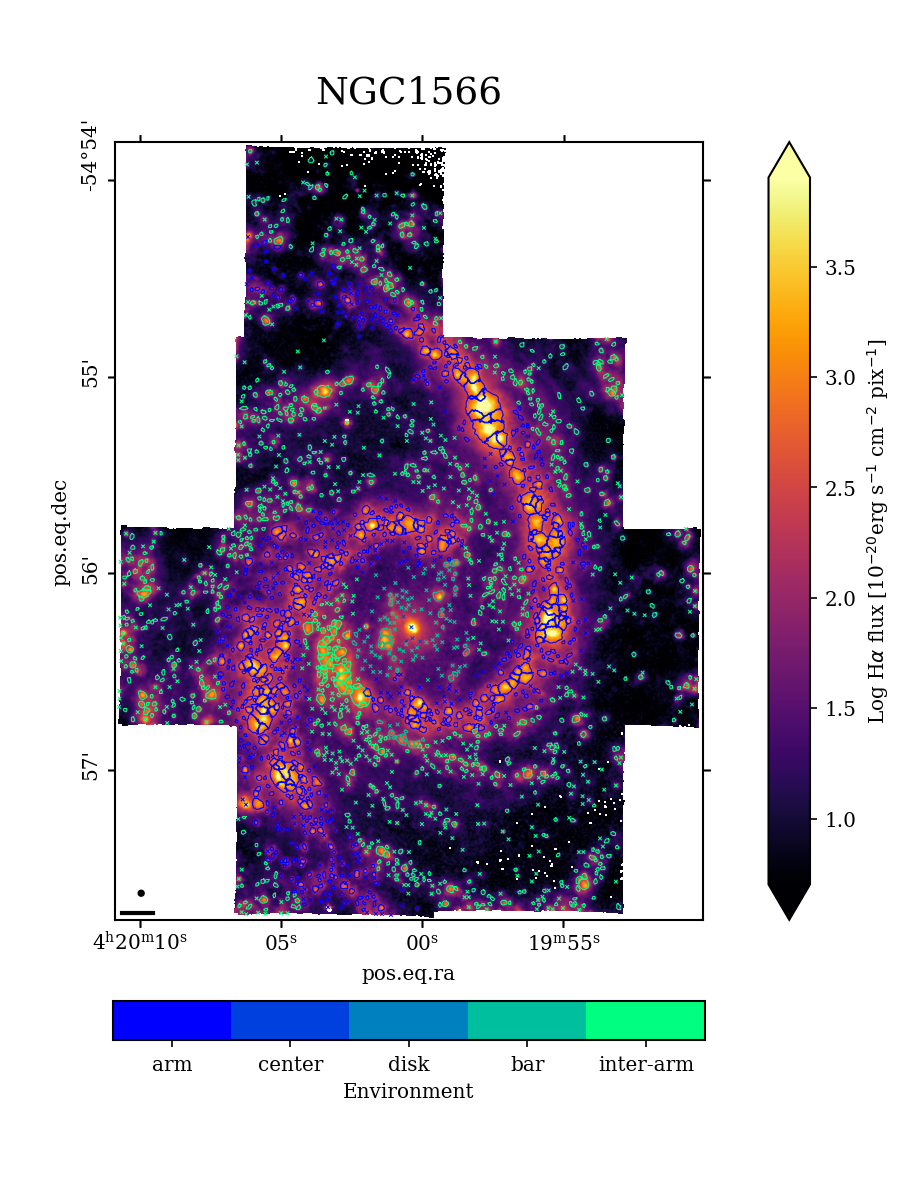}
\caption{\HII\ regions and environments for \object{NGC1566}. As in Fig.~\ref{fig:IC5332_regions}.
\label{NGC1566_regions} }
\end{center}
\end{figure*}

\begin{figure*}
\begin{center}
\includegraphics[angle=90,origin=c,width=0.7\textwidth]{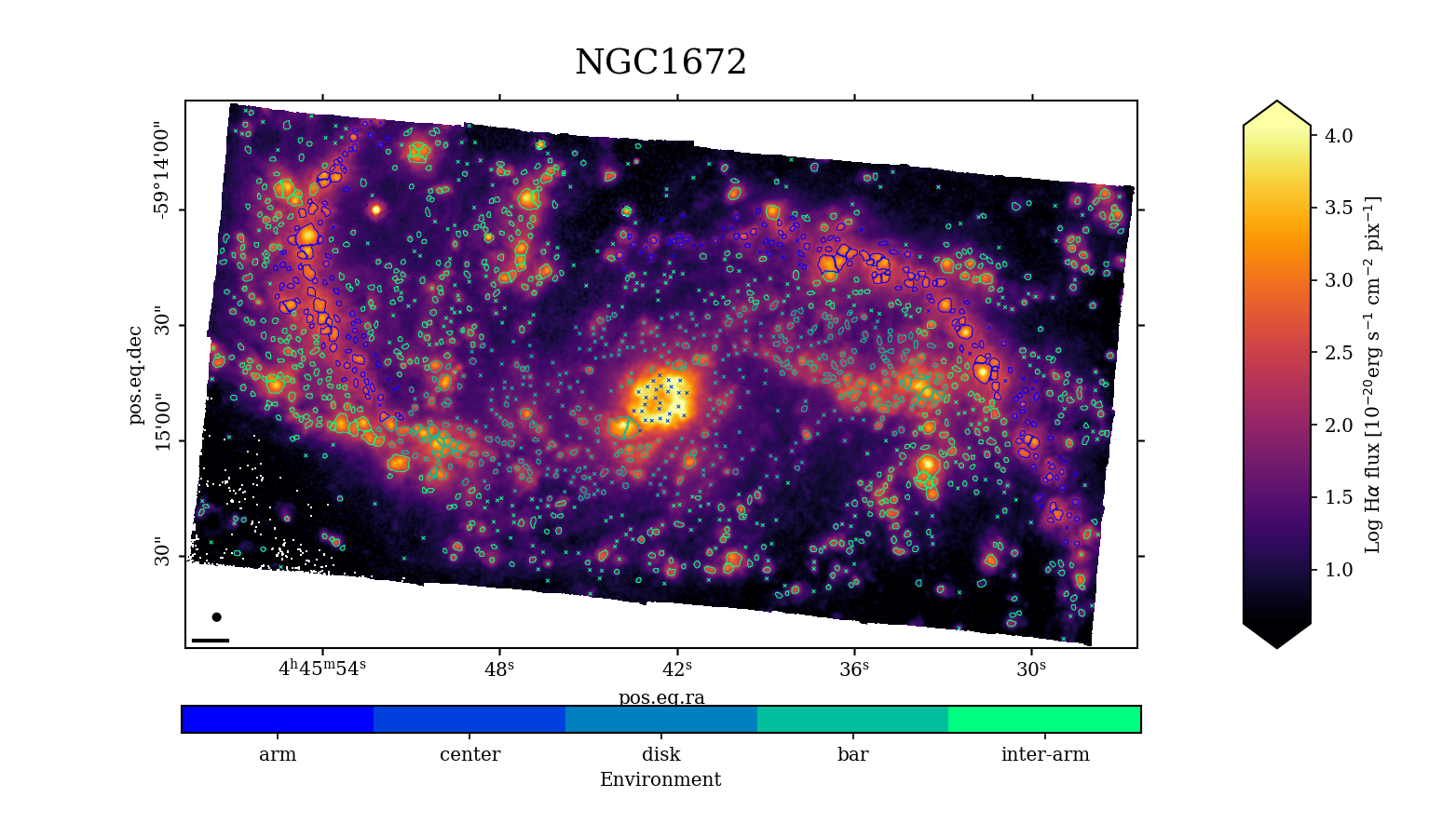}
\caption{\HII\ regions and environments for \object{NGC1672}. As in Fig.~\ref{fig:IC5332_regions}.
\label{NGC1672_regions} }
\end{center}
\end{figure*}

\begin{figure*}
\begin{center}
\includegraphics[width=1.0\textwidth]{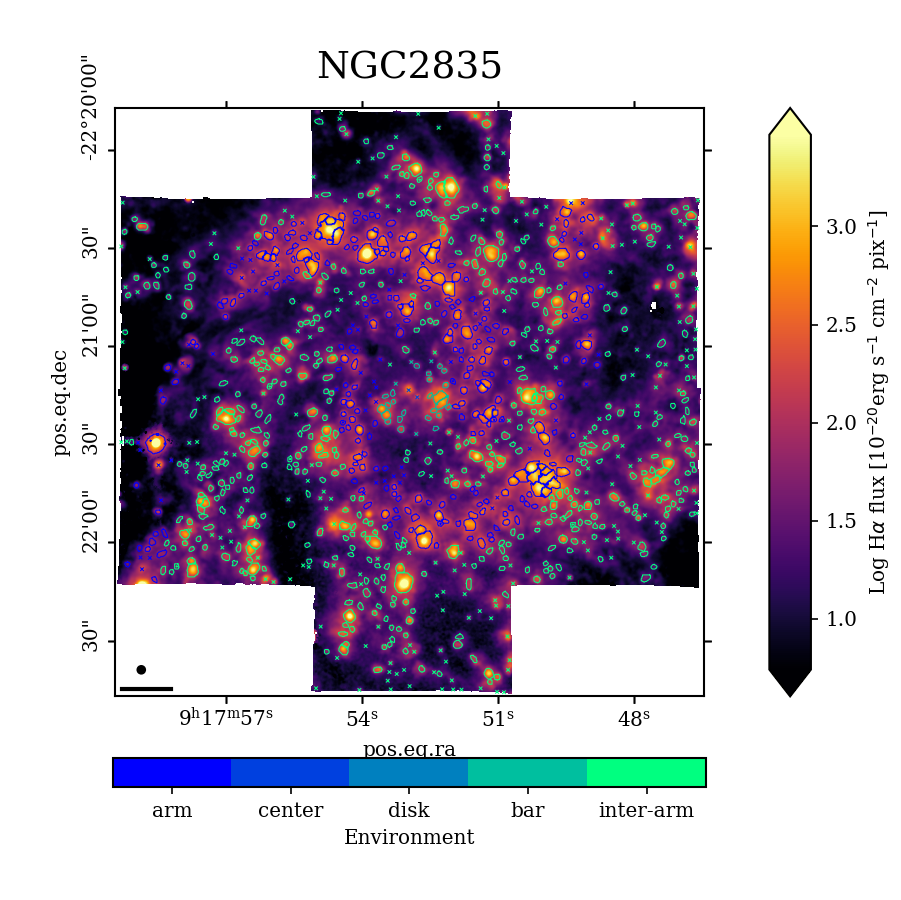}
\caption{\HII\ regions and environments for \object{NGC2835}. As in Fig.~\ref{fig:IC5332_regions}.
\label{NGC2835_regions} }
\end{center}
\end{figure*}

\begin{figure*}
\begin{center}
\includegraphics[width=1.0\textwidth]{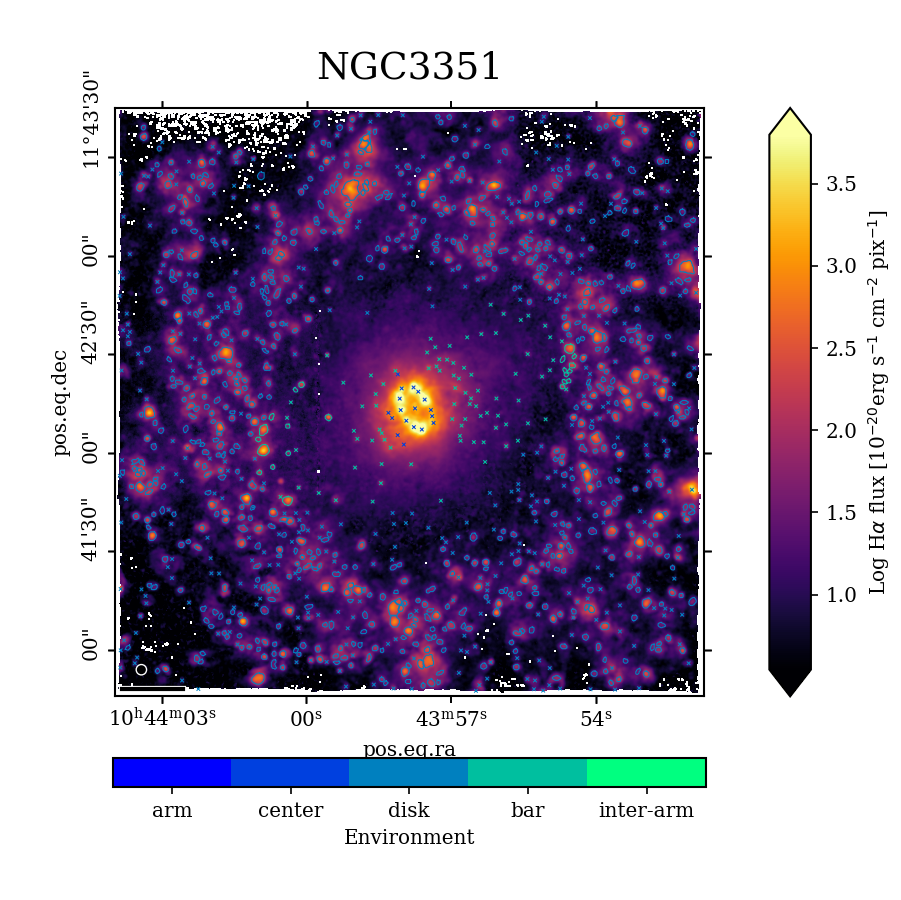}
\caption{\HII\ regions and environments for \object{NGC3351}. As in Fig.~\ref{fig:IC5332_regions}.
\label{NGC3351_regions} }
\end{center}
\end{figure*}

\begin{figure*}
\begin{center}
\includegraphics[width=0.6\textwidth]{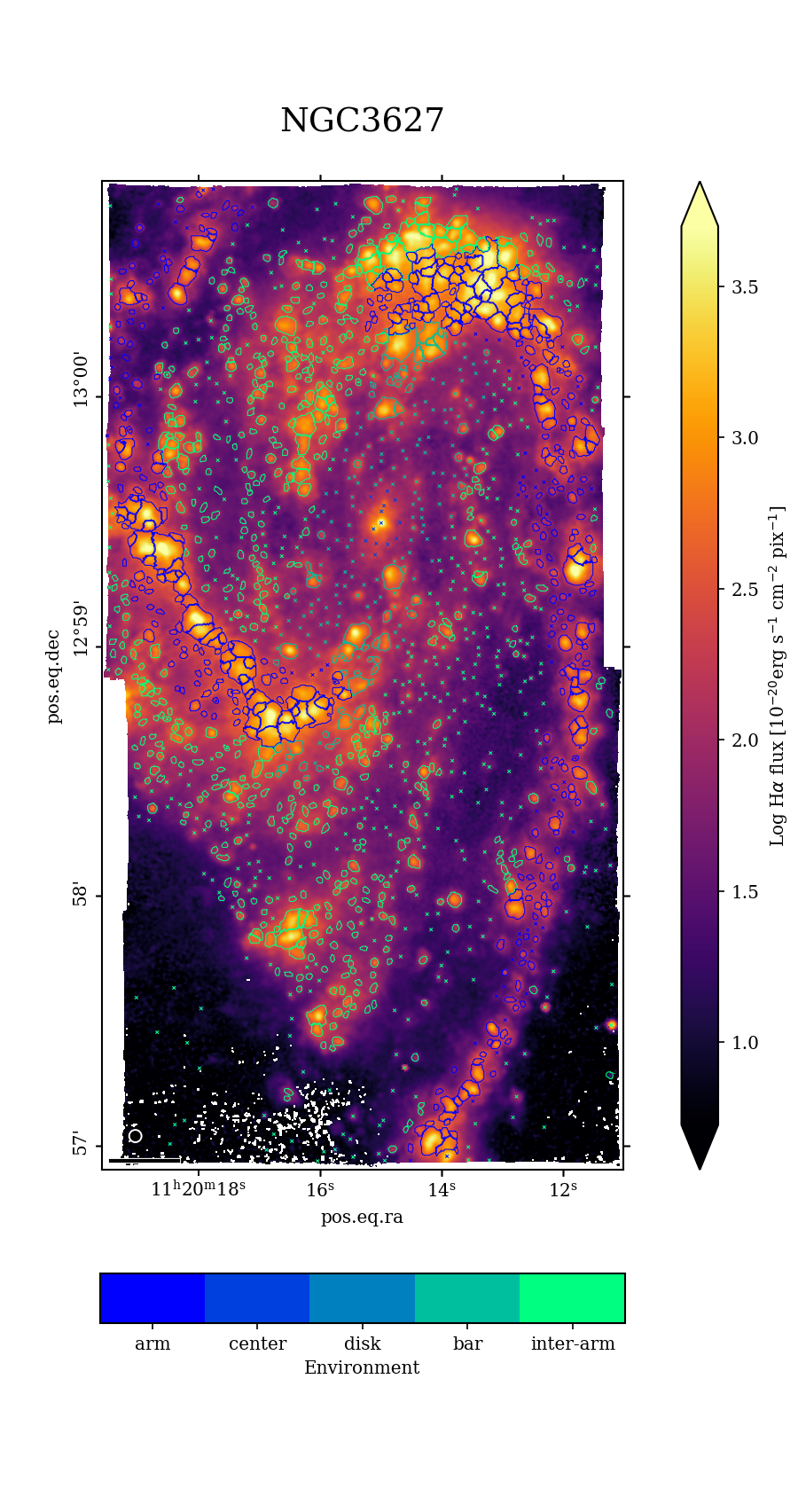}
\caption{\HII\ regions and environments for \object{NGC3627}. As in Fig.~\ref{fig:IC5332_regions}.
\label{NGC3627_regions} }
\end{center}
\end{figure*}

\begin{figure*}
\begin{center}
\includegraphics[width=1.0\textwidth]{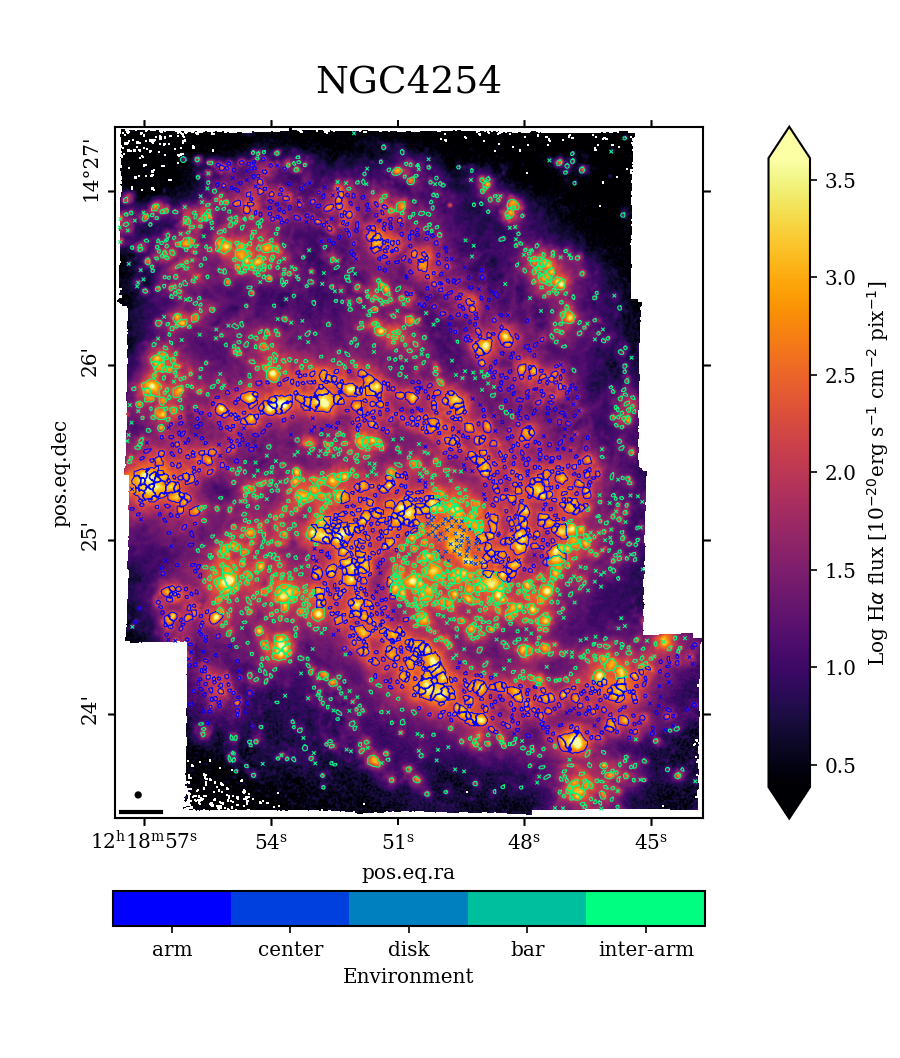}
\caption{\HII\ regions and environments for \object{NGC4254}. As in Fig.~\ref{fig:IC5332_regions}.
\label{NGC4254_regions} }
\end{center}
\end{figure*}

\begin{figure*}
\begin{center}
\includegraphics[width=1.0\textwidth]{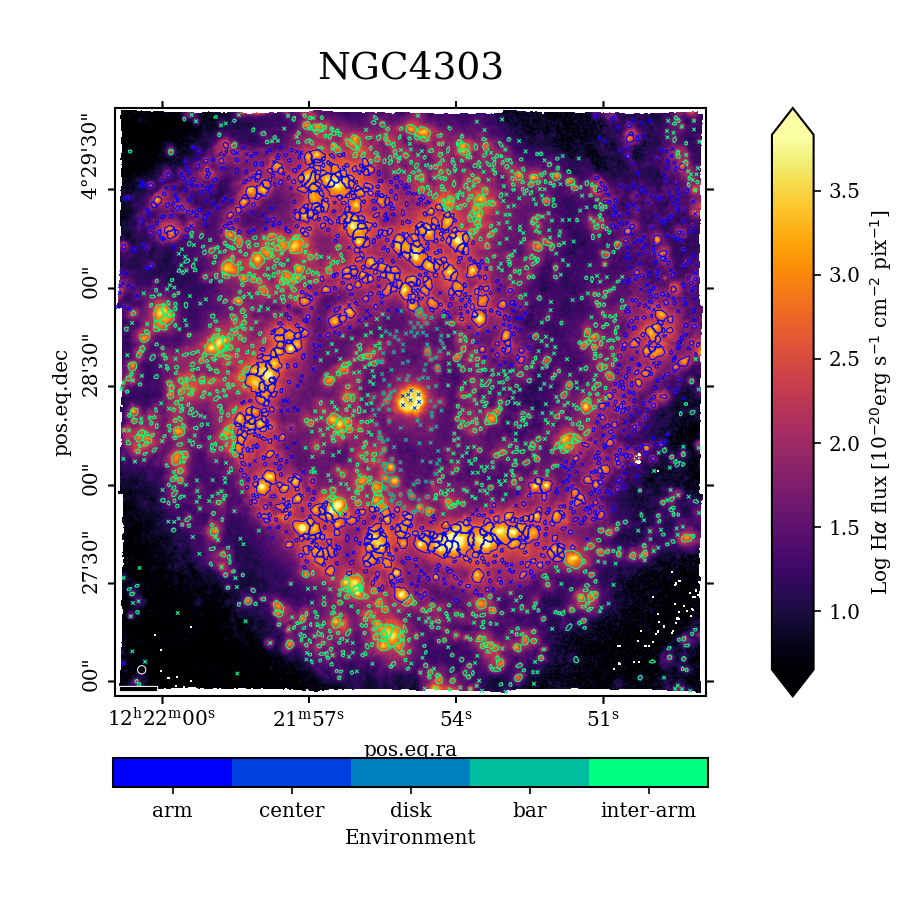}
\caption{\HII\ regions and environments for \object{NGC4303}. As in Fig.~\ref{fig:IC5332_regions}.
\label{ NGC4303_regions} }
\end{center}
\end{figure*}

\begin{figure*}
\begin{center}
\includegraphics[width=0.8\textwidth]{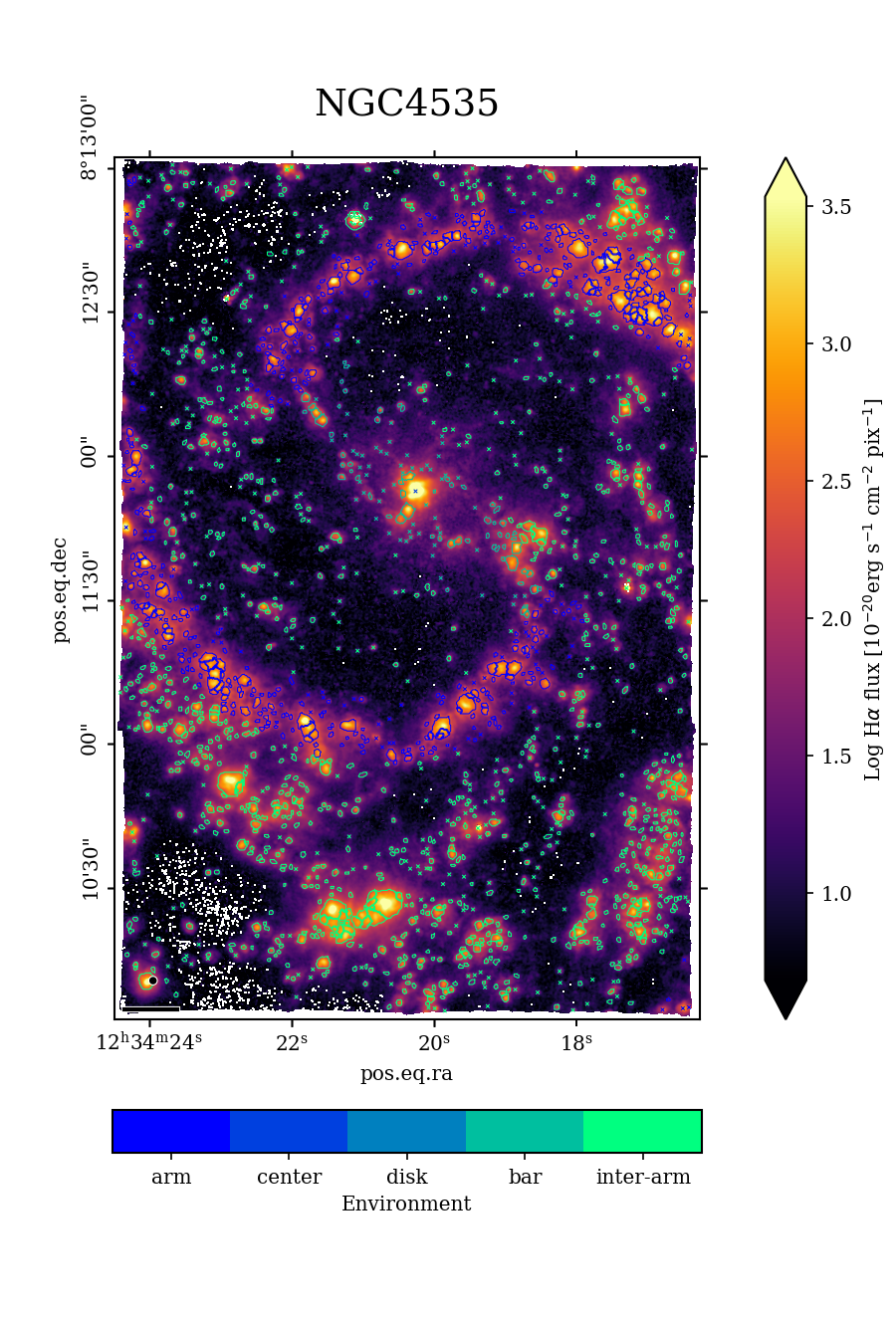}
\caption{\HII\ regions and environments for \object{NGC4535}. As in Fig.~\ref{fig:IC5332_regions}.
\label{ NGC4535_regions} }
\end{center}
\end{figure*}

\begin{figure*}
\begin{center}
\includegraphics[width=1.0\textwidth]{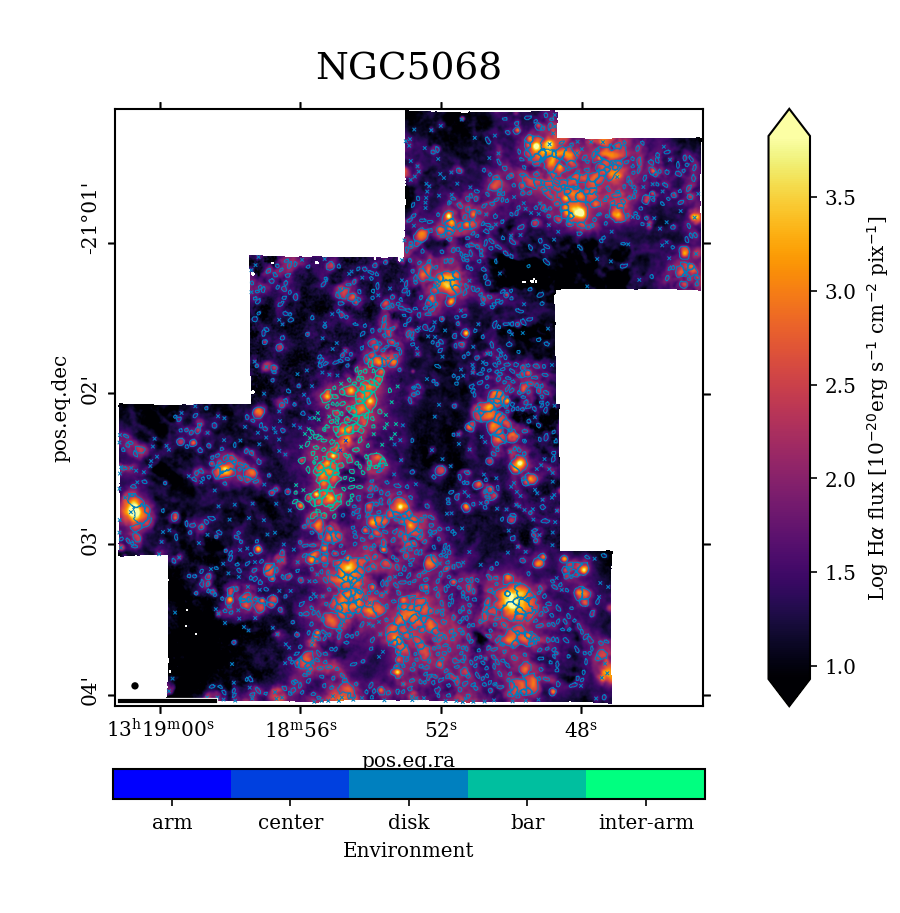}
\caption{\HII\ regions and environments for \object{NGC5068}. As in Fig.~\ref{fig:IC5332_regions}.
\label{ NGC5068_regions} }
\end{center}
\end{figure*}

\begin{figure*}
\begin{center}
\includegraphics[width=0.7\textwidth]{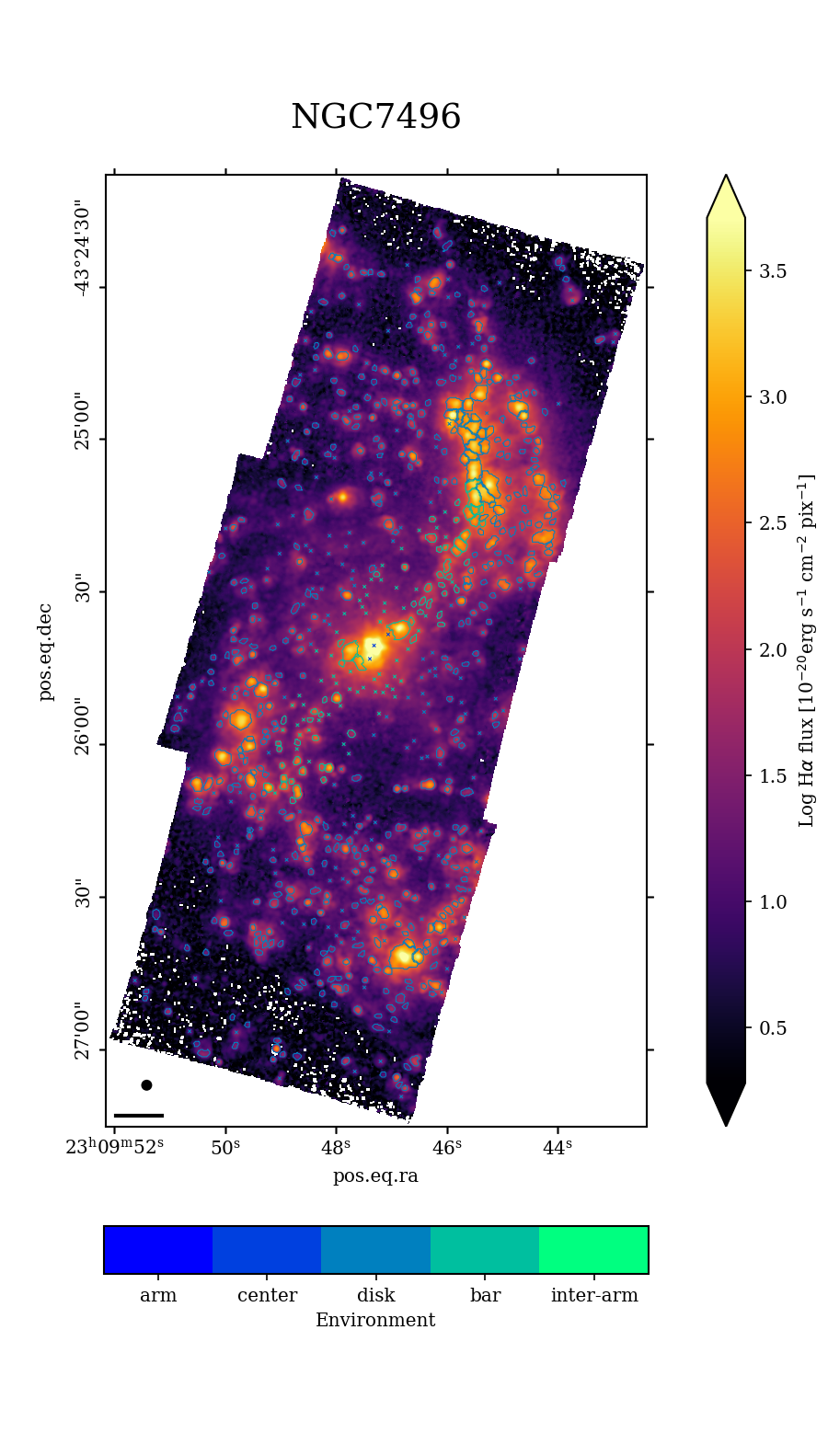}
\caption{\HII\ regions and environments for \object{NGC7496}. As in Fig.~\ref{fig:IC5332_regions}.
\label{ NGC7496_regions} }
\end{center}
\end{figure*}

\begin{figure*}
    \centering
    \includegraphics[width=0.85\textwidth]{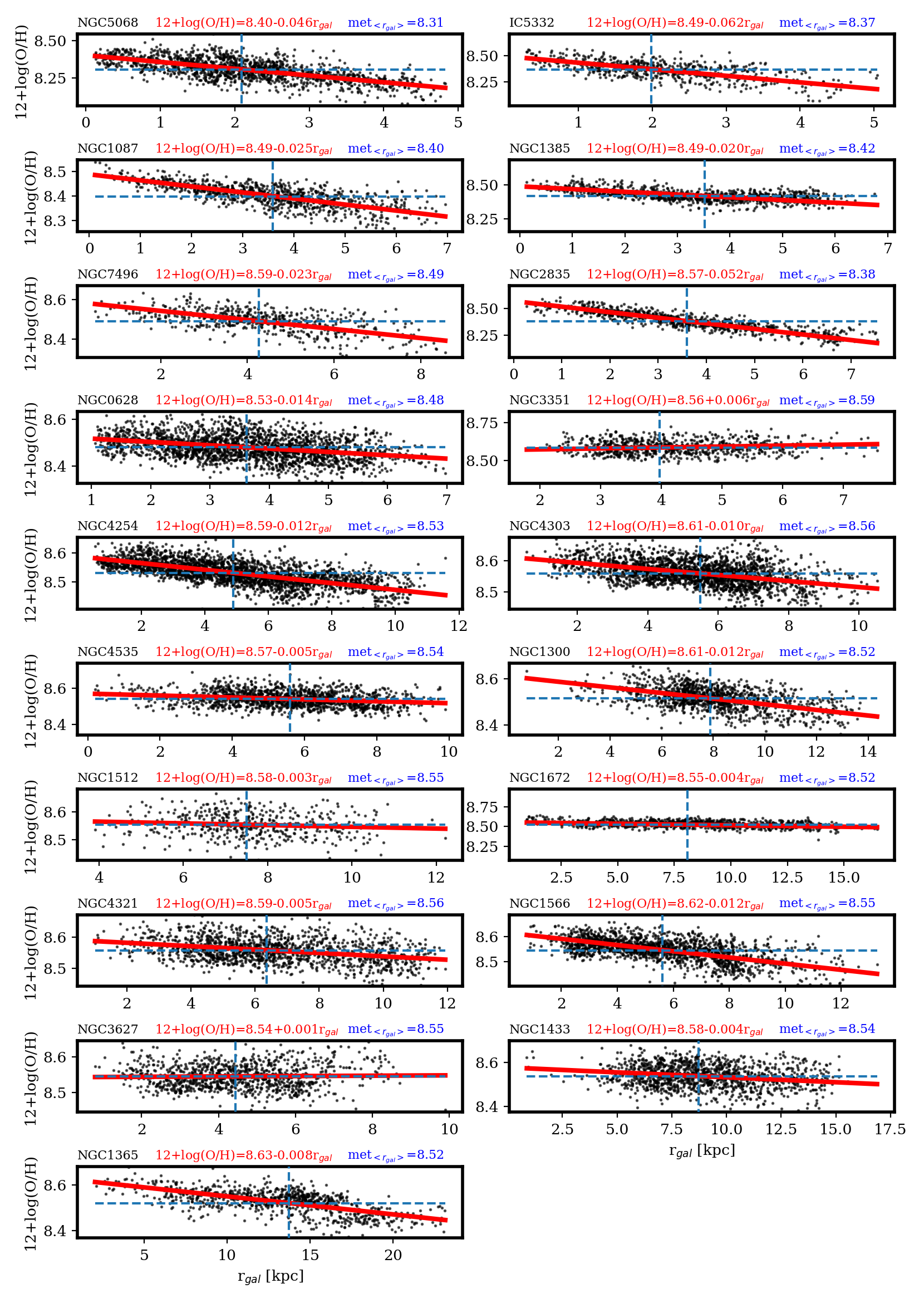}
    \caption{Fit of the radial metallicity gradients for the galaxies in the PHANGS--MUSE sample. Galaxies are ordered according to increasing stellar mass from top left to bottom right and their name is indicated on the top right of each panel. The \HII\ region metallicity $12+\log(\mathrm{O/H})$ is plotted against its de-projected galactocentric radius $r_\mathrm{gal}$ measured in kpc (black points). The best linear unweighted least-square fit relation reported above each panel (red label) is shown using a red solid line. The blue dashed vertical and horizontal lines intercept at the metallcity given by the best-fitting at the mean \HII\ region galacrocentric radius $\langle r_\mathrm{gal} \rangle$, the latter value $\mathrm{met}_{\langle r_\mathrm{gal} \rangle}$ is reported above each panel (blue label).
    \label{fig:met_radial_gradients}}
\end{figure*}

In this section, we present the spatial masks of the \HII\ regions for 18~galaxies of the PHANGS--MUSE sample (from Fig.~\ref{fig:IC5332_regions} to Fig.~\ref{ NGC7496_regions}) highlighting the \HII\ region footprints and the different galaxy environments. 
At the end of this section in Fig.~\ref{fig:met_radial_gradients}, we also show the best-fitting radial metallicity gradients for all the galaxies in our sample. They provide the measurement of the metallicity at the median galactocentric radius covered by the MUSE observations used in Sec.~\ref{sec:LF variations between galaxies}.

\newpage

\section{Variations of the LF within single galaxies}\label{appendix2}

\begin{figure*}
\begin{center}
    \includegraphics[trim=1cm 3cm 2cm 3cm, clip, width=1\textwidth]{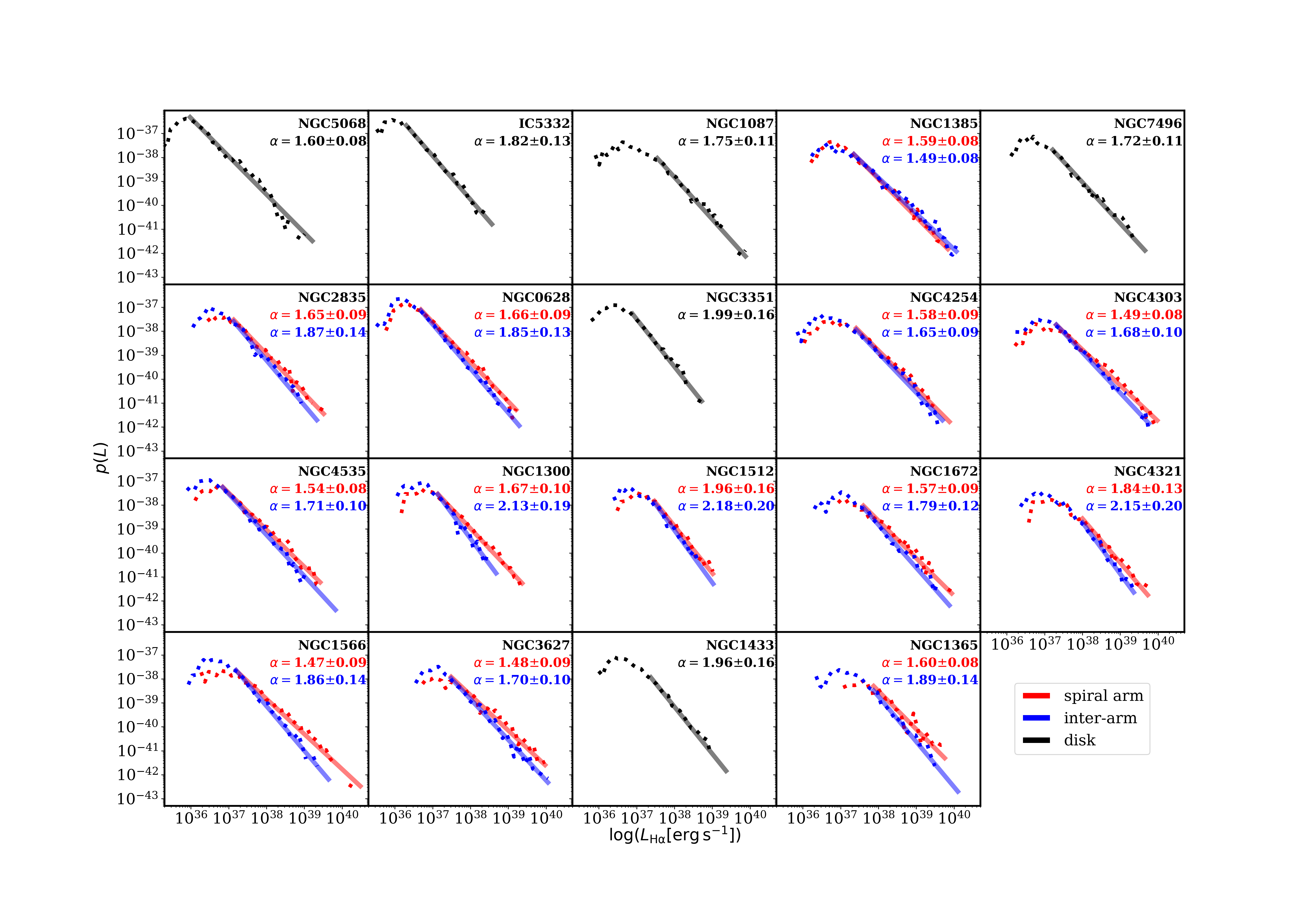}
    \caption{LFs for \HII\ regions in spiral arm and inter-arm areas. Galaxies are ordered according to increasing stellar mass from top left to bottom right and their names are indicated within each panel. For the 13 galaxies showing evident spiral arms, the LF and the best-fitting model are shown in red and blue color for \HII\ regions located in spiral arms and inter-arms areas, respectively. For the remaining galaxies, the LF and the best-fitting model refer to the \HII\ regions located in the entire disk, excluding the areas occupied by the bars, and are drawn in black color. The dashed and solid lines indicate the empirical LF and the best-fitting model, respectively. The LF slopes are reported in the top right corner of each panel following the same color scheme.}
    \label{Figure:LF_arm_interarm}
\end{center}
\end{figure*}

\begin{figure*}
\begin{center}
    \includegraphics[trim=1cm 3cm 2cm 3cm, clip,width=1\textwidth]{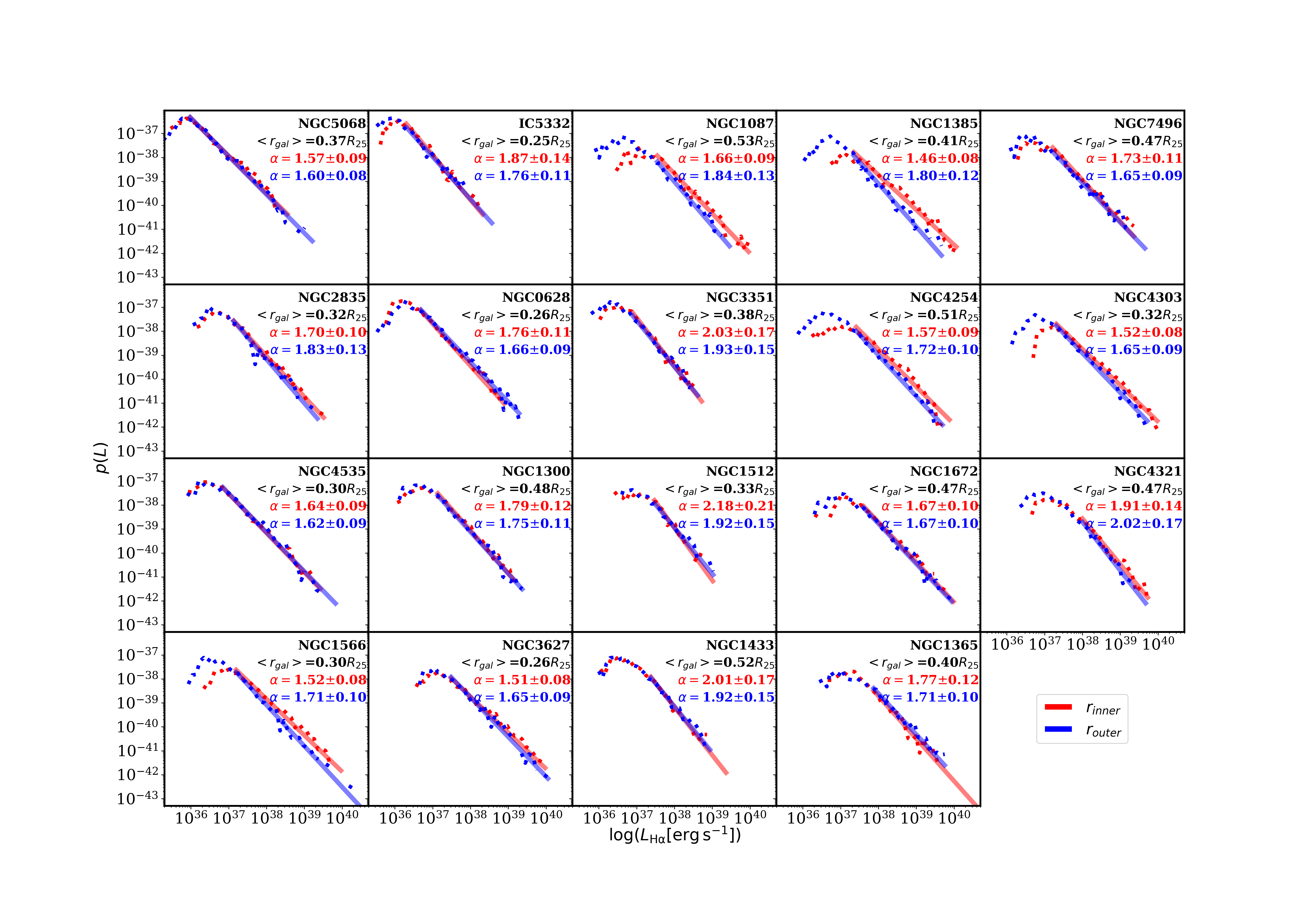}
    \caption{LFs for \HII\ regions located in the inner (red colors) and outer disk (blue colors) areas. Galaxies are ordered according to increasing stellar mass from top left to bottom right and their names are indicated within each panel. The dashed and solid lines indicate the empirical LF and the best-fitting model, respectively. The LF slopes are reported in the top right corner of each panel following the same color scheme. The median galactocentric radius $\langle r_\mathrm{gal} \rangle$ of the \HII\ region parent sample, used to separate inner and outer disks, is indicated below each galaxy name in units of $R_{25}$.}
    \label{Figure:LF_radial}
\end{center}
\end{figure*}

\begin{figure*}
\begin{center}
    \includegraphics[trim=1cm 3cm 2cm 3cm, clip,width=1\textwidth]{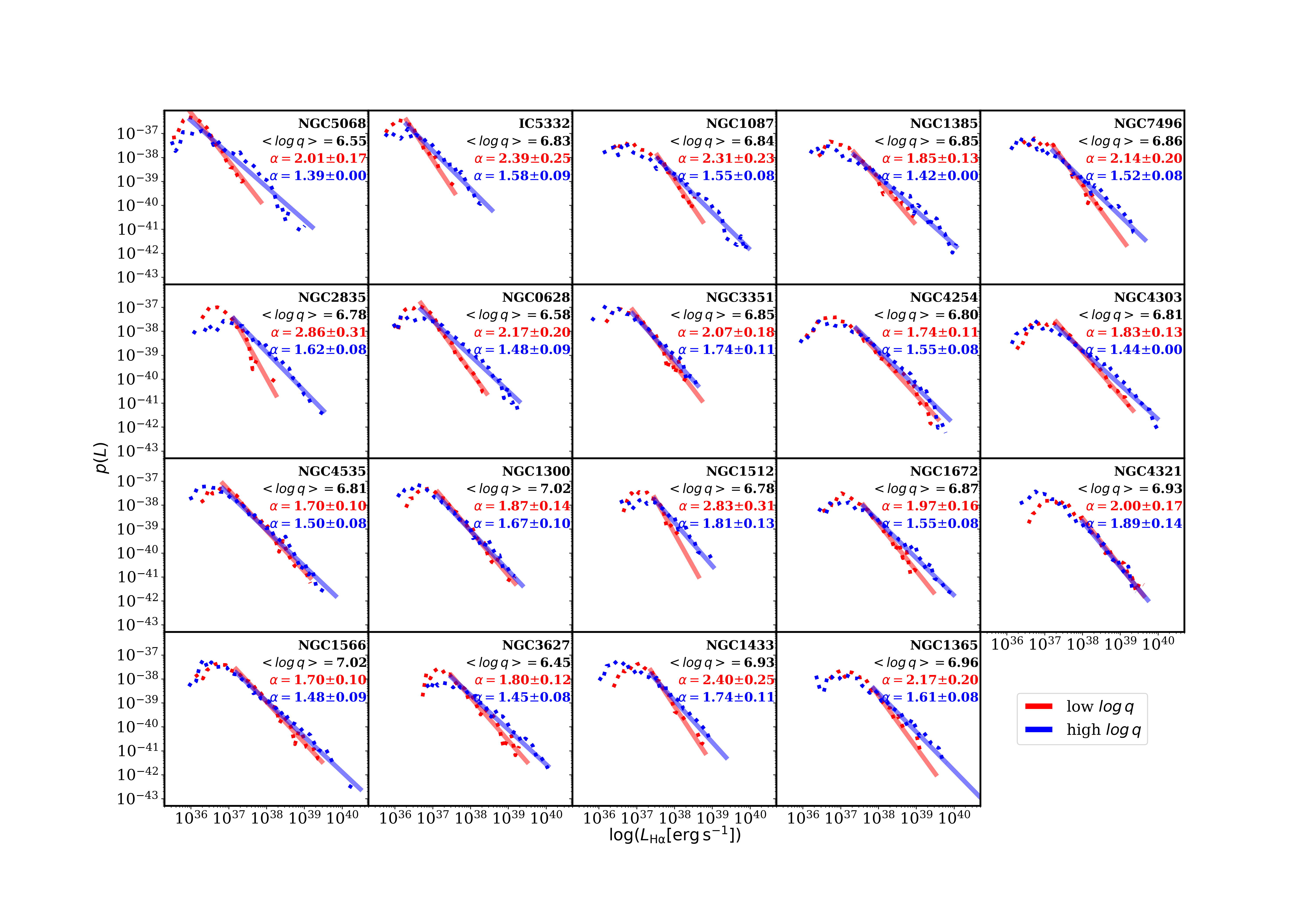}
    \caption{Best-fitting model of the \HII\ region LF for regions with high (blue colors) and low (red colors) gas ionization parameter~$q$. Galaxies are ordered according to increasing stellar mass from top left to bottom right and their names are indicated within each panel. The dashed and solid lines indicate the empirical LF and the best-fitting model, respectively. The LF slopes are reported in the top right corner of each panel following the same color scheme. The median ionization parameter $\langle \log q \rangle$ of the \HII\ region parent sample, used to separate young and old \HII\ regions, is indicated below each galaxy name in logarithmic units.}

    \label{Figure:LF_Q}
\end{center}
\end{figure*}

In this section, we show the LFs and their fits for the \HII\ region sub-samples described in the main text in Sec.~\ref{sec:LF variations within galaxies}.  Fig.~\ref{Figure:LF_arm_interarm} shows the LFs of the spiral arm and inter-arm areas, six of our 19~galaxies do not show evident spiral arms, for these galaxies we show the LF and best-fitting models of the \HII\ regions across the entire disk but excluding bars (i.e.\ this is the reason for slight variations with respect to the fits shown in Fig.~\ref{Figure:global_LF}).
Fig.~\ref{Figure:LF_radial} shows the LF of the inner and outer parts of the star-forming disk while Fig.~\ref{Figure:LF_Q} shows the LFs of \HII\ regions with high and low ionization parameter $\log q$.

\section{The effect of dust correction and BPT cuts on the LF slopes}\label{appendix3}

\begin{figure*}
    \centering
    \includegraphics[trim=1cm 3cm 2cm 3cm, clip,width=1\textwidth]{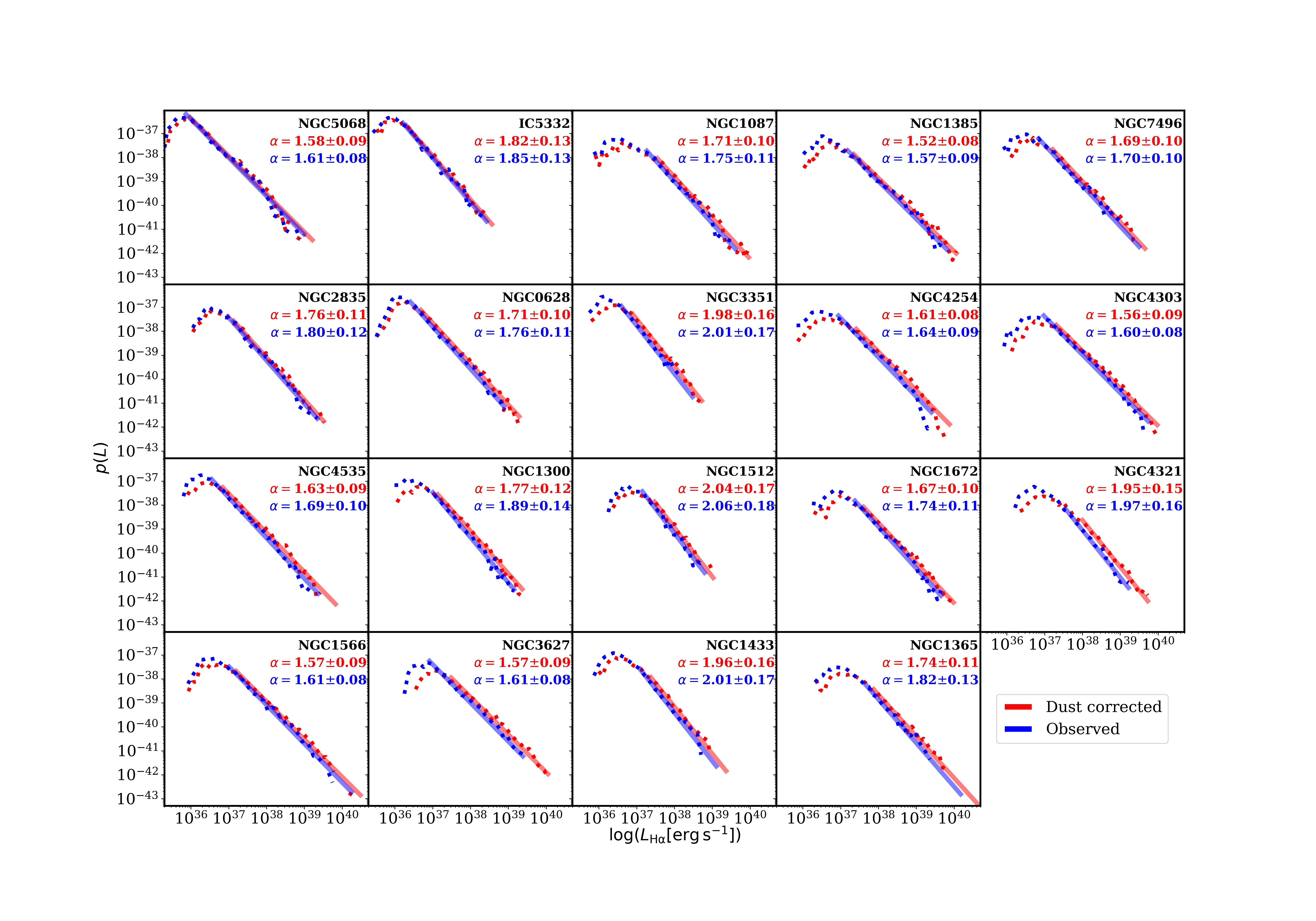}
    \caption{Fit of the LF for \HII\ regions obtained from the dust-corrected (same as in Fig.~\ref{Figure:global_LF}, red colors) and observed (blue colors) \Ha\ fluxes. Galaxies are ordered by increasing stellar mass from top left to bottom right and their names are indicated within each panel. The dashed and solid lines indicate the empirical LF and the best-fitting model, respectively. The LF slopes are reported in the top right corner of each panel following the same color scheme. }
    \label{Figure:LF_slope_dustcorrection}
\end{figure*}

\begin{figure}
    \centering
    \includegraphics[width=0.5\textwidth]{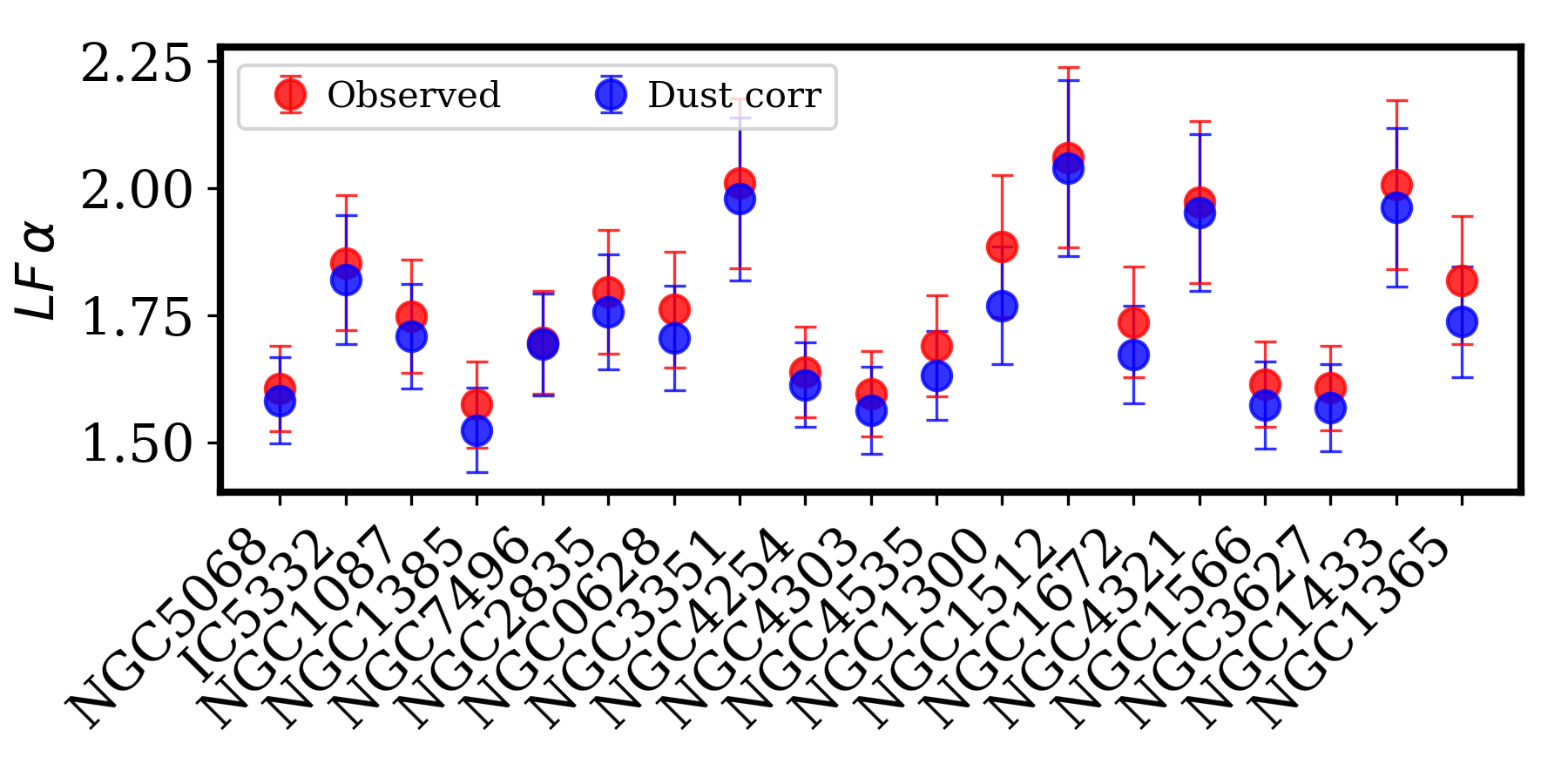}
    \caption{Slope of the LF for \HII\ regions obtained from the dust-corrected (same as in Fig.~\ref{Figure:global_LF}, red colors) and observed (blue colors) \Ha\ fluxes. Galaxies are ordered by increasing stellar mass from left to right and their names are indicated along the abscissa.}
    \label{Figure:LF_slope_dustcorrection_small}
\end{figure}

In this section, we discuss in more detail the effect that dust extinction and selection criteria might play on the LF slopes presented in our paper.
In Sec.~\ref{sec:HII regions luminosity function}, we discuss LFs built from extinction corrected \Ha\ luminosities of \HII\ regions. The spectroscopic nature of the MUSE data allow us to perform extinction correction via the \Ha/\Hb\ Balmer decrement, however, in some cases (i.e.\ when the nebular LF is obtained from narrow-band \Ha\ imaging) this is not possible. To understand the effect of dust extinction, we fit the LF built with observed \Ha\ luminosities and in Fig.~\ref{Figure:LF_slope_dustcorrection} and Fig.~\ref{Figure:LF_slope_dustcorrection_small}, we compare the results with what we presented in Sec.~\ref{sec:HII regions luminosity function}. 
Overall, we see only a small change in the LF slope if we do not correct for dust extinction. The variations always remain within the uncertainty estimated for the LF slope. The dust-corrected LF slopes are, as one would expect, shallower due to the presence of more luminous \HII\ regions after the dust correction is performed. This guarantees that our results can be compared to nebular LF studies, performed with e.g.\ narrow-band \Ha\ observations, that do not correct for the effect of dust extinction. As can be noted in Fig.~\ref{Figure:LF_slope_dustcorrection} another effect of the lack of a dust correction is the lowering of \Lmin\ due to the fact that we detect a higher number of faint \HII\ regions. 

\begin{figure*}
    \centering
    \includegraphics[trim=1cm 3cm 2cm 3cm, clip,width=1\textwidth]{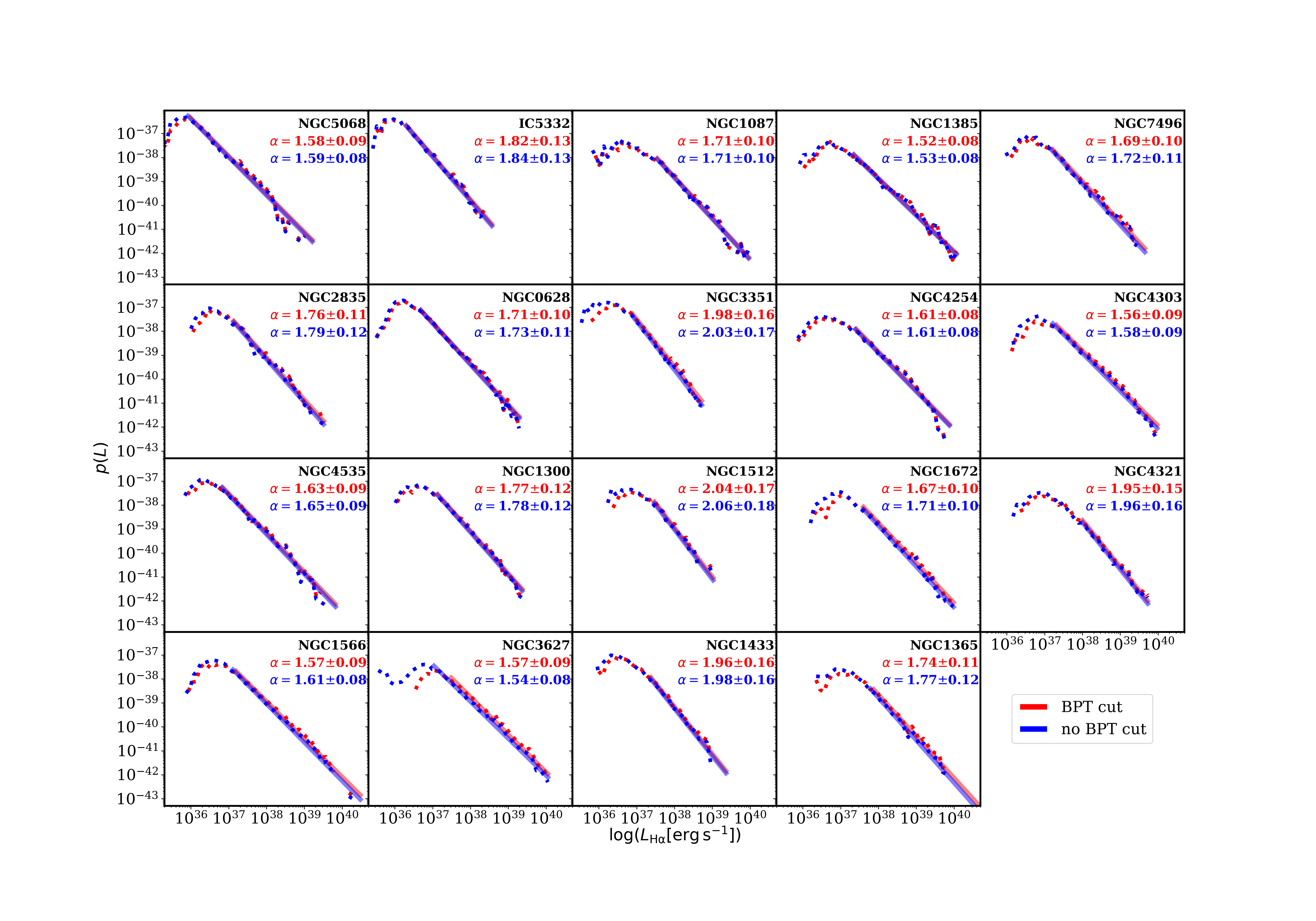}
    \caption{Fit of the LF for \HII\ regions obtained applying the BPT cut (same as in Fig.~\ref{Figure:global_LF}, red colors) and not applying the BPT cut (blue colors) on our nebulae catalogs. Galaxies are ordered according to increasing stellar mass from top left to bottom right and their names are indicated within each panel. The dashed and solid line respectively indicate the empirical LF and the best-fitting model. The models slopes are reported in the top right corner of each panel following the same color scheme. }
    \label{Figure:LF_slope_BPT}
\end{figure*}

\begin{figure}
    \centering
    \includegraphics[width=0.5\textwidth]{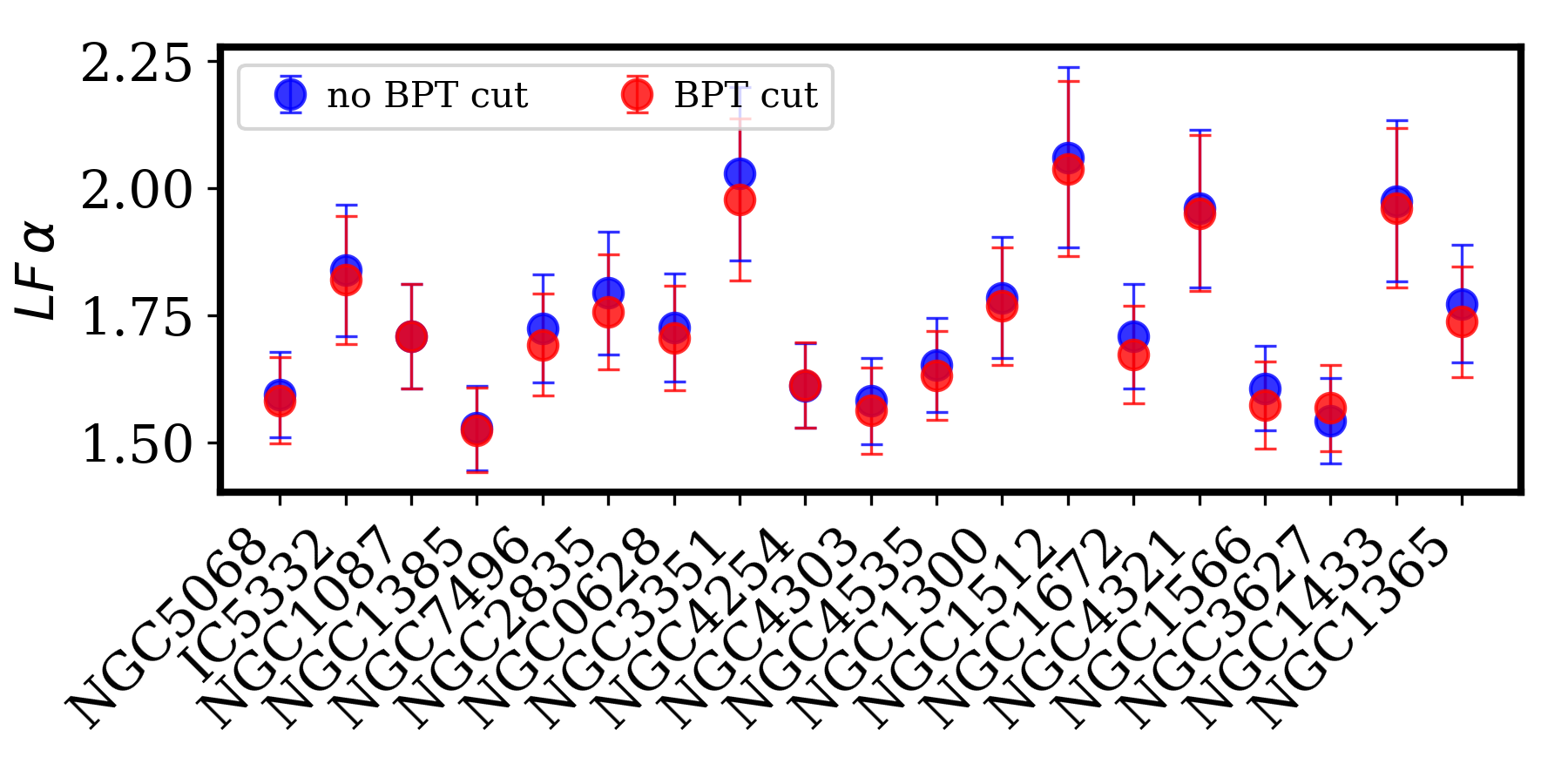}
    \caption{Slope of the LF for \HII\ regions obtained applying the BPT cut (same as in Fig.~\ref{Figure:global_LF}, red colors) and not applying the BPT cut (blue colors). Galaxies are ordered according to increasing stellar mass from left to right and their names are indicated along the abscissa.}
    \label{Figure:LF_slope_BPT_small}
\end{figure}

The most significant drop in the number of ionized nebulae entering our \HII\ region catalog is due to the cuts on the line ratios applied using semi-empirical demarcation lines in the three classical BPT diagrams (``BPT cuts'' hereafter; see Sec.~\ref{sec:The final HII region catalogs}). It is known that, depending on their gas-phase metallicity and ionization parameter, \HII\ regions can lie outside the boundaries that are commonly adopted to have a clean \HII\ region sample and avoid nebulae whose gas is ionized by mechanisms other than photoionization from young OB~stars (e.g.\ AGN photoionization/\linebreak[0]{}shocks).
As shown by Fig.~\ref{Figure:LF_slope_BPT} and Fig.~\ref{Figure:LF_slope_BPT_small}, if we drop the BPT cuts when cleaning our ionized nebula catalogs and rerun the LF fit, we find no significant variation in the measured LF slope. Remarkably, this is true also for  \object{NGC1365} for which there is clear evidence of an AGN ionization cone extending further out the central region \citep{Belfiore2021}.
As can be seen by comparing the empirical LFs, the regions that are flagged by the BPT cuts tend to populate the fainter end of the LF and thus have negligible effects on the LF shape at the bright end. In addition, we note that the LF \Lmin\ remains substantially unchanged aside from the case of \object{NGC3627}, the only strongly interacting galaxy in our sample.

In conclusion, we find that both the dust correction and selection effects do not have a significant impact on the slopes of the LFs that we measure in Sec.~\ref{sec:Results}.

\end{appendix}

\end{document}